\documentclass[11pt,letter]{article}
\usepackage[margin=1in]{geometry}

\usepackage{comment}
\usepackage{dsfont, amssymb,amsmath,amscd,latexsym, amsthm, amsxtra,amsfonts}

\newcommand{\A}{\mathcal{A}}
\newcommand{\F}{\mathscr{F}}
\newcommand{\D}{\mathrm{d}}               
\newcommand{\mP}{\mathbb{P}}
\newcommand{\define}{\triangleq}
\newcommand{\E}{\mathrm{exp}}
\newcommand{\e}{\mathbb{E}}
\newcommand{\G}{\mathcal{G}}

\newcommand{\St}{\mathbb{S}^{T_1}_t}
\newcommand{\supp}{\mathop{\text{sup}}\limits_}

\newcommand{\tauy}{\hat{\tau}_{y^*}}

\newcommand{\R}{\mathbb{R}}
\newcommand{\Q}{\mathbb{Q}}
\newcommand{\W}{\mathcal{W}}

\newcommand{\Pmeasure}{\mathbb{P}}
\newcommand{\tauop}{\hat{\tau}_y}

\usepackage{xcolor}
\usepackage{booktabs}
\usepackage{csquotes}
\usepackage{amssymb}
\usepackage{mathrsfs}
\usepackage[caption=false]{subfig}
\usepackage{float}
\usepackage{overpic}
\usepackage{enumitem}
\usepackage{appendix}
\usepackage{epstopdf}
\usepackage{verbatim}
\usepackage{mathtools}
\numberwithin{equation}{section}

\usepackage{lipsum}
\usepackage{amsfonts}
\usepackage{graphicx}
\usepackage{epstopdf}
\usepackage{algorithmic}
\ifpdf
  \DeclareGraphicsExtensions{.eps,.pdf,.png,.jpg}
\else
  \DeclareGraphicsExtensions{.eps}
\fi

\usepackage{hyperref}
\usepackage{cleveref}

\newtheorem{remark}{Remark}
\newtheorem{theorem}{Theorem}[section]

\newtheorem{lemma}[theorem]{Lemma}

\theoremstyle{definition}
\newtheorem{definition}[theorem]{Definition}

\theoremstyle{definition}

\theoremstyle{definition}
\theoremstyle{definition}
\newtheorem{assumption}{Assumption}

\Crefname{lemma}{Lemma}{Lemmas}
\Crefname{corollary}{Corollary}{Corollaries}
\Crefname{remark}{Remark}{Remarks}
\Crefname{definition}{Definition}{Definitions}
\Crefname{example}{Example}{Examples}
\Crefname{assumption}{Assumption}{Assumptions}
\Crefname{problem}{Problem}{Problems}

\allowdisplaybreaks

\title{Retirement decision with addictive habit persistence in a jump diffusion market\thanks{This work was funded by the National Natural
Science Foundation of China (Grant No.11901574, No.11871036, and No.12271290).}}

\author{Guohui Guan \thanks{Center for Applied Statistics and School of Statistics, Renmin University of China, Beijing, China. Email:
\href{mailto:guangh@ruc.edu.cn}{guangh@ruc.edu.cn}.
} 
\and Qitao Huang\thanks{Department of Mathematical Sciences, Tsinghua University, Beijing, China. Email: 
\href{maito:huangqt22@mails.tsinghua.edu.cn}{huangqt22@mails.tsinghua.edu.cn}.}
\and Zongxia Liang\thanks{Department of Mathematical Sciences, Tsinghua University, Beijing, China. Email: \href{mailto:liangzongxia@mail.tsinghua.edu.cn}{liangzongxia@mail.tsinghua.edu.cn}.}
\and Fengyi Yuan\thanks{Department of Mathematical Sciences, Tsinghua University, Beijing, China, Email:
\href{mailto:yfy19@mails.tsinghua.edu.cn}{yfy19@mails.tsinghua.edu.cn}.}}

\usepackage{amsopn}

\date{}

\begin{document}

\maketitle

\begin{abstract}
This paper investigates the optimal retirement decision, investment, and consumption strategies in a market with jump diffusion, taking into account habit persistence and stock-wage correlation. Our analysis considers multiple stocks and a finite time framework, intending to determine the retirement boundary of the ``wealth-habit-wage" triplet $(x, h, w)$. To achieve this, we use the habit reduction method and a duality approach to obtain the retirement boundary of the primal variables and feedback forms of optimal strategies. {  When dealing with the dual problem, we address technical challenges in the proof of integral equation characterization of optimal retirement boundary using a $C^1$ version of It$\hat{\rm o}$'s formula.} Our results show that when the so-called ``de facto wealth" exceeds a critical proportion of wage, an immediate retirement is the optimal choice for the agent. 
Additionally, we find that the introduction of jump risks allows for the possibility of discontinuous investment strategies within the working region, which is a novel and insightful finding. Our numerical results effectively illustrate these findings by varying the parameters.

\end{abstract}

\section{Introduction}
Retirement decision-making has always been an important topic in research fields like quantitative finance, economics and actuarial sciences. In the scenario of voluntary early retirement, it becomes a trade-off between more labour income and more leisure gains for the agent facing the retirement-decision problem. There are more and more works combining voluntary early retirement, optimal investment and optimal consumption. The agent is faced with two problems coupled together: the optimal stopping problem and the stochastic control problem (Stopping-Control Problem, or StopCP for short). In this paper, we focus on the retirement decision-making problem in a jump diffusion market, with a specific focus on voluntary early retirement and linear habit persistence. Our objective is to provide a quantitative and visual understanding of retirement boundaries, optimal consumption and investment strategies. In other words, we investigate this important problem: how do certain factors, such as wealth, the standard of life and wage, influence the retirement decision?

Our study focuses on the optimal consumption and investment portfolio problem with early retirement options, incorporating habit persistence and a finite time horizon. We consider a market with $n$ stocks exhibiting jump diffusions, and model the agent's wage before retirement as a stochastic process correlated to these stocks. This class of problems, known as the StopCP problem, has received significant attention in recent decades. To address the interaction within the StopCP framework, Karatzas and Wang (2000)\cite{karatzas2000utility} develop a dual approach. This approach transforms the problem into a pure-stopping dual problem, which is then analyzed using superharmonic characterization or variational methods. This approach has been widely used in related problems. While many studies in the StopCP field consider infinite time spans, such as Jeanblanc, Lakner, and Kadam (2004) \cite{jeanblanc2004optimal}, Choi and Koo (2005) \cite{choi2005preference}, and Liang, Peng, and Guo (2014) \cite{liang2014optimal}, our focus is specifically on the finite time horizon. This choice is motivated by the need to consider mandatory retirement requirements, which makes the dual free boundary problem not have an explicit form. Yang and Koo (2018) \cite{yang2018optimal} investigate consumption and investment problem with mandatory retirement dates and early retirement options. They provide a characterization of the threshold of wealth as a function of time, and  show that it is strictly decreasing near the mandatory retirement date. In this paper, we aim to consider a more realistic model by introducing additional factors such as habit persistence, jump diffusion market model, and labour income uncertainty.

The concept of habit persistence has been a subject of interest in finance and economics, as it indicates that past consumption behavior influences future consumption expectations. This idea has been used to address the equity premium puzzle (Sundaresan and Suresh (1989) \cite{sundaresan1989intertemporally}, Constantinides (1990) \cite{constantinides1990habit}, Campbell and Cochrane (1999) \cite{campbell1999force}) and integrated into asset pricing theory (Abel (1990) \cite{abel1990asset}, Schroder and Skiadas (2002) \cite{schroder2002isomorphism}). Recent research has extensively explored linear habit persistence in optimal investment and consumption problems (Yu (2015) \cite{yu2015utility}) as well as target-based insurance management (He, Liang, and Yuan (2020) \cite{he2020optimal}). Therefore, it is interesting to investigate how this persistence affects retirement decisions, consumption patterns, and investment strategies. However, there is limited literature on incorporating consumption habits into retirement problems due to the added complexity introduced by habit formation and the additional state variable $w$ in the dual problem. Unlike the approach taken by Yang and Koo (2018) \cite{yang2018optimal}, our dual problem is two-dimensional and requires a change of measure to reduce its dimension. In contrast to Yang and Koo (2018) \cite{yang2018optimal}, where a negative linear constant penalty is used, we incorporate habit persistence and time-dependent leisure preference into our model. This results in a two-dimensional dual problem that requires a change of measure to reduce its dimension. The derived obstacle-type free boundary problem includes a time-evolving inhomogeneous term (see (\ref{PDE})), and we develop a novel approach to overcome the challenge of time monotonicity.

Jump diffusion models, on the other hand, have been widely used in recent years to explain the asymmetric leptokurtic features of asset prices and the volatility smile. Kou (2002) \cite{kou2002jump} proposes a double exponential jump-diffusion model for option pricing, which has demonstrated empirical effectiveness. Therefore, it is natural to investigate the application of jump diffusion in our retirement problem model. Some studies have examined the strategy of an agent operating in an incomplete market driven by a diffusion process with jumps (Bellamy (2000) \cite{bellamy2000incompleteness}). Existing papers related to jump diffusion primarily focus on either pure optimal control problems or pure stopping problems due to the complexity of addressing the intricate interaction between stopping times and stochastic control in the presence of jump diffusion. Moreover, the discontinuities introduced by jump diffusion invalidate many theoretical derivations in the presence of stopping time discontinuities (see \Cref{th_dual}), and the nonlocality of function introduced by jump diffusion necessitates the use of Pham's result \cite{pham1998optimal} to establish the relation between the value function and some obstacle-type problem. These factors present considerable difficulties. Nonetheless, considering the StopCP in a jump diffusion market is necessary, as many realistic problems involve both stopping and control. To the best of our knowledge, our paper is the first to investigate the incorporation of jump diffusion into a retirement problem.

To conclude, we investigate the retirement problem within a jump diffusion market and incorporate habit persistence in a finite time horizon. In this paper, we address the challenges previously discussed and make significant contributions to the field of early retirement research. First, we 
address technically intricate stopping-control mixed problems by employing a duality approach. We establish a dual theorem in a more complicated model for general utility, especially with a stochastic wage. To ensure the equivalence of the original problem and dual problem, we prove the continuity of the stopping time in a jump diffusion market (\Cref{taucontinuous}). {  Second, due to the introduction of jump risks, the global regularity of the value function and the regularity of the free boundary are challenging, and Pham (1998) \cite{pham1998optimal} only provides the interior regularity inside the continuation region. To address this, we reveal certain ``quasi-monotonicity" (\Cref{lem_m1} and \Cref{lem_m2}), and prove the continuity of free boundary directly from the optimality of $\bar{F}$, which is a transformed value function (see \Cref{th:m_conti} and proofs in \Cref{AB}). Third, based on the global $C^1$-regularity of $\bar{F}$ and the continuity of the free boundary, we use a new change of variable formula from Cai and  De Angelis (2023) \cite{cai2023change} to obtain an integral equation characterization of the free boundary, hence the optimal retirement boundary. We emphasize that, due to the introduction of habit persistence, we do not have the monotonicity of the boundary as required by Cai and  De Angelis (2023) \cite{cai2023change}. The ``quasi-monotonicity" in \Cref{lem_m2} again plays a crucial role.}
In addition to the technical contributions, we obtain intriguing economic results. We discover that the retirement boundary is expressed by a linear relation among wealth, habit, and wage, which evolves with time. Also, we observe discontinuities in optimal strategies, not only at the boundary but also in the working region, which is due to the introduction of jump risks.

The rest of this paper is organized as follows. In \Cref{section:model}, we set up the model and characterize the admissible set of strategies as well as the allowed region of the state variables, and describe the optimization problem. \Cref{section_tr} establishes the dual theorem in a general setting. In \Cref{section:free_bound}, with a specification of utility functions, we provide rigorous results about dual retirement boundaries. In \Cref{section_primal}, we obtain the results of retirement boundary in primal variables and provide a semi-explicit form of optimal consumption and portfolio. \Cref{section_nume} includes the numerical results. \Cref{section_conclude} concludes the paper. Detailed proofs are in the appendices.

\section{The Financial Model and Optimization Problem}\label{section:model}
\subsection{The financial market and labour income} Consider a $\mathbb{R}^d$-valued Brownian motion $B = \{(B_1(t), ..., B_d(t))^T : 0 \leq t \leq T\}$ on a complete probability space $(\Omega^B, \F^B, \mP^B)$,  {and an $(n - d)$-dimensional Poisson process $N(t)=\left(N_1(t), \ldots\right.$, $\left.N_{n-d}(t)\right)^T$ on a complete probability space $(\Omega^N, \F^N, \mP^N)$. Define $(\Omega, \F, \mP)$ as the complete product probability space of these two, i.e., $\Omega = \Omega^B \times \Omega^N, \F = \sigma(\F^B \cup \F^N), \mP = \mP^B \otimes \mP^N$. $B$ and $N$ are independent on $(\Omega, \F, \mP)$. Thus there are $n$ sources of risk in our model. The time horizon is $[0, T]$.}
 {
We assume that $N_k$ admits intensity $\lambda_{k}$.  To simplify our equation, we define $\tilde{N}_j (t) = N_j(t)- \lambda_{j} t$. }

We suppose that the financial market consists of one risk-free asset (bond) $S_0 = \{ S_0(t): 0 \leq t \leq T \}$ and $n$ risky assets (stock) $S_i = \{ S_i(t): 0 \leq t \leq T  \}, 1 \leq i \leq n$. We assume that the
prices of these $n + 1$ assets are formulated as follows
\begin{align*}
&\begin{cases}
    \D S_0(t) &= r S_0(t) \D t ,\\
    S_0(0) &= s_0,  \\
\end{cases}
\\
    & \begin{cases}
             \D S_i(t) &= \mu_i S_i(t) \D t + \sum_{j = 1}^d \sigma_{i,j} S_i(t) \D B_j(t) + \sum_{j = d + 1}^n {\gamma_{i, j}S_i(t) \mathrm{d}\tilde{N}_j(t)},\\
    S_i(0) &= s_i,
     \end{cases}
\end{align*}
where $r, \mu_i, \sigma_{i, j}$ and $s_i$ are all constants. In general, the price of an asset is not negative, we assume $\gamma_{i,j} > -1$. We define as usual the exponential martingale $Z = \{Z(t):0 \leq t \leq T\}$ and the pricing density  $\xi = \{\xi(t):0 \leq t \leq T\}$ by 
\begin{align*}
\begin{cases}
	Z(t) \define &  \E \bigg\{
	- \sum_{j = 1}^{d}\theta_{0, j} B_j(t) - \frac{1}{2} t\sum_{j = 1}^{d}\theta^2_{0, j}  + t\sum_{j = d + 1}^{n} \big( \text{ln}(1 - \theta_{1, j}) + \theta_{1, j} \big) \lambda_{j}  \\
	&+ \sum_{j = d + 1}^{n}\int_{0}^{t}\text{ln}(1 - \theta_{1, j})\mathrm{d}\tilde{N}_j(t)\bigg\}, \\
	\xi (t) \define& \frac{Z(t)}{S_0(t)},
 \end{cases}
\end{align*}
where $\forall ~ 1\leq j \leq d$, $\theta_{0, j}$ and $\forall~ d\leq j \leq n$, $\theta_{1, j}$ are constants satisfying
\begin{equation}
    \sum_{j = 1}^{d} \sigma_{i, j} \theta_{0, j} +  \sum_{j = d + 1}^{n} \gamma_{i, j} \theta_{1, j} \lambda_j = \mu_i - r , \quad \text{a.s.}\quad  (s, w) \in [0, T] \times \Omega.
\end{equation}
The risk-neutral measure in the financial market is
given by $\Q$: $\left.\frac{\D \mathbb{Q}}{\D \mathbb{P}}\right|_{\F_T} \define Z(T)$. Based on Girsanov’s theorem, $B_{j, \mathbb{Q}} = \{B_{j}(t) + t \theta_{0, j} :0\leq t \leq T\}$ is a standard Brownian motion on $(\Omega, \F, \Q)$, $\tilde{N}_{j, \Q}(t) = \tilde{N}_j(t)\ + t\theta_{1, j} \lambda_j$ is a compensated Poisson process on $(\Omega, \F, \mathbb{Q})$. 

We assume that the process of wage rate $\mathcal{W} = \{\mathcal{W}(t) : 0 \leq t \leq T\}$ satisfies 
\begin{align*}
	\D \mathcal{W}(t) = \mathcal{W}(t)\big[\mu_w \D t + \sum_{j = 1}^{d} \sigma_{w, j} \D B_j(t) + \sum_{j = d + 1}^{n} \gamma_{w, j}  \mathrm{d}\tilde{N}_j(t) \big],
\end{align*}
where $\mu_w$ and $\sigma_{w, j}$ are constants representing increasing rate and volatility, respectively. Throughout this paper, we consider $\mathcal{W}^{t, w}(s) = w \frac{\mathcal{W}(s)}{\mathcal{W}(t)}$ for some wage rate at present $w > 0$.  { Similarly, we define $\xi^{t}(s) =  \frac{\xi(s)}{\xi(t)}$.} 

For the agent, there is a mandatory retirement time $T_1 \in [0, T]$. After retirement, the agent has no labour income while more leisure gains. The present value of the future labour income is given by
{ 
\begin{equation}\label{eq_b}
 	b(s) =  \e^{s, w}\big[\int_{s}^{T_1} \xi^s(u) \mathcal{W} (u) \D u \big] = q(s) w,  s \in [t, T_1].
\end{equation}}
Here  {$\e^{s, w}$ represents the conditional expectation given $\mathcal{W}(s) = w$. Similar superscripts in the subsequent context convey a similar meaning.} 
 {$q(\cdot)$ is a deterministic function, which is interpreted as the implied discount rate of labour income at some future time $s \geq t$.} $q(\cdot)$ can be expressed explicitly as follows
\begin{equation} \label{eq_q}
	q(s) = 
	\begin{cases}
		\frac{e^{\vartheta(T_1 - s)} - 1}{\vartheta}, \vartheta \neq 0, \\
		T_1 - s, \vartheta = 0,
	\end{cases}
\end{equation}
with $\vartheta \define -r + \mu_w - \sum_{j = 1}^{d}\sigma_{w, j}\theta_{0, j} - \sum_{j = d + 1}^{n} \gamma_{w, j} \theta_{1, j} \lambda_j$. 
 {
For theoretical reasons, we assume that
\begin{assumption}  \label{as_up}
	$\vartheta < 0$.
\end{assumption}
This assumption indicates that the log-increase rate of the wage cannot be too large. Intuitively, the agent may choose to work the whole life in the case of a high increasing rate of wage. Then $\tau = T_1$, and the problem is a standard stochastic control problem, which is well-studied. In Dybvig and Liu (2011) \cite{dybvig2011verification}, the same requirement is also introduced.}
\subsection{The admissible set of consumption, investment and retirement decision} 
The agent chooses the retirement time within $[0, T_1]$, and we define 
\begin{equation*}
    \St \define \{\tau : \tau \text{ is }\mathbb{F}\text{-stopping time, and } t \leq \tau \leq {T_1} \text{ almost surely} \}.
\end{equation*}
At time $t$, the agent we study will choose a stopping time $\tau \in \St$ as retirement time before
the mandatory retirement time $T_1 < T$. The agent is faced with an optimal portfolio problem
after the retirement decision is made. The agent will invest in the financial market and consume within the whole time horizon $ [0, T]$. When the consumption rate $c = \{c(s) : t \leq s \leq T\}$ and
the wealth allocated to $i$-th risky asset $\pi_i = \{\pi_i(s) : t \leq s \leq T \}$ are determined, the wealth process satisfies the following stochastic differential equation (abbr. SDE)
 {
\begin{equation}\label{eq_x}
	\begin{cases}
		\begin{aligned}
			\D X(s)=  &\big[rX(s)  - c(s)  + \mathcal{W}(s) \chi_{\{0 \leq s \leq \tau \}} \big] \D s \\
			&+\sum_{i = 1}^{n} \pi_i(s)\big\{ \sum_{j = 1}^{d}\sigma_{i, j} \D B_{Q, j}(s) + \sum_{j = d + 1}^{n}\gamma_{i, j}\mathrm{d}\tilde{N}_{Q, j}(s)\big\},\\	   
		\end{aligned}\\
		X(t) = x,
	\end{cases}
\end{equation}
}
where $\chi$ is the indicator function.

The agent aims to maintain the standard of living during both the working and retirement periods, and exhibits habit persistence. To capture this, we formulate the standard of living, or the consumption habit level, as a weighted average of past consumption levels, with a decaying term. To be precise,
we define the habit level $h^{t, c, h}(s)$ for $s  \in [t, T]$ and $h > 0$ as 
\begin{equation}\label{eq_h}
	\begin{cases}
		\D h^{t,c,h}(s) = [\alpha c(s) - \beta h^{t,c,h}(s)] \D s,\\
		h^{t,c,h}(t) = h,
	\end{cases}
\end{equation}
where $\alpha$ and $\beta$ are positive constants. The goal of the agent is to determine the optimal investment and consumption strategies and the retirement time. The strategies are restricted in
the admissible set $\mathcal{A}(t, x, h, w)$ for $(t, x, h, w) \in [0, T] \times \R \times \R_+^2$ defined below. Throughout this
paper, whenever there is no confusion, we will drop all the superscripts.
{ 
\setlist[enumerate,1]{left=15pt}
\begin{definition}\label{def}
	$(\tau, c, \pi) \in \mathcal{A}(t, x, h, w)$ if and only if 
	\begin{enumerate}[label=(\arabic*), ref=\arabic*]
	    \item $\tau \in \St$.
	    \item $c$ and $\pi$ are predictable processes with respect to $\mathbb{F}$.
	    \item SDE \cref{eq_x} has a unique solution, denoted by $ X^{\tau, c, \pi, t, x}$ with $X^{\tau, c, \pi, t, x}(T) \geq 0 \text{ a.s.}$.
	    \item  \begin{equation}\label{eq_ad}
 c(s) - h(s) \geq 0 \quad a.s.,
 \end{equation}
 which indicates that the consumption is greater than or equal to the habit consumption.
 \item \begin{equation}
	    \int_{t}^{T} \pi^2(s) + c(s) \D s < \infty \quad a.s..
	\end{equation}
	\end{enumerate}
\end{definition}}
\begin{remark}
    The requirement in (\ref{eq_ad}) is called ``addictive" as the consumption is restricted to be always beyond the standard of living, see Yu (2015) \cite{yu2015utility}. Define $\hat{c} \define c - h$. Obviously, $h(\cdot)$ can also be derived from $\hat{c}$
    \begin{equation}
        \D h(t) = (\alpha \hat{c}(t) + (\alpha - \beta) h(t)) \D t.
    \end{equation}
    The agent searches for the optimal strategies in the admissible set. We define the set of all addictive consumption plans, i.e., all $c = \{c(s) : t \leq s \leq T\}$ that satisfy (2), (3), (4), and (5) of Definition \ref{def}, as $\mathcal{U}^T_t$. Also, define the set of $\pi = \{\pi(s) : t \leq s \leq T\}$ that satisfy (2), (3), and (5) as $\mathcal{V}^T_t$.
\end{remark}
\subsection{Utility functions and preference change}
After retirement, the agent loses labor income but gains more leisure time. To distinguish between preferences during the working and retirement periods, we define the preferences as follows: Let $U_i(t, x)$, $i = 1, 2$, be strictly concave utility functions of $x$ that satisfy the following assumption.
\begin{assumption}\label{as:utility_function}
	
	(1) dom $U_i = [0, T] \times \mathbb{R}_+$ and $U_i \in C^{1,2}([0, T] \times \mathbb{R}_+)$.
	
	(2) $\lim_{x \to \infty} \sup_{t\in[0, T]}\partial_x U_i(t, x) = 0, \lim_{x \to 0+} \inf_{t\in[0, T]} \partial_x U_i(t, x) = + \infty \text{ for } i = 1, 2$.
 
	{ 
	(3) There exist constants $K > 0$ and $C > 0$ such that 
 \begin{equation*}
     I_{U_i}(t, x) \leq C (1 + x^{-K}).
 \end{equation*}
	Here $I_f(t, x)$ satisfies the following equation for any strictly concave function $f$ here and throughout this paper
	\begin{align*}
	    \frac{\partial f}{\partial x} (t, I_f(t, x)) = x.
	\end{align*}
 
	(4)\quad $U_1(t, x) < U_2(t, x)$ for any $(t, x) \in [0, T] \times \mathbb{R}_+$.}
\end{assumption}
 In this paper, $U_1$ represents the utility function before retirement, while $U_2$ represents the utility function after retirement. It can be verified that many of the most famous and well-studied utility functions in the literature listed above satisfy the assumption made here.

\subsection{The allowed region of state variables}\label{sb_allow_region}
In this paper, we consider a model that incorporates stochastic labor income and habit formation. As a result, our system has three state variables: current wealth, habit level, and labor income. Before presenting and solving our optimization problem, we first discuss and define the allowed region of the state triplet $(x, h, w)$. Define
\begin{equation} \label{df_G_t}
	\mathcal{G}_t \define \{(x, h, w) \in \R \times \R_+^2 : \mathcal{A}(t, x, h, w) \neq \varnothing\}.
\end{equation}
$\G_t$ is called the allowed region of state variables. For a fixed level of wealth $w$, we denote $\G_t^w = \{(x, h) : (x, h, w) \in \G_t\}.$ After the strategies are chosen, $(X(s), h(s), \W(s))$ should be contained in $\G_s$ for $s \geq t$. 
Within this allowed region, we formulate the preferences of the agent with respect to retirement choice, consumption, and investment strategies. The agent must search for the optimal strategies that maximize the preference of the whole life. This optimization problem involves both an optimal stopping problem and a stochastic control problem. To maintain the standard of living, the agent cares about consumption over habit. We define the objective function for the agent as follows
 {
\begin{align*}
	&J(t, x , h, w; \tau , c, \pi)\define \\
	&\begin{cases}
		\e^{t, x, h, w}\big [\int_{t}^{\tau}\!\! &\!\!e^{-\rho(s - t)} U_1\big(s , c(s) - h(s)\big)\D s  +\int_{\tau}^{T} e^{-\rho(s - t)} U_2\big(s , c(s)\!- \!h(s)\big) \big] \D s,\\
		&\text{if }\e^{t, x, h, w}\left[\int_{t}^{\tau} U_1^-\big(s, c(s)-h(s)\big)\D s+\int_{t}^{\tau} U_2^-\big(s, c(s)\!-\!h(s)\big) \D s\right]< \infty,\\
		-\infty, &\text{otherwise},
	\end{cases}
\end{align*}}
where $U^-_{i} = \max\{-U_{i}, 0\}$ for $i = 1, 2$. Then the agent primarily wants to solve the following optimization
problem
\begin{equation}\label{stop-CP}
	V(t, x, h, w) = \supp{(\tau, c, \pi) \in \mathcal{A}(t, x, h, w)}J(t,x, h, w; \tau ,c, \pi).
\end{equation}
Problem (\ref{stop-CP}) involves both optimal stopping and stochastic control, which brings mathematical difficulties. To address this, we first transform Problem (\ref{stop-CP}) into a regular form in \Cref{section_tr}. Next, we establish a dual relation between the primal value function and the value function of a pure optimal stopping problem. The value function of the optimal stopping problem will be characterized by an obstacle-type free boundary problem in \Cref{section:free_bound}. The optimal retirement time can then be determined by the first entrance time of retirement (stopping) region of the state triplet $(x, h, w)$. The dual relation relies on a novel habit reduction method and the continuity of the optimal stopping time.
\section{Transformation of the Original Problem} \label{section_tr} { 
In this section, our objective is to convert our problem into a pure stopping problem. In \Cref{section_hr}, we employ a novel habit reduction method (\Cref{df_habit}) to reduce the variable $h$ and reshape our problem into a new form concerning a de facto wealth process $X_F$ (\Cref{sde:x_f}), a de facto consumption process $c_F$ and the portfolio process $\pi$. In \Cref{sec:regular_form}, we claim that the post-retirement value (\Cref{eq_U}) is a function of the retirement time $t$ and de facto wealth $x_F$. Based on this property, we can prove the replication property of Lemma \ref{lm_rep} in \Cref{sec:replication}. We reduce our problem to an equivalent problem with only one control variable $c_F$ and an objective function that depends solely on the value of $c_F(t)$ at each time $t \in [0,T]$. Finally, we establish the dual pure stopping problem (\ref{stop1}) in \Cref{sec:dual_problem}.}

 \subsection{Habit-reduction method}\label{section_hr}  {To capture the standard of living, we introduce habit persistence as in \Cref{eq_h}, which increases the dimension of our financial system. The agent's preferences also depend on the past consumption path. However, based on the isomorphism idea proposed by Schroder and Skiadas (2002) \cite{schroder2002isomorphism}, we can reduce the original problem to one that does not involve habit formation or path-dependent properties.}

\begin{definition}\label{df_habit}
	(1) \quad $\forall t \geq 0$, define de facto wealth $X_F(\cdot)$ as
	\begin{equation}\label{def_p}
		X_F(t) \define X(t) - h(t)p^T(t),
	\end{equation}
	where $p^T(t) = \frac{1 - e^{-(r + \beta - \alpha)(T - t)}}{r + \beta - \alpha}$ represents cost of subsistence consumption per unit of habit.
	
	(2) \quad $\forall t \geq 0$, define the de facto consumption $c_F(\cdot)$ as
	\begin{equation}
		c_F(t) \define a(t)(c(t) - h(t)),
	\end{equation}
	where $a(t) = \frac{r + \beta}{r + \beta - \alpha} - \frac{\alpha}{r + \beta - \alpha}e^{-(r + \beta - \alpha)(T - t)}$.
	
	(3) \quad $\forall t \geq 0$, define the de facto utility $U_{F, i}(\cdot)$ as
	\begin{equation}\label{new_utility}
		U_{F, i}(t, c_F) \define U_i\left(t, \frac{c_F}{a(t)}\right) = U_i(t, c(t) - h(t)).
	\end{equation}
\end{definition}

Then the de facto wealth process $X_F$ satisfying the following SDE{ 
\begin{equation}\label{sde:x_f}
	\begin{cases}
		\begin{aligned}
			\D X_F(s) &= \bigg[rX_F(s)  - c_F(s)  + \mathcal{W}(s) \chi_{\{0 \leq s \leq \tau \}} \bigg]\D s \\
			&+\sum_{i = 1}^{n} \pi_i(s)\bigg\{ \sum_{j = 1}^{d}\sigma_{i, j} \D B_{Q, j}(s) + \sum_{j = d + 1}^{n} \D N_j(s)\bigg\},\\	            
		\end{aligned}\\
		X_F(t) =X(t)- h(t)p^T(t).
	\end{cases}
\end{equation}}
The processes $X_F(\cdot)$ and $X(\cdot)$ in \Cref{eq_x} have the same form, except for the inclusion of habit persistence $h(\cdot)$. Therefore, we have shown that for each admissible wealth process $X(\cdot)$ with consumption process $c(\cdot)$ and habit process $h(\cdot)$, there exists an admissible de facto wealth process $X_F(\cdot)$ with the de facto consumption process $c_F(\cdot)$.

Additionally, we can derive the wealth process $X(\cdot)$ with consumption process $c(\cdot)$ and habit process $h(\cdot)$ from the admissible de facto wealth process $X_F(\cdot)$ with the de facto consumption $c_F(\cdot)$ and initial habit level $h(0) = h_0$ by the following expressions
\begin{equation}
    \begin{cases}
        \D h(s) = \big\{\alpha\frac{c_F(s)}{a(s)} + (\alpha - \beta)  h(s) \big \}\D s  ,\\
        h(0) = h_0  ,\\
        X(t) = X_F(t) + h(t)p^T(t) , \\
        c(t) = \frac{c_F(t)}{a(t)} + h(t).
    \end{cases}
\end{equation}

Therefore, we have shown that the correspondence above is bijective, which means that we can focus on the admissible de facto wealth process $X_F(\cdot)$ with de facto consumption process $c_F(\cdot)$ and initial habit level $h(0)$ rather than the wealth process $X(\cdot)$ with consumption process $c(\cdot)$ and habit process $h(\cdot)$. We denote the corresponding admissible sets by $\A_F$, $\mathcal{U}^T_{F, t}$ and $\mathcal{V}^{T}_{F, t}$, and the corresponding allowed regions of state variables by $\G_{F, t}$ and $\G^{w}_{F,t}$, the corresponding value function $V_F$.


\subsection{Regular form of Problem (\ref{stop-CP})}\label{sec:regular_form}
To represent the integrated utility after retirement in a simpler form, we first consider an auxiliary problem, which is a standard optimal consumption-portfolio problem with habit formation but without any labor income or retirement choice.  {After retirement, the post-retirement value function is defined as follows}
 {\begin{equation} \label{eq_U}
	V_{F, p}(t, x_F) \define \supp{(\tau, c_F, \pi) \in \A_F(t, x_F, w)} \e^{t, x_F, w} \left[ \int_{t}^{T} e^{-\rho(s - t)} U_{F, 2}(s, c_F(s)) \D s\right].
\end{equation}}
If there is no voluntary retirement choice (or if the retirement time is chosen before $t$) and no labor income, Problem (\ref{stop-CP}) reduces to Problem (\ref{eq_U}). By using the post-retirement value function $V_{F, p}(t, x_F)$, we transform the original problem into a regular form that does not involve strategies after retirement. The following results establish the equivalence of these two problems and illustrate the connection between $V_{F, p}(t, x_F)$ and strategies after retirement. Similar arguments can be found in Park and Jeon (2019) \cite{park2019finite}, Liang, Peng, and Guo (2014) \cite{liang2014optimal}, Choi, Shim, and Shin (2008) \cite{choi2008optimal}, and Jeanblanc, Lakner, and Kadam (2004) \cite{jeanblanc2004optimal}. The proofs are similar to those in the previously mentioned literature and are omitted here.
{ {
\begin{theorem} \label{regularform}
Suppose $(x_F, w) \in \G_{F,t}, t \in [0, T_1]$. Denote $V_{F, p}^- = \max \{ - V_{F, p}, 0\}$. Then

    (1)\quad For any $\tau \in \St$ and $\F_\tau$-measurable random variable $D$ such that $D(\omega)\in \R_+$ for almost sure $\omega \in \Omega$ as well as
    \begin{equation*}
        \e^{t, x_F, w} \big [ V_{F, p}^-(\tau, D) \big ] < \infty,
    \end{equation*}
there exists a strategy $(c_F, \pi) \in \mathcal{U}_{F, t}^{T} \times \mathcal{V}_{F,t}^{T}$ such that the solution to SDE (\ref{eq_x}) with initial condition $X_F(\tau) = D$ satisfies
\begin{equation*}
    X_F^{(c_F, \pi)}(s) \geq 0,\quad \forall s \in [\tau, T],
\end{equation*}
and 
\begin{equation} \label{eq_u2ub}
    \e^{t, x_F, w} \big[ \int_{\tau}^{T} e^{-\rho(s - t)} U_{F, 2}(s, c_F(s)) \D s \big] = \e ^{t, x_F, w}\big[ e^{-\rho(\tau - t)} V_{F, p}(\tau, D)\big].
\end{equation}

(2)\quad For $V$ defined in Problem (\ref{stop-CP}), we have
\begin{equation}\label{eq_vu}\begin{aligned}
    \hspace{-5mm}V_F(t, x_F, w)  = \sup_{(\tau, c_F, \pi) \in \A_F(t, x_F, w)}\e^{t, x_F, w} \bigg[ 
    \begin{array}{c}
       \displaystyle \int_{t}^{\tau} e^{-\rho(s - t)} U_{F, 1}\big(s, c_F(s)\big) \D s  \\
         \displaystyle   + e^{-\rho(\tau - t)} V_{F, p}\big(\tau, X_F(\tau)\big)
          \\
    \end{array}\bigg]&.
\end{aligned}
\end{equation}
\end{theorem}
}}
\Cref{eq_u2ub} establishes the connection between the post retirement value function $V_{F, p}(t, x_F)$ and strategies after retirement. Intuitively, the agent regards the preference after retirement as a standard stochastic control problem due to the Markovian property.  \Cref{eq_vu} shows that the original problem is equivalent to the regular form as presented in Karatzas and Wang (2000) \cite{karatzas2000utility}. Problem (\ref{stop-CP}) with retirement choice and strategies throughout the entire life is equivalent to the optimization problem with retirement choice, strategies before retirement, and claims at retirement. Although we have obtained a regular form, the optimization problem is still very complex compared to Karatzas and Wang (2000) \cite{karatzas2000utility} due to the inclusion of two state variables $(x_F, w)$ and the involvement of habit formation. To solve Problem (\ref{eq_vu}) and derive the optimal strategies for the agent, we use a habit reduction method and a dual method.

\subsection{The budget constraints and replication}\label{sec:replication}
Before applying the dual method, we present the budget constraints in our financial system. We will prove that the budget constraint is a sufficient and necessary condition for our problem. Based on the standard local martingale argument and the optional stopping theorem in Karatzas and Wang (2000) \cite{karatzas2000utility}, we obtain the following budget constraint in the following lemma.
\begin{lemma}\label{lm_comp}
    For any  random variable $D$ at retirement time $\tau$ with corresponding consumption process $c_F(\cdot)$, we have the following constraint
     {\begin{equation}\label{ineq_rep}
  	\e_{\Q}^{t, x_F, w} \bigg[\underbracket{e^{-r(\tau - t)}D}_{\text{wealth left at retirement}} + \int_{t}^{\tau} \underbracket{e^{-r(u - t)} c_F(u)  \D u}_{\text{wealth to consume}}\bigg] \leq \underbracket{x_F}_{\text{current wealth}} + \underbracket{b(t)}_{\text{future labour income}}
.  \end{equation}}
\end{lemma}
Here $\e_{\Q}[\cdot]$ represents the expectation under measure $\Q$. Conversely, using the method presented in Bardhan and Chao (1995) \cite{bardhan1995martingale}, we can prove the following results.
\begin{lemma}\label{lm_rep}
    For any $(\tau, c_F) \in \St \times \mathcal{U}_{F, t}^T$ satisfying
     {
 \begin{equation}\label{eq_rep}
	\e_{\Q}^{t, x_F, w}\big[e^{-r(\tau - t)}(D + b(\tau)) + \int_{t}^{\tau} c_F(u) e^{-r(u - t)} \D u\big] = x_F + b(t),
\end{equation}   }
there exists a  portfolio $\pi$ replicating wealth $D$ at retirement time $\tau$ and consumption process $c(\cdot)$.
\end{lemma}
\begin{proof}
We have the equation
\begin{align*}
     \D \big(X_F(s) + b(s)\big) &=  \bigg[r\big(X_F(s) + b(s) \big) - c_F(s)\bigg] \D s \\
     &+ \sum_{i = 1}^{n} (\pi_i(s) + q(t)) \bigg\{ \sum_{j = 1}^{d}\sigma_{i, j} \D B_{Q, j}(s) + \sum_{j = d + 1}^{n}\gamma_{i, j}\D \tilde{N}_{Q, j}(s)\bigg\},
\end{align*}
which means that we can consider $X_F(\cdot) + b(\cdot)$ as the fictitious wealth process before retirement. This wealth process contains a fictitious portfolio process $\pi(\cdot) + q(\cdot)$ and consumption process $c_F(\cdot)$, but no wage process. After this transformation, this lemma coincides with Lemma 4.2 in Bardhan and Chao (1995) \cite{bardhan1995martingale}.
\end{proof} 
In the proof of this lemma, the wage rate is reduced by employing the expression {$X_F(\cdot)+q(\cdot)w$. Although this reduction method bears similarities to the elimination of consumption habits, there are substantial differences between them. The impact of consumption habits on wealth remains independent of retirement, meaning that regardless of the future retirement time, an agent with consumption habits $h$ at time $t$ always needs to allocate a wealth of $p^T(t)h$ to maintain the consumption habits. However, the time of retirement directly influences the agent's expected wage income. Thus, in the original problem, we cannot treat $X_F(\cdot)+q(\cdot)w$ as fictitious wealth to replace $X_F(\cdot)$ in the same manner as with consumption habits.}

 { \Cref{lm_rep} shows that we only need to search for the optimal $c_F(\cdot)$ first. The optimal investment strategy $\pi$ can be derived by replication. Additionally, \Cref{lm_rep} also presents the explicit form of allowed regions}.
{
\begin{lemma}\label{G_Ft_def}
We have $\G_{F,t} = \big\{(x_F, w)|x_F + q(t) w \geq 0\big\}$ and $\G^w_{F, t} = [-q(t)w, \infty)$. 
\end{lemma}}
\begin{proof}
Using Lemma 4.2 in Bardhan and Chao (1995) \cite{bardhan1995martingale}, for any $(\tau, c_F, \pi) \in \mathcal{A}_F$, we have $X_F(\tau) \geq 0$ a.s..

On the one hand, for any $(x_F, w) \in \G_{F, t}$, using \Cref{lm_comp}, we have
\begin{equation*}
    x_F + b(t) \geq \e_{\Q}^{t, x_F, w}\left[e^{-r(\tau - t)}X(\tau) + \int_{t}^{\tau} c_F(u) e^{-r(u - t)} \D u\right]\geq 0.
\end{equation*}
Here we use $X(\tau) \geq 0$ and $c_F(u) \geq 0$ a.s.. 

On the other hand, for any $(x_F, w) \in \big\{(x_F, w)|x_F + q(t) w \geq 0\big\}$, consumption process $c_F(\cdot)$, retirement time $T_1$ and retirement wealth $D$, there exists a $\lambda$ such that  
\begin{equation*}
    x_F + b(t) = \e_{\Q}^{t, x_F, w}\left[e^{-r(\tau - t)}\lambda D + \int_{t}^{T_1} \lambda c_F(u) e^{-r(u - t)} \D u\right].
\end{equation*}
Using \Cref{lm_rep}, there exists a  portfolio $\pi$ replicating wealth $\lambda D$ at retirement time $T_1$ and consumption process $\lambda c(\cdot)$. Therefore, $(x_F, w) \in \G_{F, t}$. Thus, $\G_{F,t} = \big\{(x_F, w)|x_F + q(t) w \geq 0\big\}$ and $\G^w_{F, t} = [-q(t)w, \infty)$.
\end{proof} 
\begin{remark}
     {The constraint $x_F + q(t) w \geq 0$ means that the agent is allowed to borrow money and holds a negative de facto wealth level, while the absolute amount of debt level he holds cannot exceed the total value of future labour income. Once the agent retires, he is prohibited from holding negative de facto wealth.}
\end{remark}
\subsection{The dual problem}\label{sec:dual_problem}

Problem (\ref{eq_vu}) is a combination of optimal stopping and stochastic control problems. To separate the stopping time and stochastic controls, we first define the dual functions of the ``running utility" $U_{F, 1}$ and ``terminal utility" $V_{F,p}$ (which is the value function of the post-retirement problem defined in \Cref{eq_U}), respectively. By using the dual approach, Problem (\ref{eq_vu}) can be transformed into a pure optimal stopping problem ; see Park and Jeon (2019) \cite{park2019finite}, Yang and Koo (2018) \cite{yang2018optimal}, Liang, Peng, and Guo (2014) \cite{liang2014optimal}, Choi, Shim, and Shin (2008) \cite{choi2008optimal}, and Karatzas and Wang (2000) \cite{karatzas2000utility} for such a transformation. 

Define the dual variable $Y^{t, y}(s) = y e^{\rho (s - t)} \xi^t(s)$ for $y > 0$. In some expressions, we will omit superscripts and use $Y(s)$ for simplification. Using \Cref{lm_rep}, we have that
{     \!\!\!\begin{align*}
    	V_F(t, x_F,& w) - y\big(x_F + b(t)\big) \\
    	 \leq &\supp{(\tau,c_F, \pi) \in \mathcal{A}_F} \e^{t, x_F, w}\left[e^{-\rho(\tau - t)}V_{F, p}\big(\tau, X_F(\tau)\big) + \int_{t}^{\tau} e^{-\rho(u - t)} U_{F, 1}\big(u, c_F(u)\big) \D u\right] \\
    	& -y\e^{t, x_F, w }\bigg[\xi^t(\tau)\big(X_F(\tau) + b(\tau)\big) + \int_{t}^{\tau} c_F(u)\xi^t(u) \D u\bigg] \\
    	 = &\supp{(\tau, c_F, \pi)\in \A_F} \e^{t, x_F, y, w}\left[e^{-\rho(\tau - t)}V_{F, p}\big(\tau, X_F(\tau)\big) - e^{-\rho(\tau - t)} Y(\tau)\big(X_F(\tau) + b(\tau)\big)\right]\\
    	&+\e^{t, x_F, y, w}\left[\int_{t}^{\tau} e^{-\rho(u - t)}\bigg(U_{F, 1}\big(u, c_F(u)\big) - Y(u) c_F(u)\bigg) \D u\right] \\
    	\leq &\supp{\tau \in \St} \e^{t, y, w}\left[e^{-\rho(\tau - t)}\left( V^{\mathrm{d}}\big(\tau, Y(\tau)\big) \!\! -\!\!  Y(\tau) q(\tau)\mathcal{W}(\tau)\right)\!\! + \int_{t}^{\tau}e^{-\rho(u - t)} U^{\mathrm{d}}_{1}\big(u, Y(u)\big) \D u\right],\label{dual_inq}
    \end{align*}
 holds with equality if and only if
\begin{equation}\label{eq:duality_eq_cond}
    X^{t, x_F, \pi, c_F}_F(\tau)=I_{V_{F, p}}\left(Y^{t, y}(\tau)\right), \quad \text {and} \quad c_F(u)=I_{U_{F,1}}\left(Y^{t, y}(u)\right), \forall ~ 0 \leq t \leq \tau.
\end{equation}}
Here,
 {
\begin{align}
	& V^{\mathrm{d}}\big(t, y\big) \define  \supp{x_F \in \mathcal{G}^w_t} \big\{V_{F, p}(t, x_F) - y x_F\big\} , \label{utility2_dual}\\ 
	&U^{\mathrm{d}}_1(t, y) \define \supp{c_F \geq 0}\big\{{U_{F, 1}(t, c_F) - y c_F}\big\},\label{utility1_dual}
\end{align}
}
and the superscript $^{\mathrm{d}}$ represents ``duality". 

As such, if we introduce the following optimal stopping problem
{ 
\begin{equation}\label{stop1}
	\begin{aligned}
	W(t, y, w) \define \supp{\tau \in \St} \e^{t, y, w}\bigg[\begin{array}{c}
	      \displaystyle e^{-\rho(\tau - t)} \bigg(V^{\mathrm{d}}\big(\tau, Y(\tau)\big) -  Y(\tau) q(\tau)\mathcal{W}(\tau)\bigg) \\
	     \displaystyle + \int_{t}^{\tau}e^{-\rho(u - t)} U^{\mathrm{d}}_{1}\big(u, Y(u)\big) \D u
	\end{array}\bigg],
	\end{aligned} 
\end{equation} }
 we have $\forall ~ t \in [0, T_1]$, $x_F \in \R_+$, $y > 0$ and $\omega > 0$, 
\begin{equation}\label{V_W_ineq}
	V_F(t,x_F, w) - y\big(x_F + q(t)w\big) \leq W(t, y, w).
\end{equation}
Thus, if we establish the relationship between Problem (\ref{eq_vu}) and the dual problem (\ref{stop1}), the dual problem contains no controls and is a pure optimal stopping problem, which is easier to solve. We have the following dual theorem to guarantee the equivalence of these two problems.

\begin{theorem}\label{th_dual}
    Let $t \in [0, T_1)$ and $(x_F, w) \in  \G_t$ be fixed. Suppose that  the optimal stopping time $\hat{\tau}_y = \hat{\tau}_{t, y, w}$ of Problem (\ref{eq_vu}) can be chosen such that $\hat{\tau}_{y + \epsilon} \to \hat{\tau}_{y}$ almost surely as $\epsilon \to 0 \pm$ and satisfying 
    { 
    \begin{equation}\label{eq:e_b_limit}
    \lim_{y \to \infty}\e \bigg[b(\hat{\tau}_{y}) \xi (\hat{\tau}_{y})\bigg] = 0.
\end{equation} }
The value function of Problem (\ref{stop-CP}) satisfies 
 	\begin{equation}\label{dual_v_w}
		V_F(t,x_F,w) = \inf_{y > 0}\big\{ W(t, y, w) + y x_F + y q(t)w
		\big\}.
	\end{equation}   
\end{theorem}
\begin{proof}
 See \Cref{append_pf_th_dual}.
\end{proof} 
\begin{remark}
    The post-retirement problem does not have the stochastic wage process and the retirement decision and is well-studied in previous literature. We directly give the optimal investment and consumption strategy after retirement in \Cref{th_op_st}. 
\end{remark}
\Cref{th_dual} shows that only the continuity of $\hat{\tau}_y$ is needed to establish the dual relation, which remains true under some mild regularities of the boundary (see the proof of \Cref{lm_bd_pr}).

To conclude this section, we present a result about $V^{\mathrm{d}}$, the dual function of the post-retirement value function. In the later part of this paper, it contributes to simplifying the problem.

\begin{theorem}\label{th_dual_retire}
	
	\begin{equation}\label{v_bar}
		V^{\mathrm{d}}(t, y) =  \e^{t, y}\left[\int_{t}^{T} e^{-\rho(u - t)} U^{\mathrm{d}}_2(u, Y(u)) \D u\right],
	\end{equation}
where
	\begin{equation}
		U^{\mathrm{d}}_2(t, y) = \sup_{c_F \geq 0}\{ U_{F, 2}(t, c_F) - y c_F \}.
	\end{equation}
\end{theorem}

\begin{proof}
   The proof is the same with Theorem 8.18 in Karatzas and Shreve (2012) \cite{karatzas2012brownian} and we omit it here. 
\end{proof}

\section{Retirement Boundary in Terms of Dual Variables}\label{section:free_bound}\Cref{section_tr} presents the theoretical results of the StopCP.  {In this section, our two main objectives are to obtain the continuity of stopping time (see \Cref{th:m_conti}) and an integral equation (\ref{eq:integral}) that characterizes the optimal stopping boundary. In \Cref{sec:change_measure}, we reduce the two-dimensional state variables $(y, w)$ into one-dimensional variable $\bar{y}$ under the probability measure $\tilde{\mathbb{P}}$.  We transform a two-dimensional stopping problem (\ref{stop1}) to obtain a one-dimensional stopping problem (\ref{stop3}) with respect to $\bar{y}$. In \Cref {sub:tran}, to simplify our problem, we define a function $F$ to transform the obstacle to be 0 in the resulting obstacle problem.  To fit the Lipschitz condition in Pham (1998) \cite{pham1998optimal}, we define a new process $M$ and get a new problem (\ref{stop5}). Therefore, we can use the results in Pham (1998) \cite{pham1998optimal} to obtain the continuity of stopping boundary in \Cref{subsection:e_and_u}. The continuity of the stopping boundary leads to the continuity of stopping time, and the continuity of stopping time in turn gives the global regularity of $F$ (\Cref{th:F_c_1}). Finally, we obtain the integral equation using a new generalized It$\hat{\rm o}$'s formula and leveraging on the global regularity of $F$ (\Cref{eq:integral}).}

\subsection{ {Utility functions}}
To obtain the forms of the retirement boundary and optimal strategies, we specify the preferences using the CRRA utility function and a time-dependent leisure model.  Inspired by Kolodinsky (1990) \cite{kolodinsky1990time} and equation (3) in Farhi and Panageas (2007) \cite{farhi2007saving}, we consider leisure time as a direct source of utility. We consider the initial preference as follows
\begin{equation}\label{utility_f}
	U(t, c) = \frac{1}{1 - \gamma} (c \cdot l ^{K(t)}) ^ {1 - \gamma},
\end{equation}
where $\gamma > 0, \gamma \neq 1$ represents the risk aversion of the agent, $l$ represents the leisure time, and $K(\cdot)$ is a smooth function that represents the leisure preference. We use this utility function to emphasize the direct utility that leisure time can bring to individuals and the substitutability of leisure time and consumption. The pre-retirement and post-retirement utility functions are then specified as follows
\begin{align}
	U_1(t, c) &= \frac{1}{1 - \gamma} (c \cdot l ^{K(t)}_1) ^ {1 - \gamma},\label{utility_f_1}\\
	U_2(t, c) &= \frac{1}{1 - \gamma} (c \cdot l ^{K(t)}_2) ^ {1 - \gamma},\label{utility_f_2}
\end{align}
where $l_1$ and $l_2$ represent a constant amount of leisure time before and after retirement. 
{ 
\begin{assumption}
$l_1 < l_2$.
\end{assumption}}
This assumption implies that the agent has more leisure time after retirement. Otherwise, the agent would not choose to retire.
In \Cref{section_hr}, we reduce the habit persistence and focus on the de facto wealth. Thus, the utility functions we consider should be the de facto utility functions  {defined in \Cref{new_utility}} as follows
\begin{align}
	U_{F, 1}(t, c_F) = a(t)^{\gamma - 1}\frac{c_F^{1 - \gamma}}{1 - \gamma} l_1^{(1 - \gamma)K(t)},\label{utility_ff}\\
	U_{F, 2}(t, c_F) = a(t)^{\gamma - 1} \frac{c_F^{1 - \gamma}}{1 - \gamma} l_2^{(1 - \gamma)K(t)}.\label{utility_ff2}
\end{align}
Their dual functions  {defined in \Cref{utility1_dual}} are given by
\begin{align}
	U^{\mathrm{d}}_1(t, y) = \frac{(a(t) y)^{-\gamma^{*}}}{\gamma^{*}} l_1^{\gamma ^ * K(t)},\label{dual_utility_ff}\\
	U^{\mathrm{d}}_2(t, y) = \frac{(a(t) y)^{-\gamma^{*}}}{\gamma^{*}} l_2^{\gamma ^ * K(t)}, \label{dual_utility_ff2}
\end{align}
where $\gamma^{*} = \frac{1 - \gamma}{\gamma}$.
\subsection{ {Change of measure}}\label{sec:change_measure}
To characterize the retirement boundary in terms of dual variables, we need to solve the dual optimal stopping problem (\ref{stop1}) without controls. However, Problem (\ref{stop1}) is two-dimensional, and the infinitesimal operator of Markov process $(Y(\cdot), w(\cdot))$ is degenerate. This brings twofold difficulties. First, establishing the existence, uniqueness, and regularity of the associated differential equations and obstacle-type free boundary problems is quite difficult due to the introduction of jump diffusion and multiple variables. Second, the retirement boundary naturally depends on the pair $(y, w)$, and it is challenging to guess and describe the retirement boundary with two state variables. To obtain the retirement boundary, we use a change of measure method to reduce the two state variables $(y, w)$ to one variable.
\begin{lemma}\label{lm_mc}
Consider the exponential martingale
{ 
		\begin{equation}
		    \begin{aligned}
		\mathcal{E} \define \Bigg\{\mathcal{E}(t) &= \exp\left(-\frac{1}{2} t\sum_{j = 1}^{d}(\theta_{0, j} - \sigma_{w, j})^2  + \sum_{j = 1}^{d}(\theta_{0, j} - \sigma_{w, j}) B_{j}(t) \right.\\
		&+ t \sum_{j = d + 1}^{n} \big( \text{ln}(1 - \theta_{1, j}) + \theta_{1, j}  + \ln (1 + \gamma_{w, j}) - \gamma_{w, j} +   \theta_{1, j} \gamma_{w, j}\big) \lambda_j\\
		&+ \left.\sum_{j = d + 1}^{n}\int_{0}^{t} \text{ln}(1 - \theta_{1, j}) + \ln( 1 + \gamma_{w, j}) \D \tilde{N}_j(s) \right): 0\leq t \leq T_1\Bigg\}.	            
		    \end{aligned}
		\end{equation}}
Let $\D \tilde{\mathbb{P}} = \mathcal{E}(T_1) \D {\mathbb{P}}$ and $\tilde{\mathbb{E}}$ be the corresponding expectation operator. Then we have
\begin{equation}
	\frac{W(t, y, w)} {y w} = \tilde{w}(t, \bar{y})
\end{equation}
with
    \begin{equation}\label{stop3}
		\begin{cases}
		    &\tilde{w}(t, \bar{y}) = \displaystyle \sup_{\tau \in \St} \tilde{\mathbb{E}}^{t,\bar{y}} \left[ \int_{t}^{\tau} e^{\vartheta(u - t)} U^{\mathrm{d}}_1\big(u, \bar{Y}(u)\big) \D u + e^{\vartheta (\tau - t)}\bigg(V^{\mathrm{d}}\big(\tau, \bar{Y}(\tau)\big) - q(\tau)\bigg)\right],\\
		    &\bar{y} = y^{\frac{1}{1 - \gamma}} w ^ {\frac{\gamma}{1 - \gamma}},\\
		    &\bar{Y} = Y^{\frac{1}{1 - \gamma}} \mathcal{W} ^ {\frac{\gamma}{1 - \gamma}}.\\
		\end{cases}
    \end{equation}
\end{lemma}
\begin{proof}
    Direction calculations show
    \begin{equation*}
        Y(t) \mathcal{W}(t) = e^{(\rho + \vartheta) t} \mathcal{E} (t).
    \end{equation*}
    Then the Bayesian rules of the measure change (see Lemma 3.5.3 of Karatzas and Shreve (2012) \cite{karatzas2012brownian}) implies that for any fixed $\tau \in \St$,
   { 
    \begin{align*}
        \e^{t, y, w} & \big[e^{-\rho (\tau - t)} V^{\mathrm{d}} \left(\tau, Y(\tau), \mathcal{W}(\tau)\big)\right]\\
        & = \e^{t, y, w} \left[e^{-\rho (\tau - t)} Y(\tau) \mathcal{W}(\tau) \bigg( V^{\mathrm{d}}\big(\tau, Y(\tau)\big) / Y(\tau) \mathcal{W}(\tau)\bigg) \right] \\
        & = \e^{t, \Bar{y}} \left[e^{-\rho(\tau - t) + (\rho + \vartheta)\tau} \mathcal{E} (\tau) \bigg( V^{\mathrm{d}} \big(\tau, \bar{Y}(\tau)\big)\bigg) \right] \\
        & = y w \tilde{\e}^{t, \bar{y}} \left[e^{\vartheta(\tau - t)}  V^{\mathrm{d}}\big(\tau, \bar{Y}(\tau)\big)\right],
    \end{align*}}
    where we have used the homogeneous property
    \begin{align*}
        V^{\mathrm{d}}(t, y) / y w = V^{\mathrm{d}}\big(t, y (y w)^{-\gamma / (\gamma - 1)}\big).
    \end{align*}
    Based on the same arguments, we also have
    \begin{align*}
        \e^{t, y, w} & \left[ \int_{t}^{\tau} e^{-\rho (u - t)} U^{\mathrm{d}}_1 \big(u, Y(u)\big) \D u \right] = y w \tilde{\e}^{t, \bar{y}} \left[ \int_{t}^{\tau} e^{\vartheta(u - t)} U^{\mathrm{d}}_1\big(u, \bar{Y}(u)\big) \D u\right].
    \end{align*}
    Combining the above calculations, the results are derived. 
\end{proof}

 \Cref{lm_mc} shows that the dual problem (\ref{stop1}) with two state variables $(y, w)$ is equivalent to the reduced dual problem (\ref{stop3}) with one state variable $\bar{y} = y ^{1/(1 - \gamma)} w^{\gamma / (1 - \gamma)}$. Using optimal stopping theory, we obtain the optimal stopping time of Problem (\ref{stop3}). We observe in Problem (\ref{stop3}) that the discount factor is $\vartheta$, which initially appears in (\ref{eq_q}). 

\subsection{Transformation}\label{sub:tran}
If we can reduce the optimal stopping problem (\ref{stop3}) to the integral form of a single function, the problem will be easy to solve. Observing that the first term of (\ref{stop3}) is similar to the expression (\ref{v_bar}) of $V^{\mathrm{d}}$, we have the following results.

\begin{theorem}
Define $F(t, \bar{y}) \define \tilde{w}(t, \bar{y}) - V^{\mathrm{d}}(t, \bar{y}) + q(t).$ Then
    \begin{equation}\label{stop4}
		F(t, \bar{y}) = \sup_{\tau \in \St} \tilde{\e}^{t, \bar{y}}\bigg[\int_{t}^{\tau} e^{\vartheta (u - t)}\big( U^{\mathrm{d}}_1(u,\bar{Y}(u)) -  U^{\mathrm{d}}_2(u, \bar{Y}(u)) + 1\big) \D u\bigg].
	\end{equation}   
\end{theorem}
\begin{proof}
    For any $\tau \in \St$, we have
    \begin{align*}
     V^{\mathrm{d}}\big(\tau, \bar{Y}(\tau)\big) =& \quad e^{-\vartheta (\tau - t)}\tilde{\e}^{t, \bar{y}}\bigg[\int_{\tau}^{T} e^{\vartheta (u - t)} U^{\mathrm{d}}_2\big(u, \bar{Y}(u)\big) \D u\bigg] \\
     =& \quad e^{-\vartheta (\tau - t)}\tilde{\e}^{t, \bar{y}}\bigg[\int_{t}^{T} e^{\vartheta (u - t)} U^{\mathrm{d}}_2\big(u, \bar{Y}(u)\big) \D u\bigg] \\
     &- e^{-\vartheta (\tau - t)}\tilde{\e}^{t, \bar{y}}\bigg[\int_{t}^{\tau} e^{\vartheta (u - t)} U^{\mathrm{d}}_2\big(u, \bar{Y}(u)\big) \D u\bigg].
    \end{align*}
$q(\tau)$ is decomposed similarly. Combining the results above, we obtain (\ref{stop4}) from (\ref{stop3}).
\end{proof}
Substituting (\ref{dual_utility_ff}) and (\ref{dual_utility_ff2}) into Problem (\ref{stop4}), we obtain the following form
\begin{equation}\label{stop5}
	F(t, \bar{y}) = \supp{\tau \in \St} \tilde{\e}^{t, \bar{y}}\bigg[\int_{t}^{\tau} e^{\vartheta (u - t)}\big( \Delta(u) \bar{Y}(u)^{-\gamma^{*}} + 1\big) \D u\bigg],\\
\end{equation}
where $\Delta(t) \define \frac{a(t)^{-\gamma*}}{\gamma^*}(l_1^{\gamma^* K(t)} - l_2^{\gamma^* K(t)})$.

 {In \Cref{stop5}, the sign of $\gamma^*$  is related to the relative position of the stopping region and the continuation region: when $\gamma^* > 0$, the continuation (working) region is above the stopping (retirement) region; when $\gamma^* < 0$, the continuation region is below the stopping region.}  
 Hence, we will transform the problem into a new form. Assume that $\bar{Y}$ satisfies the following equation

\begin{equation*}
\frac{\D \bar{Y}(u)}{ \bar{Y}(u^-)} = \mu_y \D u + \sum_{j=1}^{d}\sigma_{y,j} \D B_j(u) + \sum_{j=d+1}^{n}\int_{\mathbb{R}}{\gamma_{y,j} \D \tilde{N_j}(u)},
\end{equation*}
where the coefficients $\mu_y, \sigma_{y,j}, \gamma_{y, j}$ can be calculated  by using It\^o's formula. Applying It\^o's formula to the new process $M(u) = \bar{Y}(u) ^ {-\gamma^{*}}$, we obtain
\begin{equation*}
	\frac{\D M(u)}{M(u^-)} = \mu _m \D u + \sum_{j = 1}^{d} \sigma_{m, j} \D B_j(u) + \sum_{j = d + 1}^{n}\int_{\R}{\gamma_{m, j} \D \tilde{N_j}(t)},
\end{equation*}
where 
\begin{align*}
	\mu_m =& (-\gamma^*) u_y + \frac{1}{2}(-\gamma)^*(-\gamma^* - 1)\sum_{j = 1}^{d} \sigma_{y, j}^2 \\
			&+ \sum_{j = d + 1}^{n}\big((1 + \gamma_{y, j})^{-\gamma^{*}} - 1 - (-\gamma^*) \gamma_{y, j}\big)\lambda_{j}, \\
    \sigma_{m, j} =& (-\gamma^*)\sigma _{y, j},\\
	\gamma_{m, j} =& ( 1 + \gamma_{y, j})^{-\gamma^*} - 1 > -1.
\end{align*}
 Thus, we can rewrite Problem (\ref{stop5}) as follows
\begin{equation}\label{stop6}
	\begin{cases}
	&\bar{F}(t, m) \define \supp{\tau \in \St} \tilde{\e}^{t, m}\bigg[\int_{t}^{\tau} e^{\vartheta (u - t)}\big(\Delta (u)  M(u) + 1\big) \D u\bigg],\\
	&\bar{F}(t, m) = F(t, m^{-\frac{1}{\gamma^*}}).\\
	\end{cases}
\end{equation}
\subsection{Optimal boundary}\label{subsection:e_and_u}The optimal stopping time problem (\ref{stop6}) satisfies assumptions $(2.1) - (2.8)$ of Pham (1998) \cite{pham1998optimal}. Then, Theorem 3.1 of Pham (1998) \cite{pham1998optimal} shows that $F$ is the viscosity solution to the following equation
\begin{equation}\label{PDE}
		\begin{aligned}
		\text{min} \bigg\{\!\!\!\!-\vartheta \bar{F}(t, m) \!\!-\!\! D_t \bar{F}(t, m) \!\!-\!\! (\Delta (t) m + 1)-\big[A(t, m, D_{m}\bar{F}, D_{mm}\bar{F}) \!\!+\!\! B(t, m,  \bar{F})\big],\bar{F}\bigg\} = 0\\
		\end{aligned}
\end{equation}
on $ \in [0, T_1) \times \R_+$ with boundary condition $\bar{F}(T_1, \cdot) = 0$, where 
{ 
\begin{align}
    & A\big(t, m, D_{m}\bar{F}, D_{mm}\bar{F}\big) = (\frac{1}{2}\sum_{j = 1}^{d}\sigma_{m, j}^2) m^2 D_{mm}\bar{F}(t, m) + \mu_m m D_{m}\bar{F}(t, m),\label{PDE_a} \\
    & B(t, m, \bar{F}) = \sum_{j = d + 1}^{n}\left\{\bar{F}(t, m + m\gamma_{m, j}) - \bar{F}(t, m) - m\gamma_{m, j}D_{m}\bar{F}(t, m)\right \}\lambda_j.\label{PDE_b}
\end{align}}
 According to Proposition 3.1 of Pham (1998) \cite{pham1998optimal}, $\bar{F}$ satisfies the dynamic programming principle. Observing the form of the optimal stopping time problem (\ref{stop6}), 
we obtain the following theorem.  
{ 
\begin{lemma}\label{th_mon}
$\bar{F}(t, m)$ decreases in $m$.
\end{lemma}}
\begin{proof}
    As the integrand in (\ref{stop6}) decreases in $m$, $\bar{F}(t, m)$ decreases in $m$.
\end{proof}


Define $\bar{M}(s) = \int_{t}^{s} e^{-\vartheta (u - t)}\big( \Delta(u) \bar{Y}^{t, \bar{y}}(u)^{-\gamma^{*}} + 1\big) \D u$. It is easy to see that $\bar{M}$ is of DL class, which guarantees the existence of the optimal stopping time. The dynamic programming principle means that we can define a continuous region and a stopping region. According to \Cref{th_mon}, if we denote $m^*(t) = \text{inf}\{m|\bar{F}(t, m) = 0\}$ (if the set is empty, $m^*(t) = \infty$), we have the continuous region $\mathcal{C}$ and stopping region $\mathcal{S}$ as follows
\begin{align*}
    \mathcal{C} &\define \{(t, m) \in (0, T] \times \R_+: m < m^{*}(t)\},\\
\mathcal{S} &\define \{(t, m) \in (0, T] \times \R_+: m \geq m^{*}(t)\}.
\end{align*}
For simplifycity, denote $\mathcal{O} \define \mathcal{C} \cup \mathcal{D}$. Besides, the optimal stopping time is given by
\begin{equation}\label{df:op_stop_time}
    \tau_{s, m}^* \define \text{inf}\{t:(t, M^{s, m}(t)) \notin \mathcal{C}\} = \text{inf}\{t: M^{s, m}(t) \geq m^*(t)\}.
\end{equation}
{ {Here, $M^{s, m}(\cdot)$ represents the process $M(\cdot)$ with an initial value $m$ at time $s$. To guarantee the continuity of stopping time in \Cref{th_dual}, and also to obtain proper characterizations of the free boundary, we need the continuity of $m^*(t)$.
\begin{theorem}\label{th:m_conti}
    $m^*(t)$ is bounded and continuous on $[0, T_1)$ and $\lim_{t \to T_1} m^*(t)\!=\! -(\Delta (T_1))^{-1}$.
\end{theorem}}

\begin{proof}
See Appendix \ref{AB}.
\end{proof}

When we transform the StopCP into the optimal stopping time problem, we require the continuity of the optimal stopping time $\tau$ to ensure \Cref{dual_v_w}, which is proved in the following theorem.
 {
\begin{theorem} \label{taucontinuous}
        For any sequence $\left\{(t_n, m_n)\right\}_{n = 1}^{\infty}\subset \mathcal{O}$ satisfying $\lim\limits_{n \rightarrow \infty} (t_n, m_n) = (t_0, m_0)\in \mathcal{O}$, we have
    \begin{equation}
        \limsup_{n \to \infty} \tau_{t_n, m_n}^*(\omega) = \liminf_{n \to \infty} \tau_{t_n, m_n}^*(\omega) = \tau_{t_0, m_0}^*(\omega) \ \text{ for $\omega$ a.s..}
    \end{equation}
    Therefore, we have  that $\tau_{s, m}^*(\omega)$ is continuous in $m$ for fixed $s$.
\end{theorem}
\begin{proof}
    See \Cref{append_pf_taucontinuous}.
\end{proof}
}
 {
Even though each path of the process $M^{s, m}(\cdot)$ may not be continuous, the process $M^{s, m}(\cdot)$ is continuous in probability. Therefore, we can obtain the continuity results for the stopping time.}

 {
Classical optimal stopping theory determines the boundary of the set $\mathcal{C}$ by imposing the smooth-fit condition. In our context, we establish a stronger result  demonstrating the continuous differentiability of $\bar{F}$ across the stopping boundary $\partial \mathcal{C}$. In other words, we verify the global continuity of the gradient of $\bar{F}$ with respect to all variables.}
 
Before we prove the $C^1$ property of $\bar{F}$, we need the following theorem first.
\begin{theorem}\label{th:F_c_1_2_c}
    We have $\bar{F} \in C^{1,2}(\mathcal{C})$ which means that $\bar{F}_t$, $\bar{F}_m$ and $\bar{F}_{mm}$ are continuous in $\mathcal{C}$.
\end{theorem}
\begin{proof}
    This can be proved similar to Proposition 5.3 of Pham (1998) \cite{pham1998optimal} by using the same technique of considering $f_v$ therein.
\end{proof}
\begin{theorem}\label{th:F_c_1}
   We have $\bar{F} \in C^1(\mathcal{C} \cup \mathcal{D})$.
\end{theorem}

\begin{proof}
    See \Cref{append_pf_th_F_c_1}.
\end{proof}
\normalcolor

 {
Summarizing the above arguments, we have \(v \in C^1(\mathcal{C} \cup \mathcal{D}) \cap C^{1,2}(\mathcal{C}) \cap C^{1,2}(\mathcal{D})\), and the optimal stopping boundary \(m^*\) is continuous. However, this is not sufficient to apply the change of variable formula developed in Peskir (2007) \cite{peskir2007change}, which is commonly used in optimal stopping problem literature to establish Itô's formula for the value function. As Peskir (2007) \cite{peskir2007change} deals with functions that are not necessarily \(C^1\), it requires \(m^*\left(t, M_t\right)\) to be a semimartingale so that the local time on the stopping boundary is well-defined. Although we are unable to prove it for our optimal boundary, we can leverage the continuous differentiability of our value function and use a generalization of Itô's formula from Cai and Angelis (2023) \cite{cai2023change}. This generalization only requires the monotonicity of the boundary. Applying a similar approach to  Cai, Angelis, and Palczewski (2022) \cite{cai2022american}, we obtain the following result for calculating the optimal stopping boundary.
\begin{theorem}\label{th:integral_eq}
Function $m^*(\cdot)$ satisfies the following integral equation
    \begin{equation} \label{eq:integral}
        \tilde{\mathbb{E}}^{t, m^*(t)}\left[\int_t^{T_1} e^{\vartheta(u-t)}\big(1+\Delta(u)M(u)\big) \chi_{\left\{M(u)<m^*(u)\right\}} \mathrm{d} u\right]=0 .
    \end{equation}
\end{theorem}
}
\begin{proof}
    See \Cref{append_pf_th_integral_eq}.
\end{proof}
\subsection{Uniqueness of solution}
Now we have established that the solution to the optimal stopping time problem (\ref{stop6}) is a classical solution to the HJB equation (\ref{PDE}). The question arises whether the converse is true. In other words, we need to determine whether a classical solution to the HJB equation (\ref{PDE}) is necessarily the solution to the optimal stopping time problem (\ref{stop6}).
\begin{theorem}
    For any viscosity solution to the HJB equation (\ref{PDE}), it must be a solution to the optimal stopping time problem (\ref{stop6}).
\end{theorem}
\begin{proof}
To show that any viscosity solution to the HJB equation (\ref{PDE}) is a solution to the optimal stopping time problem (\ref{stop6}), it suffices to prove the uniqueness of the solution to the PDE (\ref{PDE}). This can be derived from Theorem 4.1 of Pham (1998) \cite{pham1998optimal}.
\end{proof}

\section{Retirement Boundaries and Optimal Strategies in Terms of Primal Variables} \label{section_primal}
In this section, we present the retirement boundary expressed in terms of the primal variables $(x, h, w)$, and provide the optimal investment and consumption strategies. The relationship between the primal problem (\ref{eq_vu}) and the dual problem (\ref{stop1}) is established in \Cref{th_dual}, which will be verified first. Based on the primal-dual relation, the retirement boundary by $(y, w)$ in \Cref{section:free_bound} can be transformed to be expressed in terms of $(x, h, w)$. Using replication, we derive the feedback forms of the optimal consumption and investment strategies, which are functions of the de facto wealth $x-p^T(t)h$ and wage $w$. We also present numerical implications of the retirement boundary and optimal strategies, and intuitively show the discontinuities of optimal strategies at the retirement boundary.
\subsection{The primal-dual relation} In this subsection, we present \Cref{lm_bd_pr} to show that the condition in  \Cref{th_dual} holds. We also demonstrate how the optimal stopping time in terms of dual variables can be obtained using the relation between $m$ and $(y, w)$.

\begin{lemma}\label{lm_bd_pr}
    Let $m^*(\cdot)$ be as in \Cref{section:free_bound}, $y > 0, w > 0$ and $t \in [0, T)$ be fixed. Define
	\begin{equation}\label{bd(y,w)}
		\hat{\tau}_{t, y, w} = 
			\inf \bigg \{   t\leq s \leq T_1 : \big(Y^{t, y}(s)\big)^{-\frac{1}{\gamma}} \big(\mathcal{W}^{t, w}(s)\big)^{-1} \leq m^*(s)\bigg \} .
	\end{equation}
	Then $\hat{\tau}_{t, y, w}$ is the optimal stopping time of Problem (\ref{stop1}), and satisfies the continuous condition and \Cref{eq:e_b_limit} in \Cref{th_dual}.
 
 {Additionally, we have 
 \begin{equation*}
     \lim_{y \to \infty}\e \bigg[b(\hat{\tau}_{t, y, w}) \xi(\hat{\tau}_{t, y, w})\bigg] = 0.
 \end{equation*}}\end{lemma}\begin{proof}
    See \Cref{append_c}.
\end{proof}
Now we characterize the retirement boundary in the primal variables $(x, h, w)$.
\begin{lemma}\label{lm_primal_map}
Define the primal map $\mathcal{P}$ as
\begin{equation*}
    \mathcal{P}_{t, w}(y) = -W_y(t, y, w),
\end{equation*}
and define  the primal and dual continuation regions, respectively, as
\begin{align*}
   & \mathcal{C}_{t, w}^p = \big\{(x, h): (x, h, w) \in \G_t, V_F(t, x - p^T(t)h, w) > V_{F, p}(t, x - p^T(t)h)\big\},\\
&\mathcal{C}_{t, w}^d = \big\{ y > 0: W(t, y, w) >  V^{\mathrm{d}}(t, Y(t))- q(t) y w\big\}.
\end{align*}
{ Here, $V_F$ is the value function defined in \Cref{eq_vu}, $V_{F, p}$ is the post retirement value function defined in \Cref{eq_U}, $W$ is the dual function of $V_F$ defined in \Cref{stop1} and $V^{\mathrm{d}}$ is the dual function of $V_{F, p}$. 
Then $\mathcal{C}_{t, w}^p = \widehat{\mathcal{C}_{t, w}^p} \define \big\{(x, h) : x - p^T(t) h + q(t) w \in \mathcal{P}_{t, w} \mathcal{C}_{t, w}^d \big\}.$}
\end{lemma}
\begin{proof}
See \Cref{append_c}.
\end{proof}
\begin{theorem}\label{th_bd_pr}
The retirement boundary can be expressed in primal variables as $\{(t, x, h, w) : (x, h) \in \partial \mathcal{C}_{t, w}^p\}$, where $\partial \mathcal{C}_{t, w}^p = \{(x, h): x - p^T(t) h + q(t) w \in \mathcal{P}_{t, w} \partial \mathcal{C}_{t, w}^d\}$. In particular, we have
\begin{equation}
    \begin{aligned}
    &\partial \mathcal{C}_{t, w}^p &= \big\{(x, h) : x - p^T(t)h = G^*(t) w \big\},\\
    &\mathcal{C}_{t, w}^p &= \big\{(x, h) : x - p^T(t) h < G^*(t) w \big\},
    \end{aligned}
\end{equation}
where $G^*(t) \define \gamma^* V^{\mathrm{d}}(t, 1) m^*(t)^{{\gamma^*}}$ and $V^{\mathrm{d}}(t, y)$ is defined by \Cref{th_dual_retire}.
\end{theorem}
\begin{proof}
    See \Cref{append_c}.
\end{proof}

We observe that the de facto wealth $x-p^T(t)h$ appears as a whole in the expression of the retirement boundary in terms of the primal variables. This phenomenon is also found in the feedback forms of the optimal investment and consumption strategies. Furthermore, \Cref{th_bd_pr} shows that the retirement boundary can be expressed as a linear relationship among $(x, h, w)$.
\subsection{Numerical illustrations of primal retirement boundary} 
\Cref{th_bd_pr} establishes the relationship between the retirement boundaries in the dual and primal variables, and provides a semi-explicit form. In this subsection, we will present some numerical results of the retirement boundary. The basic parameters are listed in  \Cref{table_para_initial}. We assume $\lambda_2 = 1$, $l_1 = 1$ and a smooth leisure function $l_2^{K(t)} = \exp\{\epsilon_0(T - t)\}$, where $\epsilon_0 = 0.02$ to simplify our numerical results. We discretize the time interval and use a recursive method to obtain the retirement boundary.
\begin{table}[ht]
	\centering
	\caption{Basic parameters in the numerical illustrations.}
	\begin{tabular}{cccccccc}
		\hline
		$\mu_1$ &$\sigma_{1, 1}$ &$\gamma_{1, 2}$ &$r$     &$\mu_w$ &$\sigma_{w, 1}$ &$\gamma_{w,2}$ &$\gamma$\\
		$0.05$  &$0.2$      &$0.1$      &$0.01$  &$0.02$  &$0.1$      &$0$        &$0.5/1.5$\\
		\hline
		$\mu_2$ &$\sigma_{2, 1}$ &$\gamma_{2, 2}$ &$\rho$  &$\alpha$&$\beta$    &$T_1$     &$T$\\
		$0.04$  &$0.1$      &$0.2$      &$0.01$  &$0.2$   &$0.4$      &$30$       &$45$\\
		\hline
	\end{tabular}
	\label{table_para_initial}
\end{table}

First, we illustrate the retirement boundary in the $(x, h, w)$-axis at time $t = 0, 15, 24$ in \Cref{3} and \Cref{4}. Since $(x, h, w)$ has a linear relationship in  \Cref{th_bd_pr}, the retirement boundaries are all planes for a fixed time $t$.

Our findings are novel in that we describe the retirement boundary using three state variables: wage, habit, and wealth. Our results reveal more economic implications than previous studies. We observe from both \Cref{3} and \Cref{4} that agents with higher habit levels tend to continue working to maintain the standard of living. Additionally, agents with larger wealth or lower wage rates are more likely to choose to retire and enjoy leisure gains. As time passes, the agent is expected to receive less future labor income, which reduces the importance of the wage rate on retirement decisions. {\Cref{3} indicates that the retirement surface moves closer to the $(h, w)$ plane over time, and the critical level of wage rate that pushes one to retire becomes larger.} To illustrate the movement of the retirement boundary in the $(x, h)$-axis with time, we fix $w=1$ and present the results in \Cref{5}. An elder agent will be more likely to retire and \Cref{5} shows that the retirement region becomes larger with time.

\begin{figure}[ht]
\begin{centering}
\subfloat[$t = 0$]{
\label{Fig.sub.3.1}\begin{centering}
\includegraphics[width=3cm]{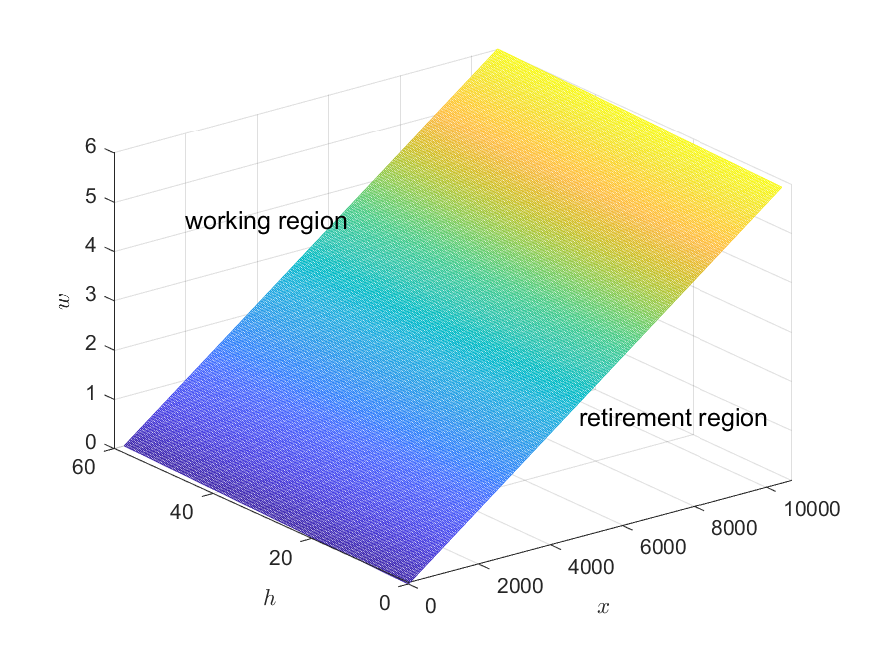}
\par\end{centering}
}
\subfloat[$t = 15$]{
\label{Fig.sub.3.2}\begin{centering}
\includegraphics[width=3cm]{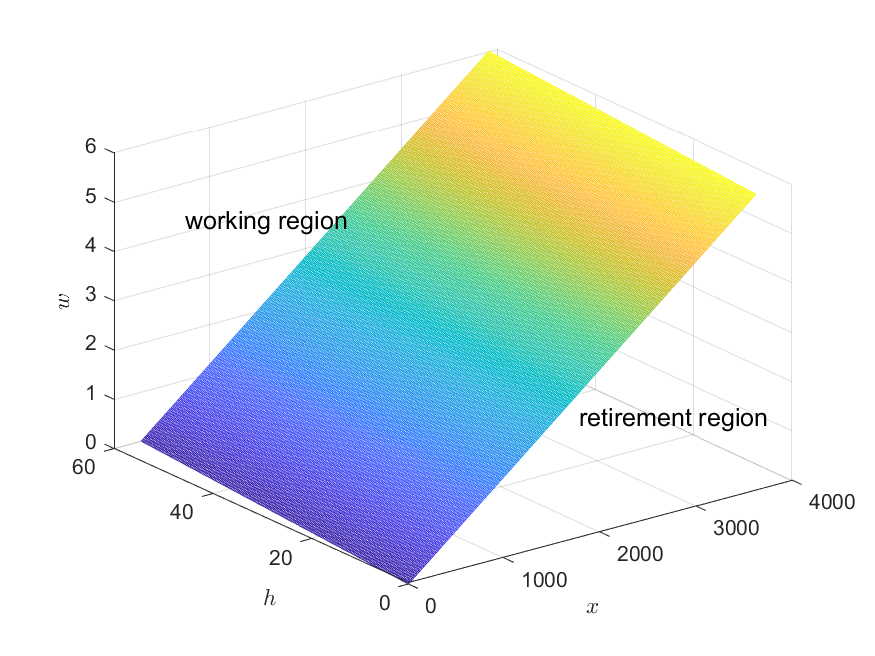}
\par\end{centering}
}
\subfloat[$t =24$]{
\label{Fig.sub.3.3}\begin{centering}
\includegraphics[width=3cm]{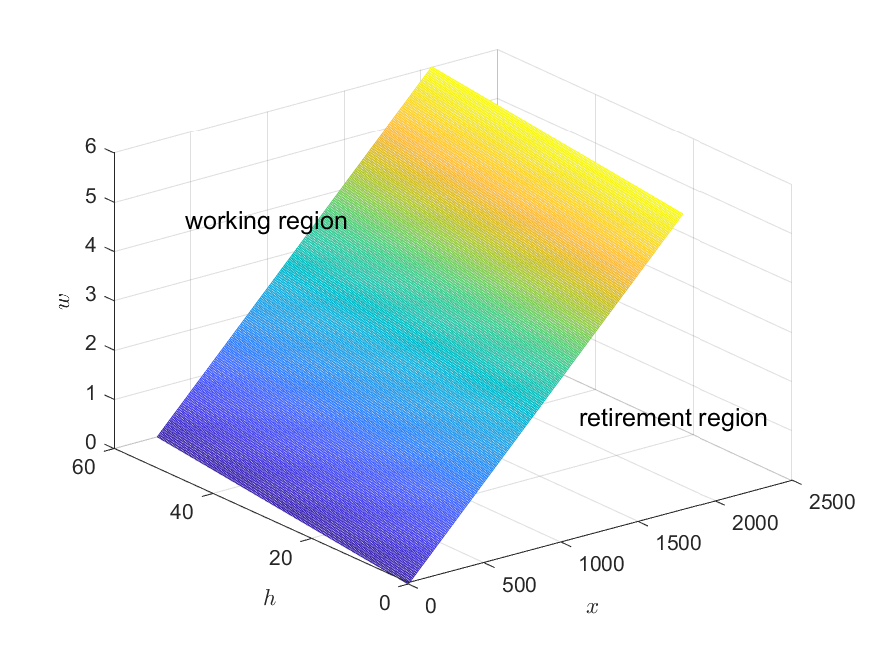}
\par\end{centering}
}
\par\end{centering}
\caption{\label{3}Retirement boundary in $(x, h, w)$-axis, $\gamma = 0.5$.}
\end{figure}

\begin{figure}[ht]
\begin{centering}
\subfloat[$t = 0$]{
\label{Fig.sub.4.1}\begin{centering}
\includegraphics[width=3cm]{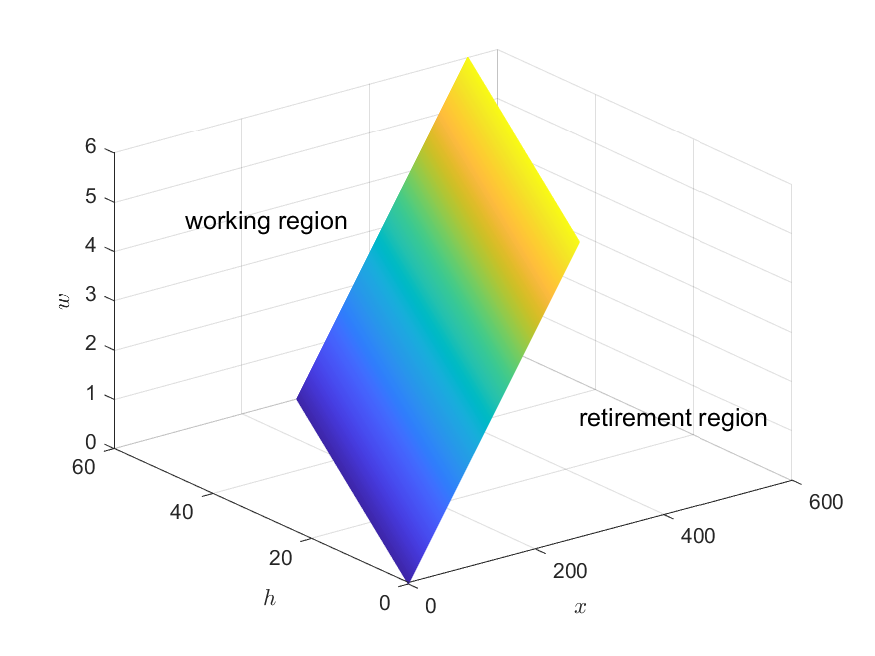}
\par\end{centering}
}
\subfloat[$t = 15$]{\begin{centering}
\includegraphics[width=3cm]{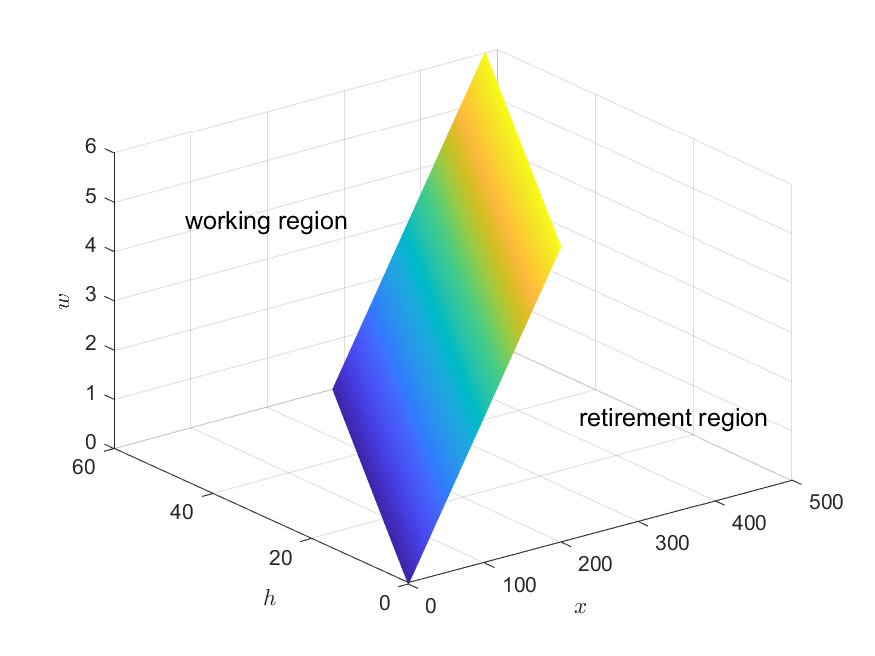}
\par\end{centering}
}
\subfloat[$t =24$]{\begin{centering}
\includegraphics[width=3cm]{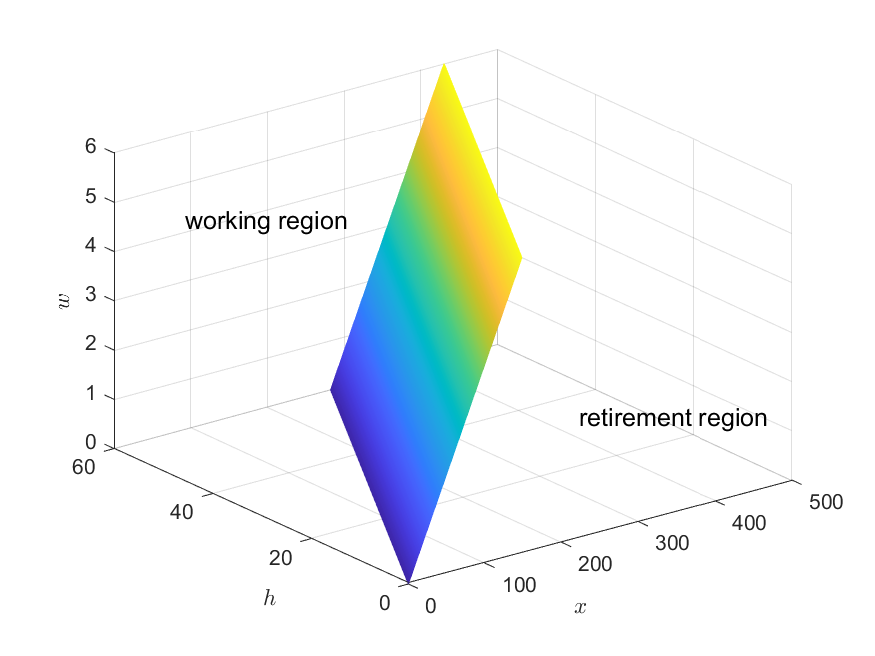}
\par\end{centering}
}
\par\end{centering}
\caption{\label{4}Retirement boundary in $(x, h, w)$-axis, $\gamma = 1.5$.}
\end{figure}

\begin{figure}[ht]
\begin{centering}
\subfloat[$\gamma = 0.5$]{
\label{Fig.sub.5.1}\begin{centering}
\includegraphics[width=5cm]{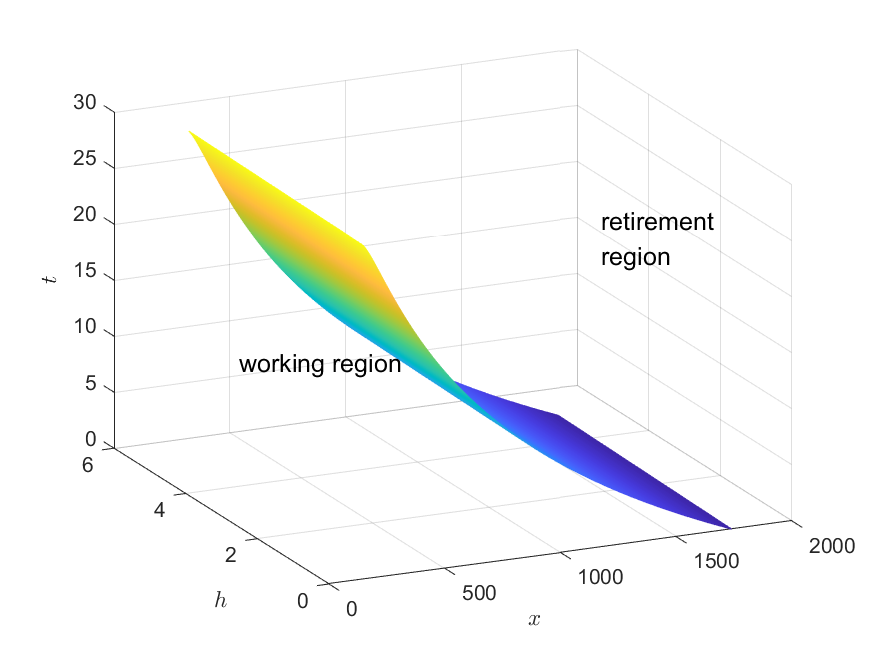}
\par\end{centering}
}
\subfloat[$\gamma = 1.5$]{\begin{centering}
\includegraphics[width=5cm]{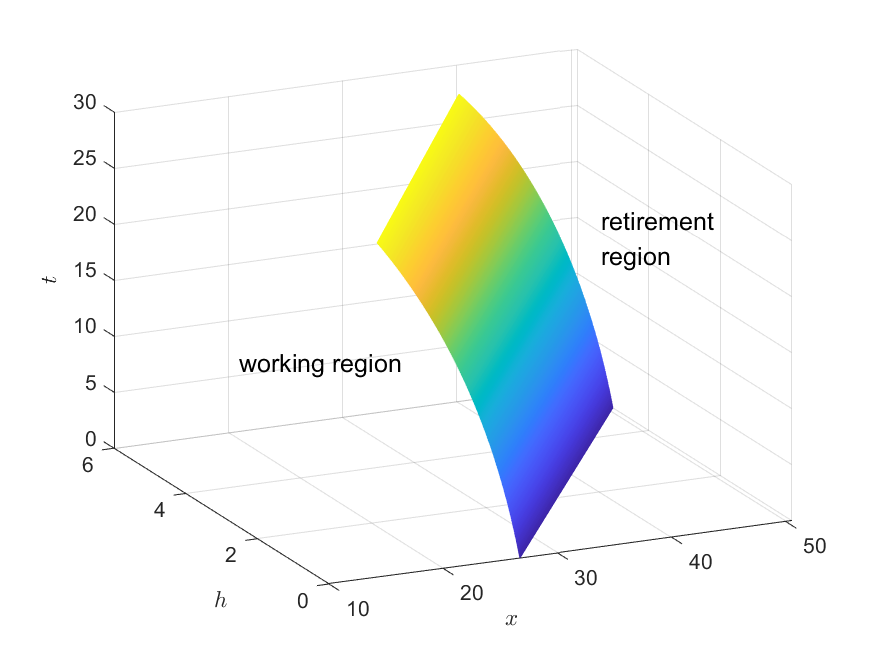}
\par\end{centering}
}
\par\end{centering}
\caption{\label{5}Retirement boundary in $(x, h, t)$-axis, $\gamma = 1.5$.}
\end{figure}

\subsection{Optimal consumption and portfolio} In this subsection, our goal is to characterize the optimal consumption and investment strategies in feedback forms in $(t, x, h, w)$. By using the replication theorem, these strategies can be computed based on the value of $W$. However, it is worth noting that $W$ is only known in a semi-explicit form with $z^*$.

Before presenting the optimal strategy, we need some notations to simplify our form. Denote
\begin{align*}
    \begin{cases}
         (\sigma_M)_{i, j} = \sigma_{i, j}, &1 \leq j \leq d,\\
 (\theta)_j = \theta_{0, j},
& 1 \leq j \leq d ,\\
 (\theta)_j = \theta_{1, j}, 
 &d < j \leq n ,\\
\big(D(W_y)\big)_j = y W_{y y}(t, y ,w)  \theta_{0, j}  -w W_{y w}(t, y, w)  \sigma_{w,j}, & 1 \leq j \leq d ,\\
\big(D(W_y)\big)_j =  W_y(t, y, w) - W_y(t, y(1 - \theta_{1, j}), w)\\
\quad+W_y(t, y, w) -  W_y(t, y, w(1 + \gamma_{w, j})),&d < j \leq n.
    \end{cases}
\end{align*}
Then the optimal strategies are given in the following theorem.

\begin{theorem}\label{th_op_st}
The optimal consumption-investment strategies $(c^*, \pi ^*)$ are expressed by $c^* = \{c^*(t) = C^*\big(t, X^*(t), h^*(t), \mathcal{W}(t)\big):0 \leq t \leq T\}$ and $\pi^* = \{\pi^*(t) = \Pi^*\big(t, X^*(t), h^*(t), \mathcal{W}(t)\big):0 \leq t \leq T\}$, where $X^*$ represents the solution to SDE (\ref{eq_x}) with strategies $\pi = \pi^*$ and $c = c^*$. The functions $C^*$ and $\Pi^*$ are  expressed by 
	\begin{align}
		&C^*(t, x, h ,w) =
		\begin{cases}
			I_{U_1}\left(a(t) y\right) + h ,\quad \quad \text{if }( x, h ) \in \mathcal{C}^{t, w}_p,  \\
			I_{U_2}\left(a(t) y\right) + h ,\quad \quad \text{otherwise}, \label{op_c}\\
		\end{cases}
	\\
		&\Pi^*(t, x, h ,w) = 
		\begin{cases}
			\sigma_M^{-1} \big(-\sigma_{wM} q(t) w + D(W_y)(t, y, w)), &\text{if }(x, h) \in \mathcal{C}^{t, w}_p, \\
			\left(\gamma^* + 1\right) \sigma_M^{-1} \theta \big(x - p^T(t)h\big), & \text{otherwise,}
		\end{cases}\label{op_pi}
	\end{align}
	where $y = \mathcal{P}^{-1}_{t, w}(x - p^T(t) h + w)$.
\end{theorem}
\begin{proof}
See Appendix D.
\end{proof}

\begin{figure}[ht]
\begin{centering}
\subfloat[$t = 0$]{
\label{Fig.sub.6.1}\begin{centering}
\includegraphics[width=3cm]{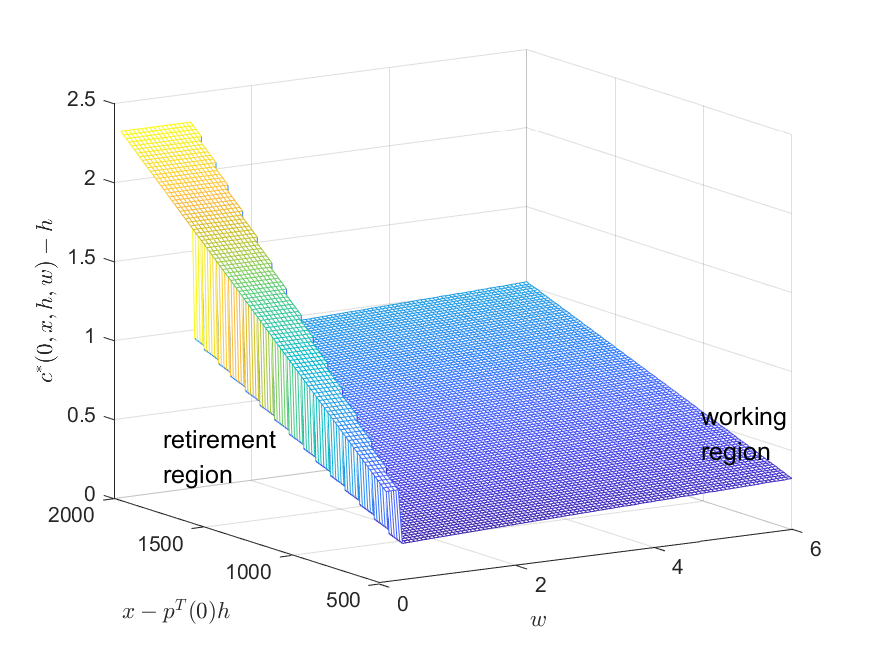}
\par\end{centering}
}
\subfloat[$t = 15$]{
\label{Fig.sub.6.2}\begin{centering}
\includegraphics[width=3cm]{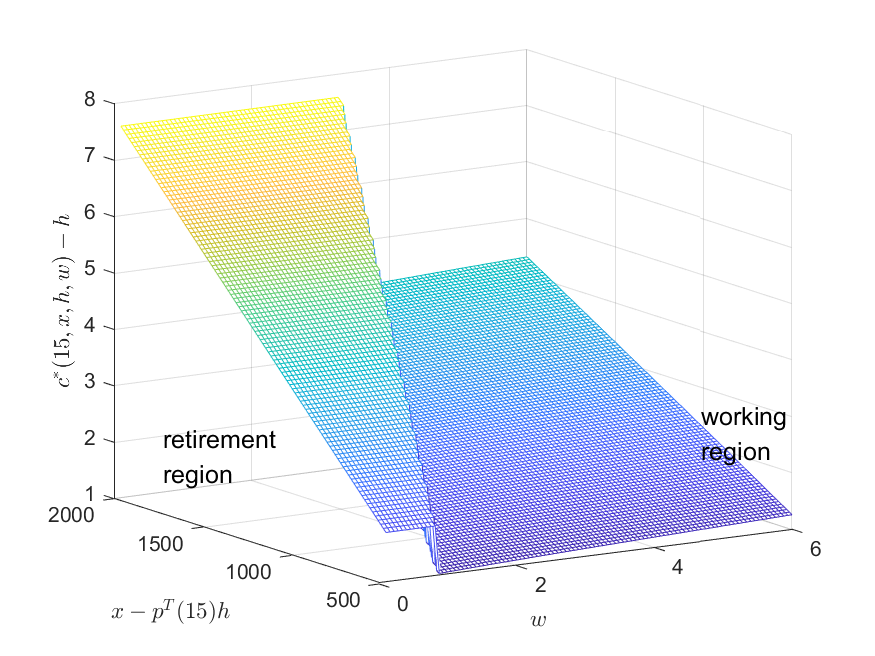}
\par\end{centering}
}
\subfloat[$t =24$]{
\label{Fig.sub.6.3}\begin{centering}
\includegraphics[width=3cm]{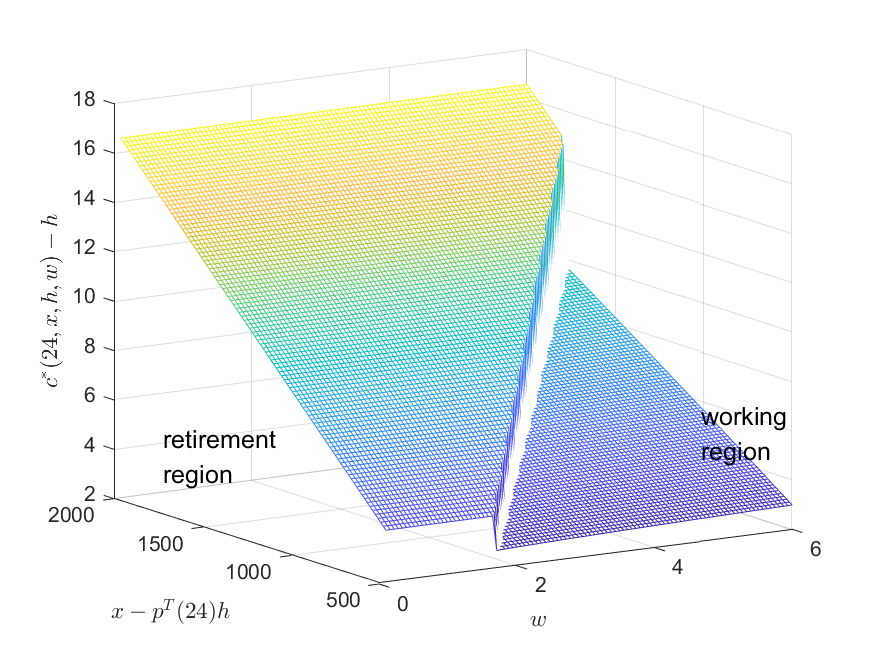}
\par\end{centering}
}
\par\end{centering}
\caption{\label{6} Optimal consumption, $\gamma = 0.5$.}
\end{figure}

\begin{figure}[ht]
\begin{centering}
\subfloat[$t = 0$]{
\label{Fig.sub.7.1}\begin{centering}
\includegraphics[width=3cm]{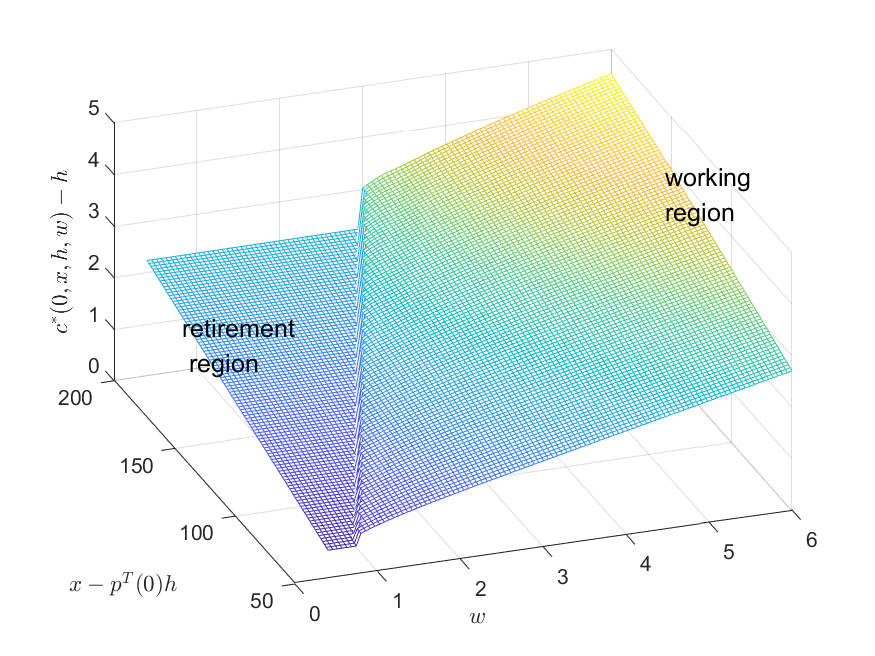}
\par\end{centering}
}
\subfloat[$t = 15$]{
\label{Fig.sub.7.2}\begin{centering}
\includegraphics[width=3cm]{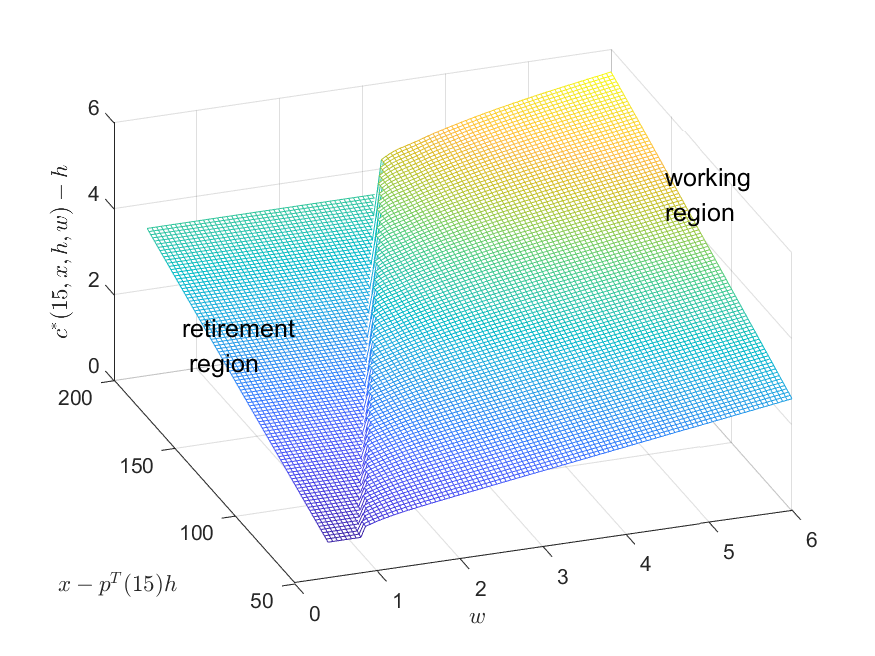}
\par\end{centering}
}
\subfloat[$t =24$]{
\label{Fig.sub.7.3}\begin{centering}
\includegraphics[width=3cm]{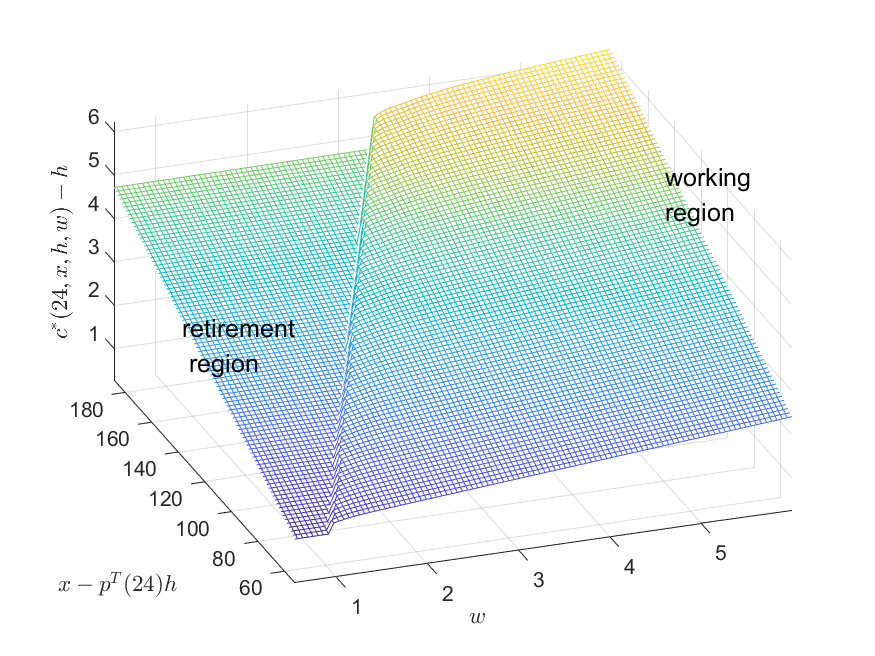}
\par\end{centering}
}
\par\end{centering}
\caption{\label{7} Optimal consumption, $\gamma = 1.5$.}
\end{figure}
\subsection{ {Optimal consumption before and after retirement}}
Using \cref{th_op_st}, we can perform a numerical sensitivity analysis of $c^*$ and $\pi^*$. The variables $x$ and $h$ appear as a combined value of the de facto wealth $(x - p^T(t)h)$ in the strategies. We plot the strategies at three different times, namely $t=0$, $t=15$, and $t=24$. The numerical results are presented in \Cref{6} and \Cref{7}.  {When $\gamma>1$, the agent experiences a consumption jump-down at retirement, which is consistent with the findings of Chen, Hentschel, and Xu (2018) \cite{chen2018optimal} and Dybvig and Liu (2010) \cite{dybvig2010lifetime}. This partially explains the retirement consumption puzzle as described in Banks, Blundell, and Tanner (1998) \cite{banks1998there}. Indeed, when $\gamma>1$, as assumed in Chen, Hentschel, and Xu (2018) \cite{chen2018optimal} and Dybvig and Liu (2010) \cite{dybvig2011verification}, the marginal utility after retirement decreases. As noted in Hurd and Rohwedder (2003) \cite{hurd2003retirement}, the agent naturally consumes less due to the decrease in marginal utility. Based on 2001 survey data, Ameriks, Caplin, and Leahy (2007) \cite{ameriks2007retirement} showed that more than $55\%$ of people expect a fall in consumption, while less than $8\%$ of people expect an increase after retirement. However, there is still a minority of people who consume more after retirement. We observe in \Cref{6} that agents with risk aversion $0<\gamma<1$ experience a consumption jump-up at retirement. In the case of $0<\gamma<1$, the marginal utility after retirement increases, implying an increase in consumption after retirement. Conversely, we may also deduce that most people have a risk aversion parameter larger than 1, as documented in Azar (2006) \cite{azar2006measuring} ($\gamma$ is between 4.2 and 5.4) and Schechter (2007).}

From \Cref{6} and \Cref{7}, we observe that when the wage or the de facto wealth increases, the agent has more wealth and will consume more. However, consumption increases at a slower rate when the ``wealth-habit-wage" triplet approaches the retirement boundary. In other words, when the agent considers retirement, he becomes more hesitant to consume more, even with more de facto wealth or a larger wage.

\begin{figure}[ht]
\begin{centering}
\subfloat[$t = 0, \pi^*_1/x$]{
\label{Fig.sub.8.1}\begin{centering}
\includegraphics[width=3cm]{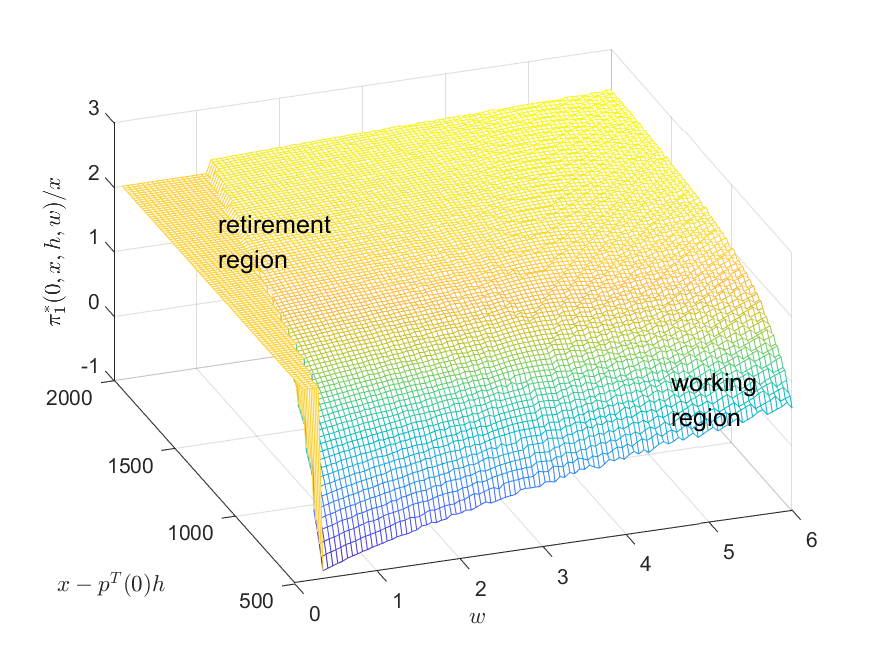}
\par\end{centering}
}
\subfloat[$t = 15, \pi^*_1/x$]{
\label{Fig.sub.8.2}\begin{centering}
\includegraphics[width=3cm]{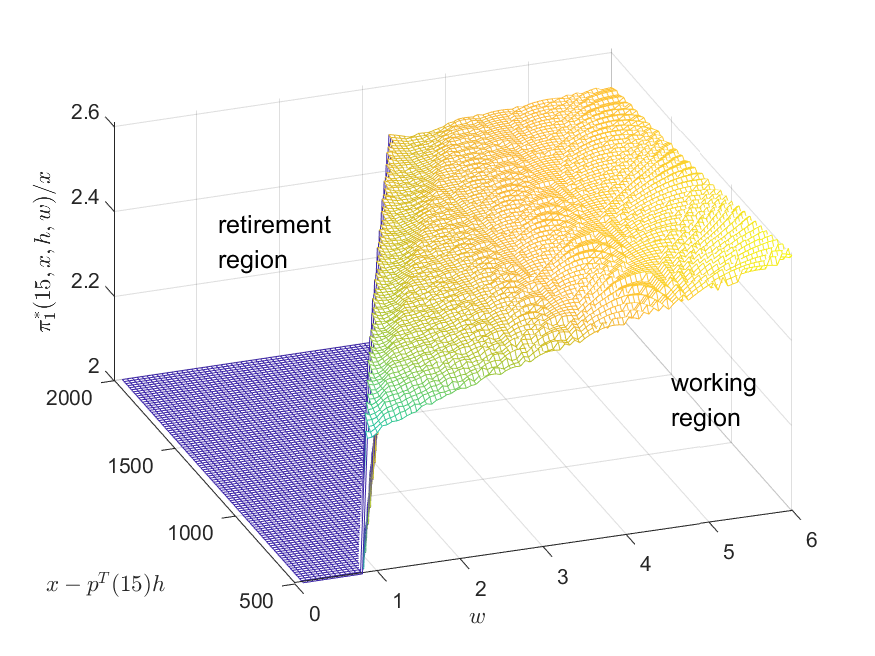}
\par\end{centering}
}
\subfloat[$t =24, \pi^*_1/x$]{
\label{Fig.sub.8.3}\begin{centering}
\includegraphics[width=3cm]{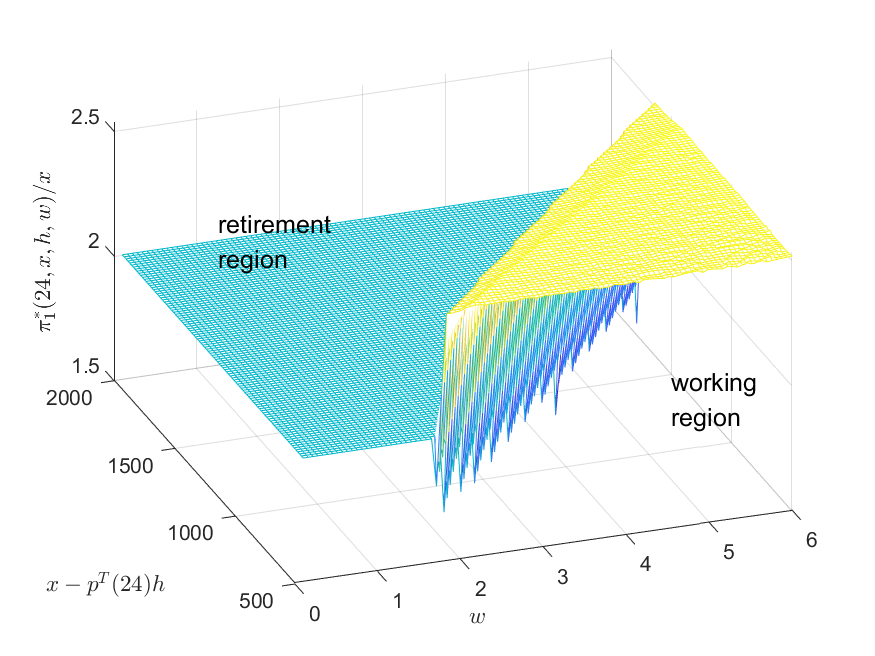}
\par\end{centering}
}

\subfloat[$t = 0, \pi^*_2/x$]{
\label{Fig.sub.8.4}\begin{centering}
\includegraphics[width=3cm]{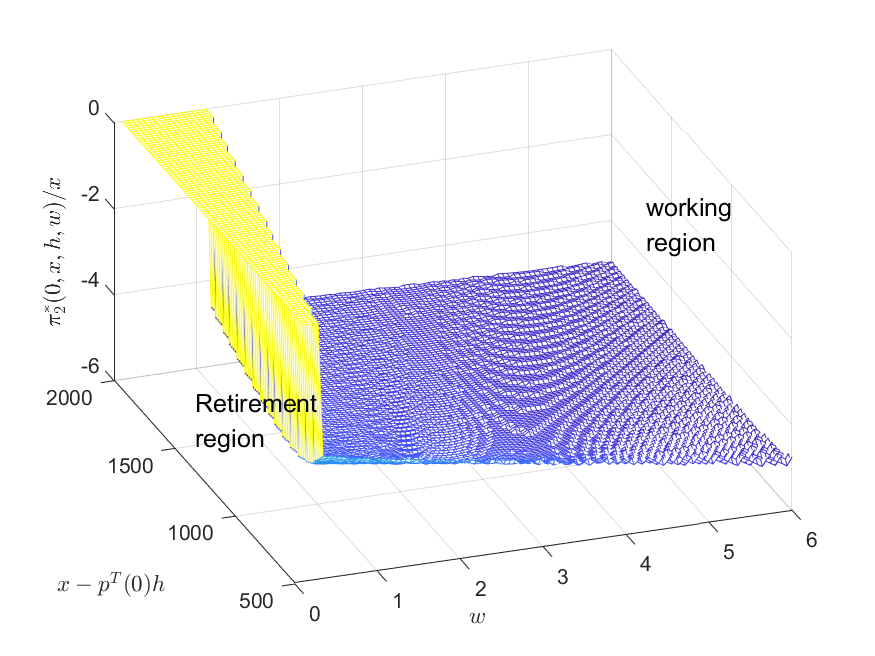}
\par\end{centering}
}
\subfloat[$t = 15, \pi^*_2/x$]{
\label{Fig.sub.8.5}\begin{centering}
\includegraphics[width=3cm]{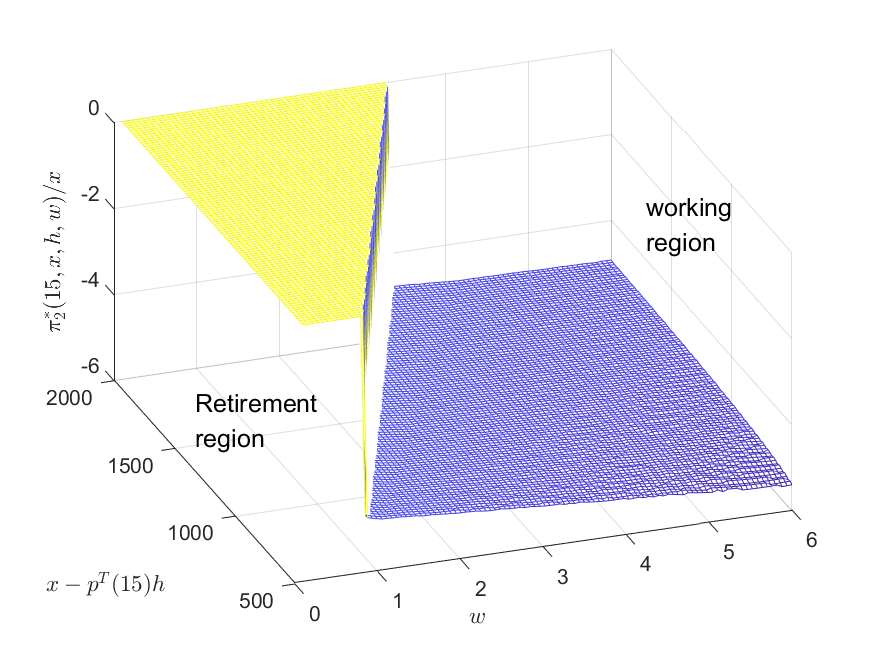}
\par\end{centering}
}
\subfloat[$t =24, \pi^*_2/x$]{
\label{Fig.sub.8.6}\begin{centering}
\includegraphics[width=3cm]{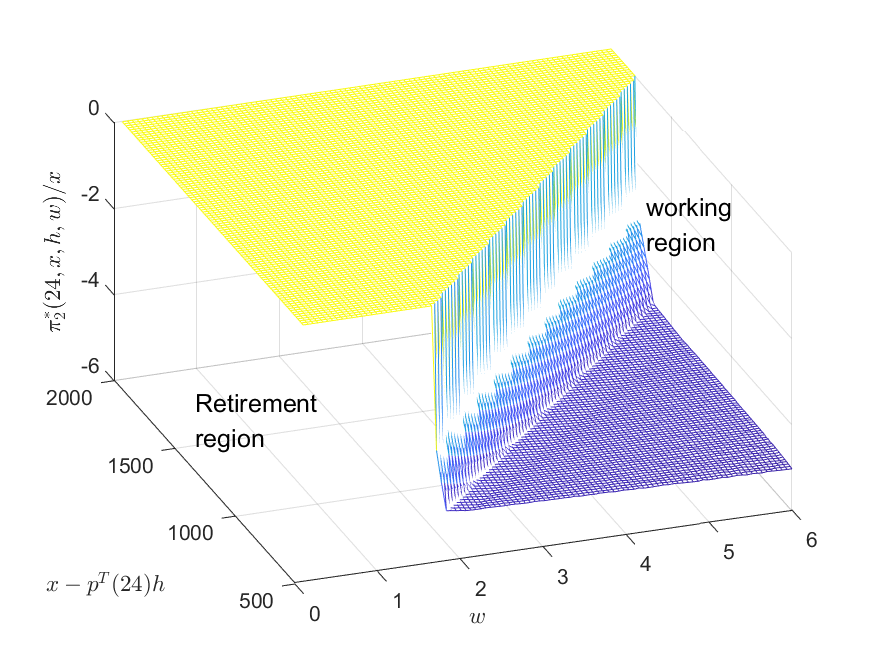}
\par\end{centering}
}
\par\end{centering}
\caption{\label{8} Optimal portfolio, $\gamma = 0.5$.}
\end{figure}

\begin{figure}[ht]
\begin{centering}
\subfloat[$t = 0, \pi^*_1/x$]{
\label{Fig.sub.9.1}\begin{centering}
\includegraphics[width=3cm]{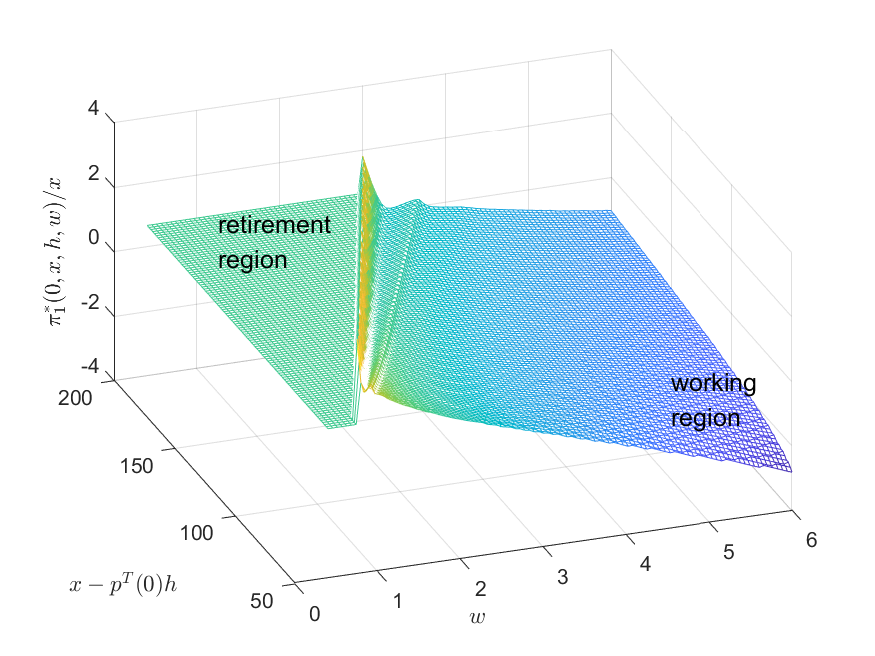}
\par\end{centering}
}
\subfloat[$t = 15, \pi^*_1/x$]{
\label{Fig.sub.9.2}\begin{centering}
\includegraphics[width=3cm]{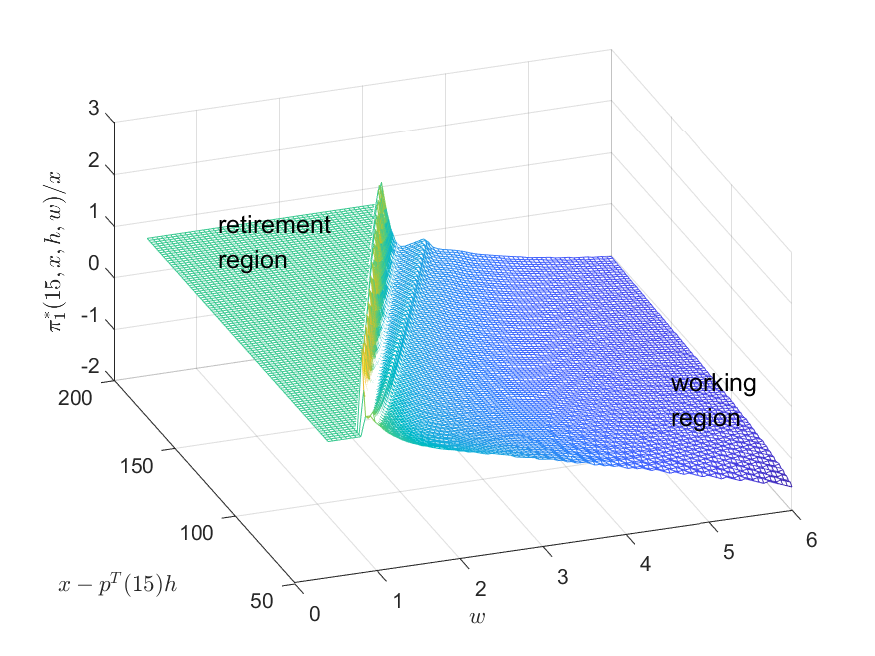}
\par\end{centering}
}
\subfloat[$t =24, \pi^*_1/x$]{
\label{Fig.sub.9.3}\begin{centering}
\includegraphics[width=3cm]{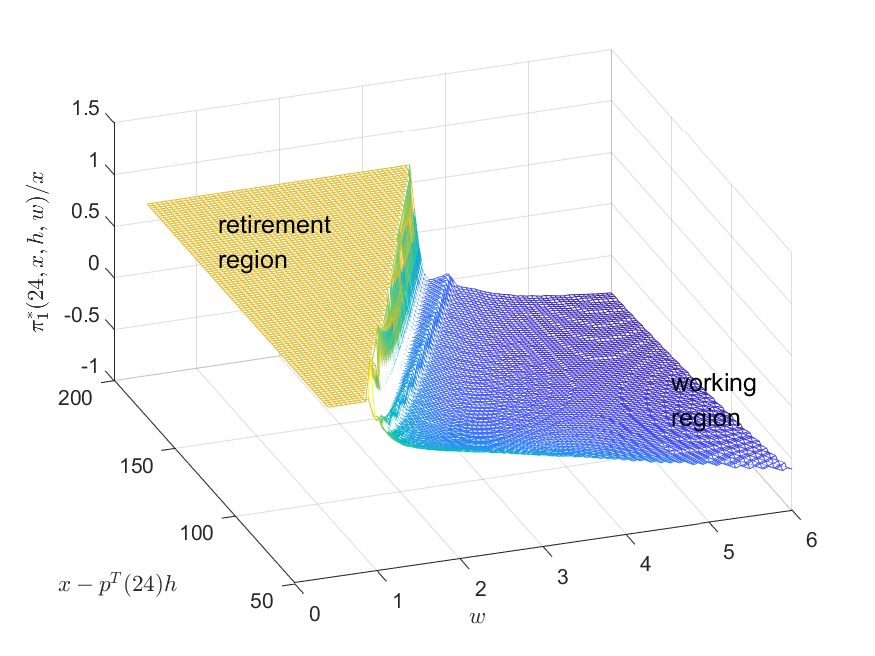}
\par\end{centering}
}

\subfloat[$t = 0, \pi^*_2/x$]{
\label{Fig.sub.9.4}\begin{centering}
\includegraphics[width=3cm]{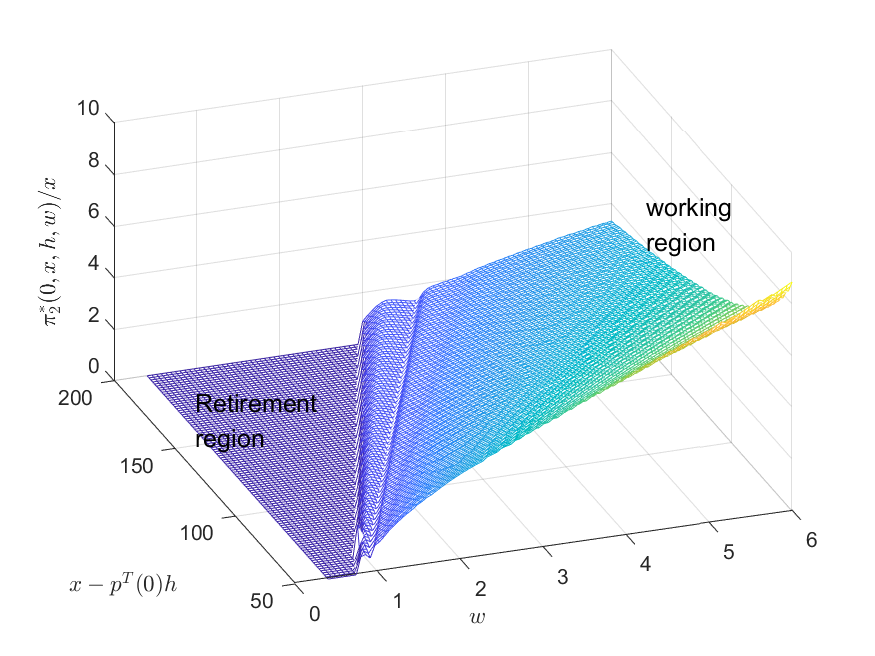}
\par\end{centering}
}
\subfloat[$t = 15, \pi^*_2/x$]{
\label{Fig.sub.9.5}\begin{centering}
\includegraphics[width=3cm]{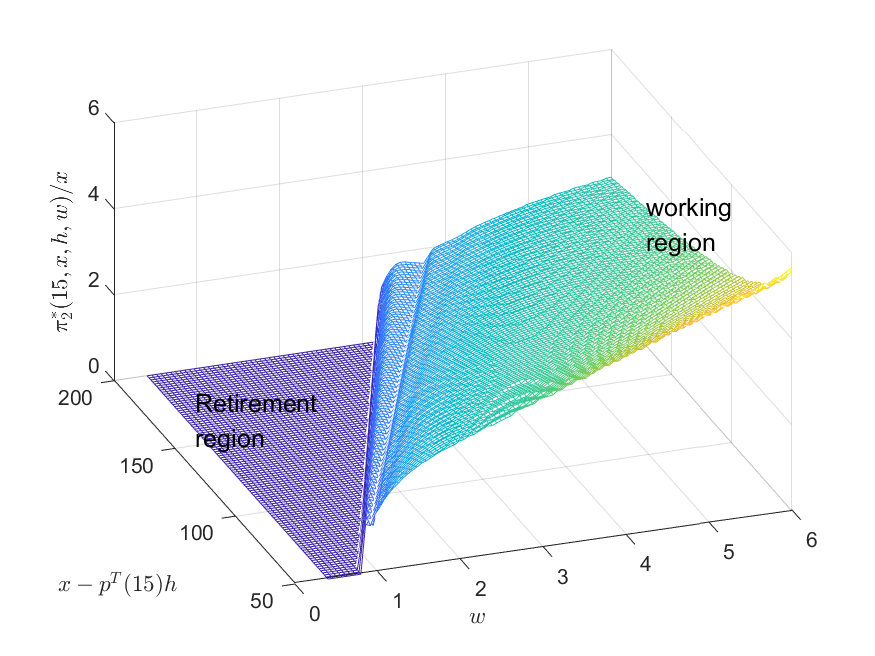}
\par\end{centering}
}
\subfloat[$t =24, \pi^*_2/x$]{
\label{Fig.sub.9.6}\begin{centering}
\includegraphics[width=3cm]{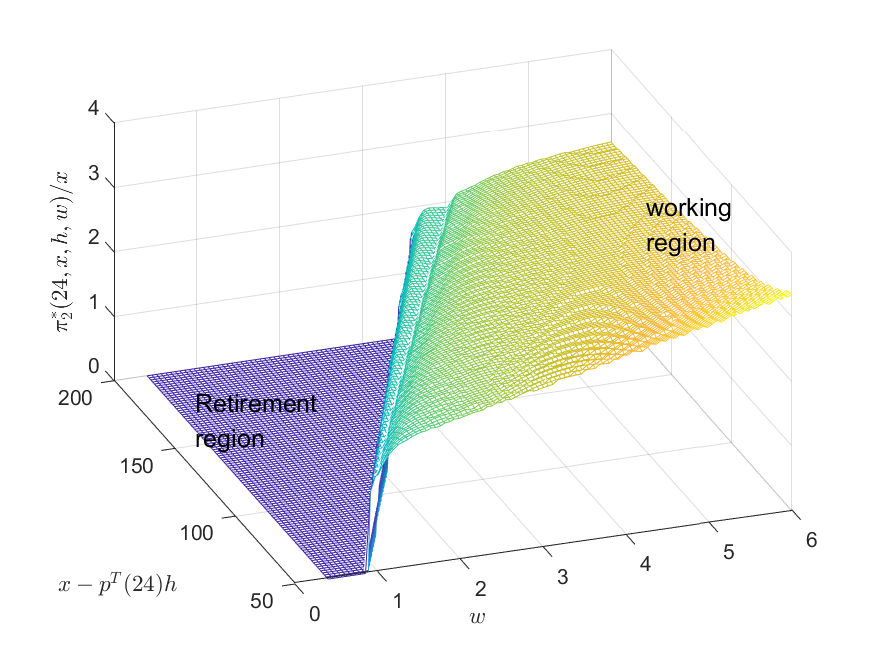}
\par\end{centering}
}
\par\end{centering}
\caption{\label{9} Optimal portfolio, $\gamma = 1.5$.}
\end{figure}

\subsection{Optimal investment before and after retirement}
\Cref{8} and \Cref{9} present the optimal proportion of risky investment $\pi_1^*(t,x,h,w)/x$ and $\pi_2^*(t,x,h,w)/x$ for $\gamma=0.5$ and $\gamma=1.5$, respectively.  {Both \Cref{8} and \Cref{9} suggest a jump in the proportion of stocks for elder agents, which is also observed in real data by Coile and Milligan (2009) \cite{coile2009household} for the US and Spicer, Stavrunova, and Thorp (2016) \cite{spicer2016portfolios} for Australia. This is because the agent's investment decisions are based on the current wealth and future wage income. When the agent retires, the future wage income becomes $0$. Thus the portfolio has a jump at retirement. We observe that the optimal investment changes significantly near retirement, especially when  $\gamma=1.5$. This phenomenon can be attributed to the fact that $D_{mm} \bar{F}$ is not continuous up to the retirement boundary, as depicted in \Cref{17}. This is because $\bar{F}$ is only $C^{1, 2} $ in $\mathcal{C}$ and $C^1$ in $\mathcal{O}$.}

\begin{figure}[ht]
\begin{centering}
\subfloat[$t = 0$]{
\label{Fig.sub.17.1}\begin{centering}
\includegraphics[width=3.8cm]{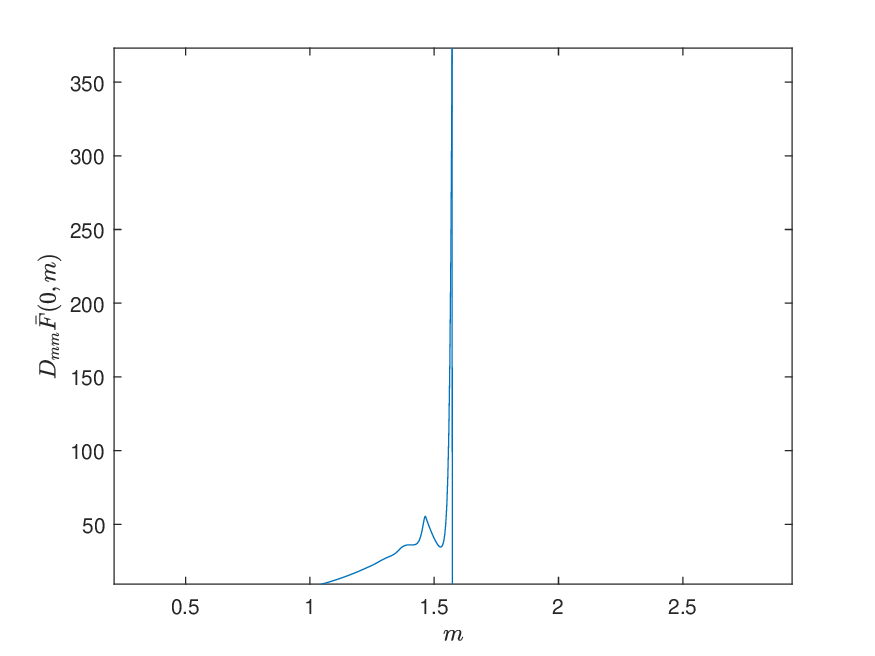}
\par\end{centering}
}
\subfloat[$t = 15$]{
\label{Fig.sub.17.2}\begin{centering}
\includegraphics[width=3.8cm]{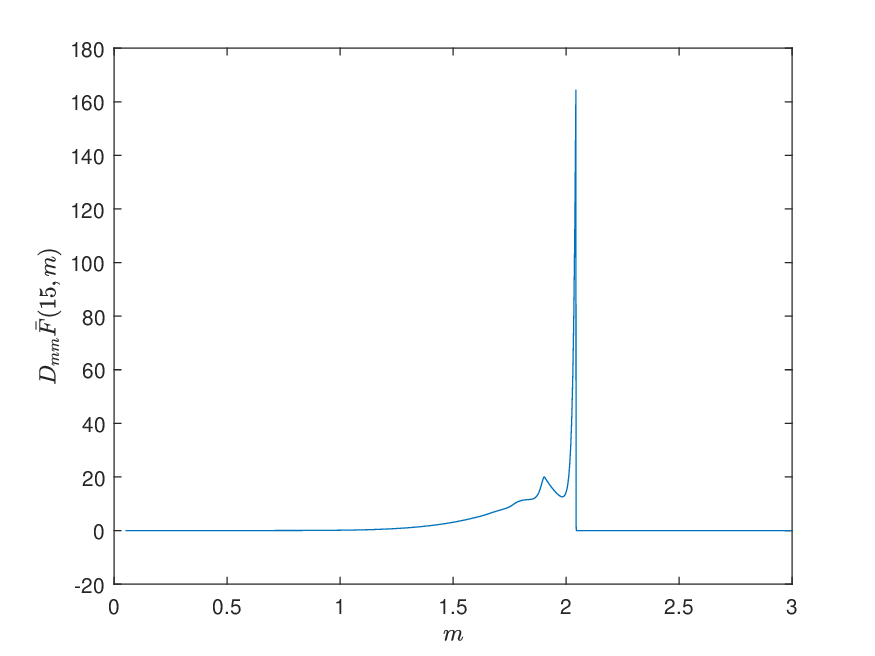}
\par\end{centering}
}
\subfloat[$t = 24$]{
\label{Fig.sub.17.3}\begin{centering}
\includegraphics[width=3.8cm]{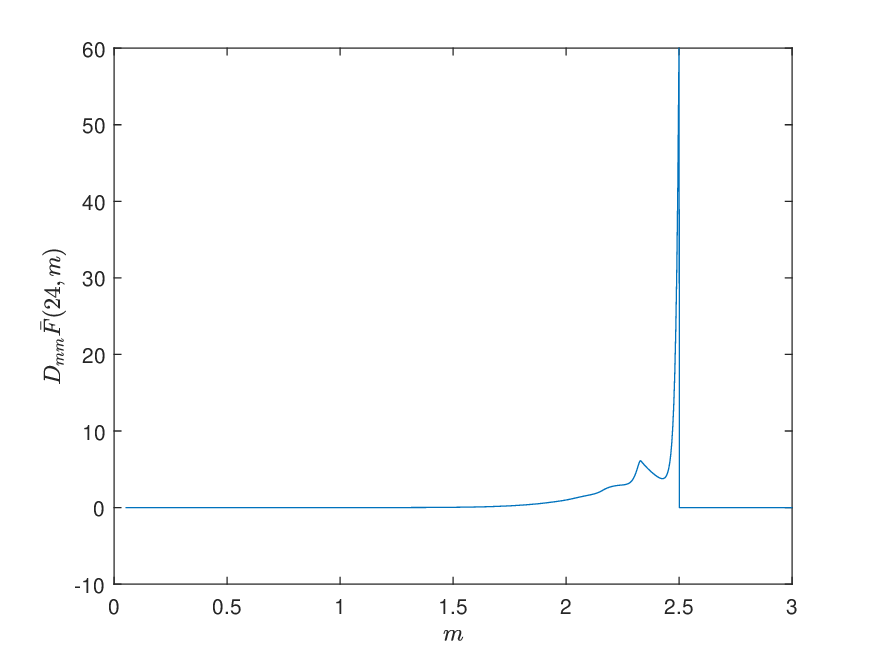}
\par\end{centering}
}
\par\end{centering}
\caption{\label{17}Figure of $D_{mm}\bar{F}$.}
\end{figure}

Moreover, we observe that for the same agent, the proportion of investments at the time of retirement is primarily determined by the retirement time rather than wealth. This is because $J(t, kx, kh, kw; \tau, kc, \pi) = k^{1-\gamma}J(t, x, h, w; \tau, c, \pi)$, which implies $\Pi^*(t, kx, kh, kw) = \Pi^*(t, x, h, w)$. Therefore, the optimal investment strategy is a function of $(t, \frac{x-p^T(t)h}{w})$. Furthermore, the retirement boundary exhibits a linear relationship in the $(x-p^T(t)h, w)$ plane. Consequently, the optimal investment strategies at retirement are identical. This observation suggests a similarity in investment strategies among agents retiring simultaneously.

\subsection{Retirement consumption puzzle} 

Our results provide an explanation for the retirement consumption puzzle. Since the last century, the standard life-cycle theory proposed by Modigliani and Brumberg (1954) \cite{modigliani1954utility} and the Permanent Income Hypothesis (LCPIH) proposed by Friedman (1957) \cite{friedman1957permanent} have become the basic framework for analyzing individuals' consumption and savings behaviors. Economists have proposed that if all future events are predictable, household consumption should exhibit smooth patterns. However, Battistin, Brugiavini, Rettore, and Weber (2009) \cite{battistin2009retirement} discovered a $9.8\%$ fall in consumption upon retirement, while Hurst (2008) \cite{hurst2008retirement} refuted their findings by suggesting that while consumption related to food and work does decline, other forms of consumption increase. Ameriks, Caplin, and Leahy (2007) \cite{ameriks2007retirement} demonstrated that $55\%$ of individuals anticipate a fall in consumption, while $8\%$ anticipate an increase. The complexity of this puzzle presents a significant challenge in terms of providing a comprehensive explanation. 

First, is the decline in consumption upon retirement truly a puzzle? Our perspective suggests that individuals prioritize their marginal utility over smoothing their consumption. Although traditional life-cycle theory proposes that consumption should be smoothed over an individual's lifetime to maintain a constant level of utility, our model shows a discontinuity in the utility function at retirement. To ensure smoothness in marginal utility, a corresponding jump in consumption is required. 

Second, it is crucial to investigate the underlying reasons for the divergent empirical findings regarding the retirement consumption puzzle. Upon examining the figure, it becomes evident that the magnitude of risk aversion plays a crucial role in determining whether a jump up or a jump down occurs in retirement consumption. Specifically, when the coefficient of risk aversion, denoted as $\gamma$, is greater than 1, consumption experiences an upward jump. Conversely, when $\gamma$ is less than 1, consumption encounters a downward jump. As the empirical studies under consideration examine agents with different levels of risk aversion, it follows that their respective findings diverge accordingly. Notably, our proposed model provides a comprehensive explanation for all of these empirical observations.

Last, it is important to investigate how wealth and standard of living impact consumption jumps in retirement. The magnitude of the jump in consumption corresponds to the changes in life patterns experienced by individuals. Notably, individuals with higher initial wealth tend to reduce working hours, resulting in less change in leisure time and life patterns before and after retirement. Additionally, the size of the consumption jump is determined by actual wealth, rather than the habit level (standard of living). Higher habitual levels lead to elevated consumption levels and a relatively smaller proportional jump. It is widely believed that individuals with greater wealth have higher habit levels. Our model suggests that retirement has a smaller impact on the consumption of wealthier individuals. This is supported by empirical studies, which show that the top $25\%$ of wealthy households experience the least decline in consumption, while the bottom $25\%$ face the most significant decline. The consistency between our model and empirical findings validates its applicability in understanding the retirement consumption puzzle.

\section{Sensitivity Analysis}\label{section_nume}
\subsection{Influence of habit} We have incorporated habit levels into our model. Then how do the parameters related to habit levels impact the retirement boundary? To provide clarity, we show the evolution of retirement boundaries with respect to $x$ over time by fixing $h = 0.5$ and $w = 1$. Our primary interest lies in the effects of habit parameters on the agent's retirement decision. We plot retirement boundaries for three sets of habit parameters: $(\alpha, \beta) = (0.2, 0.4)$ (benchmark), $(0.1, 0.4)$, and $(0.3, 0.4)$. The results are presented in \Cref{11}.
\begin{figure}[ht]
\begin{centering}
\subfloat[$\gamma = 0.5$]{
\label{Fig.sub.11.1}\begin{centering}
\includegraphics[width=4.5cm]{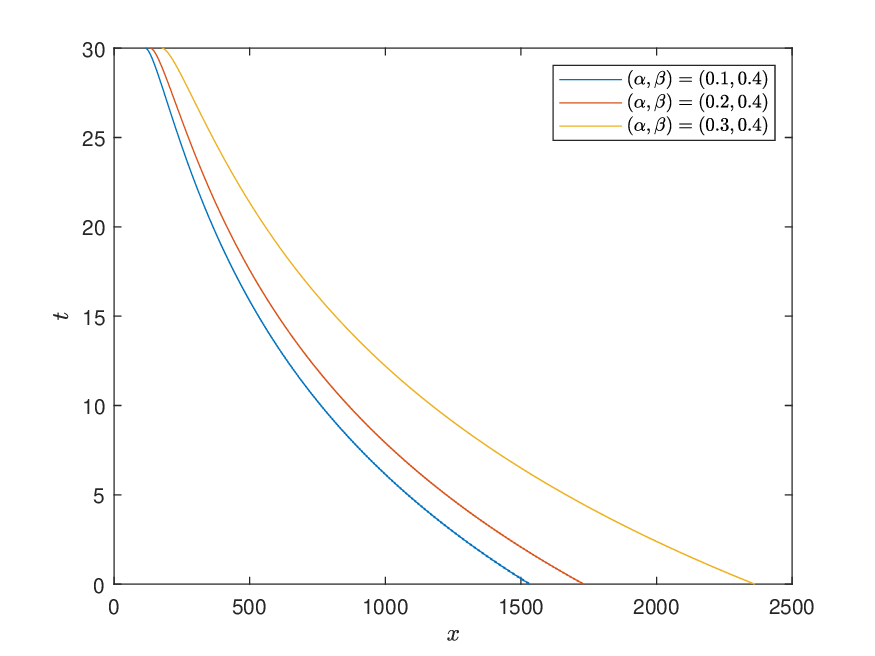}
\par\end{centering}
}
\subfloat[$\gamma = 0.6$]{
\label{Fig.sub.11.2}\begin{centering}
\includegraphics[width=4.5cm]{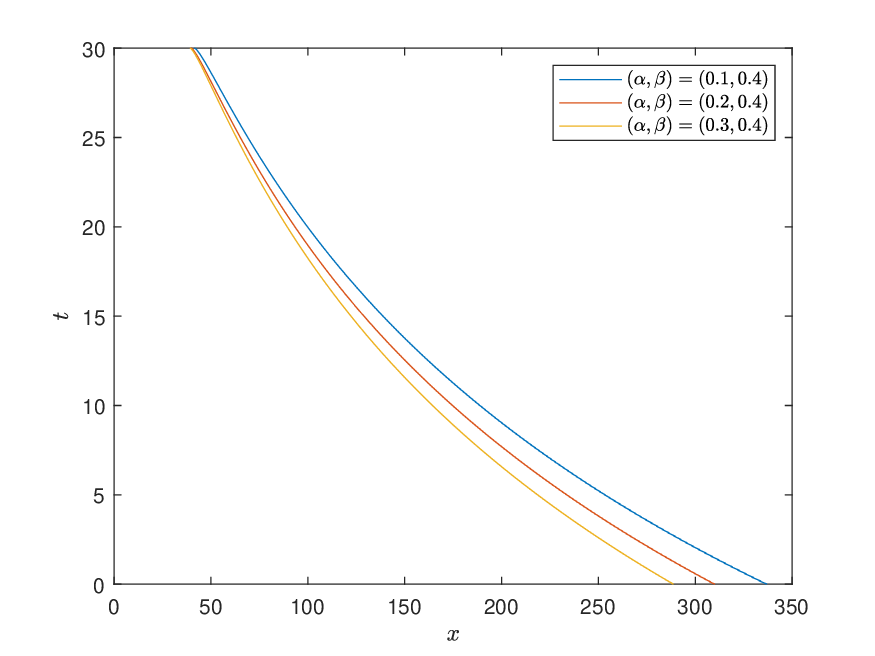}
\par\end{centering}
}
\subfloat[$\gamma = 1.5$]{
\label{Fig.sub.11.3}\begin{centering}
\includegraphics[width=4.5cm]{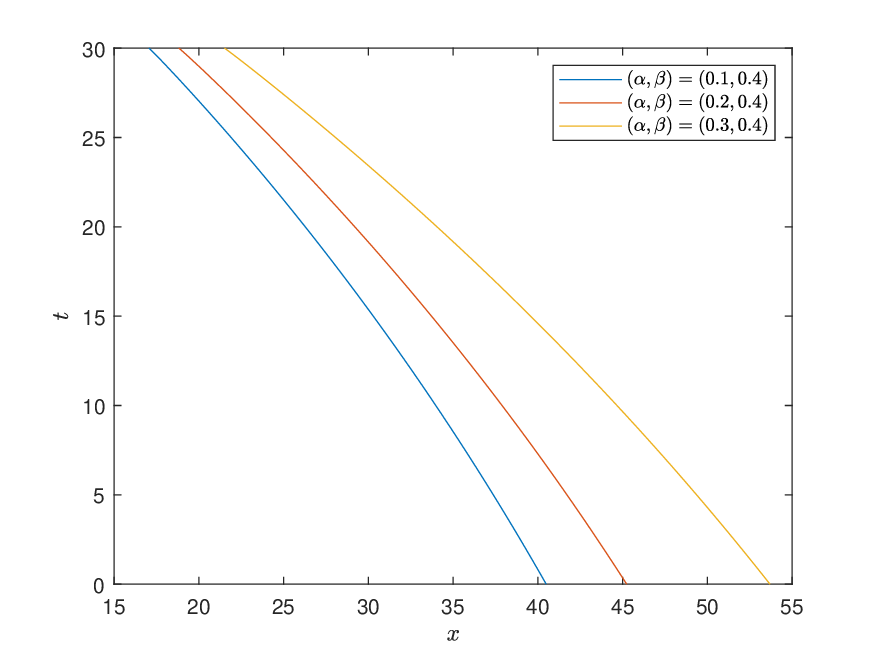}
\par\end{centering}
}
\par\end{centering}
\caption{\label{11}Retirement boundary in $(x, t)$-axis, $h = 0.5, w = 1$.}
\end{figure}
A common assumption is that, while holding other conditions constant, if the agent has a stronger desire to maintain her previous standard of living (reflected by a larger value of $\alpha$), he would retire earlier. While this phenomenon holds true in certain cases ($\gamma=0.5, 1.5$ in \Cref{11}), our numerical results suggest otherwise. Specifically, our results indicate that when $\gamma = 0.6$, the required wealth for retirement actually decreases as $\alpha$ increases. 
\subsection{Introduction of Jump diffusion}

In our model, we introduce jump diffusion. While similar to Brownian motion in many calculations and proofs presented in this paper (such as the expression for $q(s)$ in (\ref{eq_q}), the duality in \Cref{th_dual} and \Cref{th_dual_retire}), jump diffusion models discontinuous changes in state.
Therefore, peaks at the boundary may potentially influence the interior due to jump effects, resulting in peaks in the optimal strategies within continuous regions as well. We conduct numerical simulations to examine the effects of various market scenarios on optimal portfolio, consumption, and retirement decisions. Specifically, we examine two scenarios: (1) all risks arise from Brownian motions, and (2) one risk arises from a jump while another risk arises from Brownian motion. The optimal consumption strategies are displayed in \Cref{15}. 

(a) Pure Brownian motion when $d = 2$.
\begin{center}
\begin{tabular}{cccccccc}
   \hline
   $\mu_1$ &$\sigma_{1, 1}$ &$\sigma_{1, 2}$ &$r$     &$\mu_w$ &$\sigma_{w, 1}$ &$\sigma_{w,2}$ &$\gamma$\\
   $0.05$  &$0.2$      &$0.1$      &$0.01$  &$0.02$  &$0.1$      &$0$        &$0.5/1.5$\\
   \hline
   $\mu_2$ &$\sigma_{2, 1}$ &$\sigma_{2, 2}$ &$\rho$  &$\alpha$&$\beta$    &$T_1$     &$T$\\
   $0.04$  &$0.1$      &$0.2$      &$0.01$  &$0.2$   &$0.4$      &$30$       &$45$\\
   \hline
\end{tabular}
\end{center}

(b) Brownian motion + jump  when $d = 1$.
\begin{center}
\begin{tabular}{cccccccc}
   \hline
   $\mu_1$ &$\sigma_{1, 1}$ &$\gamma_{1, 2}$ &$r$     &$\mu_w$ &$\sigma_{w, 1}$ &$\gamma_{w,2}$ &$\gamma$\\
   $0.05$  &$0.2$      &$0.1$      &$0.01$  &$0.02$  &$0.1$      &$0$        &$0.5/1.5$\\
   \hline
   $\mu_2$ &$\sigma_{2, 1}$ &$\gamma_{2, 2}$ &$\rho$  &$\alpha$&$\beta$    &$T_1$     &$T$\\
   $0.04$  &$0.1$      &$0.2$      &$0.01$  &$0.2$   &$0.4$      &$30$       &$45$\\
   \hline
\end{tabular}
\end{center}

\begin{figure}[ht]
\begin{centering}
\subfloat[pure Brownian motion]{
\label{Fig.sub.15.3}\begin{centering}
\includegraphics[width=4.5cm]{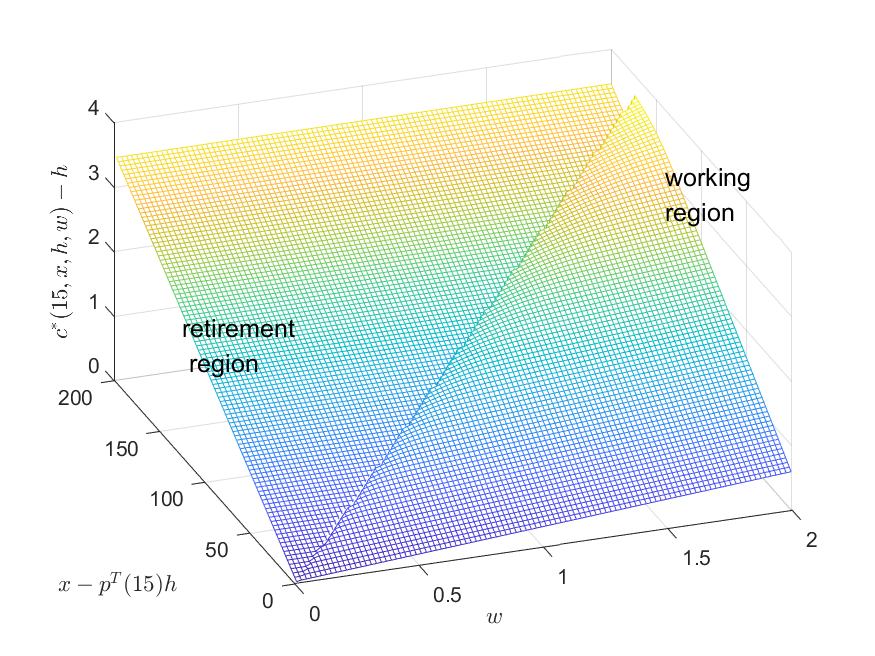}
\par\end{centering}
}
\subfloat[Brownian motion and Jump]{
\label{Fig.sub.15.1}\begin{centering}
\includegraphics[width=4.5cm]{image/gamma=1.5_c_star_t=15.eps}
\par\end{centering}
}
\par\end{centering}
\caption{\label{15} Optimal consumption $c^*$, $\gamma = 1.5$, $t = 15$.}
\end{figure}

\begin{figure}[ht]
\begin{centering}
\subfloat[pure Brownian motion]{
\label{Fig.sub.13.3}\begin{centering}
\includegraphics[width=4.5cm]{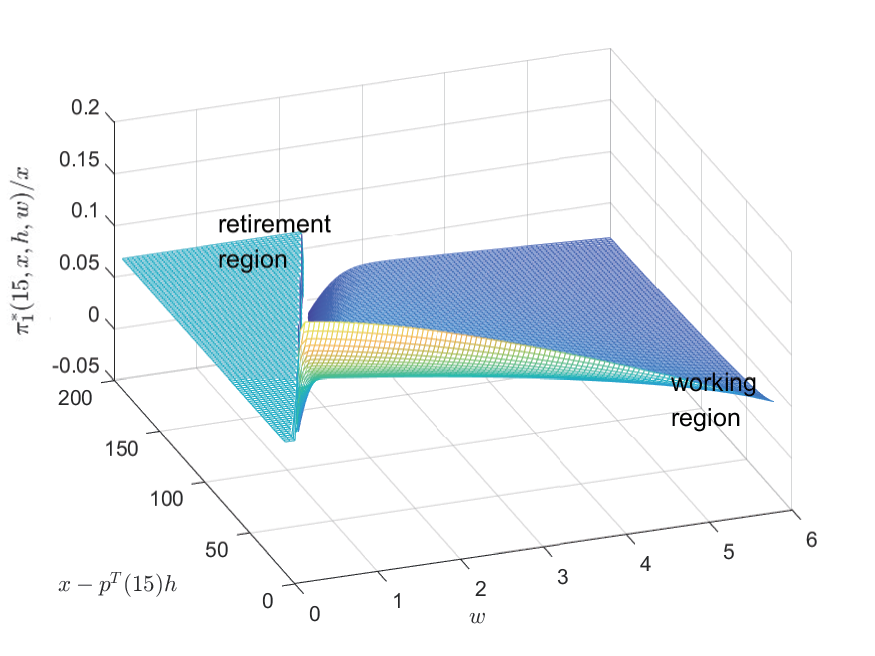}
\par\end{centering}
}
\subfloat[Brownian motion and Jump]{
\label{Fig.sub.13.1}\begin{centering}
\includegraphics[width=4.5cm]{image/gamma=1.5_pi1_t=15.eps}
\par\end{centering}
}
\par\end{centering}
\caption{\label{13} Optimal portfolio $\pi_1^* / x$, $\gamma = 1.5$, $t = 15$.}
\end{figure}

\begin{figure}[ht]
\begin{centering}
\subfloat[pure Brownian motion]{
\label{Fig.sub.14.3}\begin{centering}
\includegraphics[width=4.5cm]{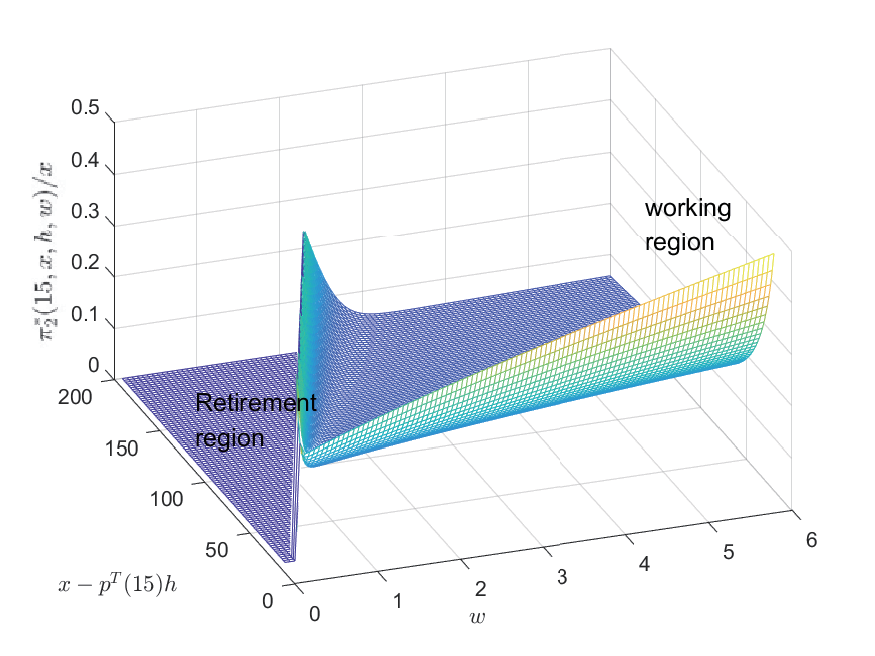}
\par\end{centering}
}
\subfloat[Brownian motion and Jump]{
\label{Fig.sub.14.1}\begin{centering}
\includegraphics[width=4.5cm]{image/gamma=1.5_pi2_t=15.eps}
\par\end{centering}
}
\par\end{centering}
\caption{\label{14} Optimal portfolio $\pi_2^* /x$, $\gamma = 1.5$, $t = 15$.}
\end{figure}
The optimal portfolios when $\gamma=1.5$ are presented in \Cref{13} and \Cref{14}. We find that the incorporation of jump produces a distinctive peak in the optimal portfolio strategy near the retirement boundary.  The reason is that the function $D_{mm}\bar{F}(t, m)$ exhibits a steep ascent near the boundary, and this ascent is transited into the interior due to the jump diffusion model, resulting in the presence of a peak in the optimal strategy. In our numerical example, the number of apparent peaks corresponds to the number of jump risks $(n - d)$. {This effect is even more pronounced at $\gamma = 1.5$.} \Cref{17} provides a visual representation of the function $D_{mm}\bar{F}(t, m)$. In summary, our findings emphasize the significant impact of risk type on the optimal strategy, which gives rise to distinct peaks near the boundaries. 

\subsection{Correlation between wage and market}
The connection between wages and markets is typically strong, with prosperous markets often leading to higher wage levels. Nevertheless, wage levels are affected by various factors, such as economic circumstances, market demand, and policy influences. Consequently, the relationship between wages and markets is intricate and necessitates a comprehensive model capable of capturing this relationship. Our model achieves this by quantifying the association between markets and wages through correlation coefficients (exposure) with respect to different risk factors. To examine the impact of the correlation between wage risk and stock risk on investment strategies, we commence by orthogonalizing the stocks. The reason for not orthogonalizing the stocks in the original problem is that the return rate of orthogonalized stocks is lower than the risk-free rate, which appears less realistic. We then proceed to examine two scenarios that pertain to optimal investment portfolios, consumption, and retirement decisions. In the first scenario, the wage risk is positively related to the first stock. In other words, the source of risk of wage and the first stock is the same. In the second case, the wage risk is positively related to the second stock. In other words, the source of risk of wage and the second stock is the same.
The parameter settings for these two scenarios are as follows. 

(I) The wage risk is positively related to the first stock.
\begin{center}
\begin{tabular}{cccccccc}
   \hline
   $\mu_1$ &$\sigma_{1, 1}$ &$\gamma_{1, 2}$ &$r$     &$\mu_w$ &$\sigma_{w, 1}$ &$\gamma_{w,2}$ &$\gamma$\\
   $0.05$  &$0.2$      &$0.1$      &$0.01$  &$0.02$  &$\frac{0.2}{\sqrt{5}}$      &$\frac{0.1}{\sqrt{5}}$        &$0.5$\\
   \hline
   $\mu_2$ &$\sigma_{2, 1}$ &$\gamma_{2, 2}$ &$\rho$  &$\alpha$&$\beta$    &$T_1$     &$T$\\
   $0.008$  &$-0.06$      &$0.12$      &$0.01$  &$0.2$   &$0.4$      &$30$       &$45$\\
   \hline
\end{tabular}
\end{center}

(II) The wage risk is positively related to the second stock.
\begin{center}
\begin{tabular}{cccccccc}
   \hline
   $\mu_1$ &$\sigma_{1, 1}$ &$\gamma_{1, 2}$ &$r$     &$\mu_w$ &$\sigma_{w, 1}$ &$\gamma_{w,2}$ &$\gamma$\\
   $0.05$  &$0.2$      &$0.1$      &$0.01$  &$0.02$  &$-\frac{0.1}{\sqrt{5}}$      &$\frac{0.2}{\sqrt{5}}$        &$0.5$\\
   \hline
   $\mu_2$ &$\sigma_{2, 1}$ &$\gamma_{2, 2}$ &$\rho$  &$\alpha$&$\beta$    &$T_1$     &$T$\\
   $0.008$  &$-0.06$      &$0.12$      &$0.01$  &$0.2$   &$0.4$      &$30$       &$45$\\
   \hline
\end{tabular}
\end{center}

\begin{figure}[ht]
\begin{centering}
\subfloat[Case I, $\pi^*_1/x$]{
\label{Fig.sub.18.1}\begin{centering}
\includegraphics[width=4.5cm]{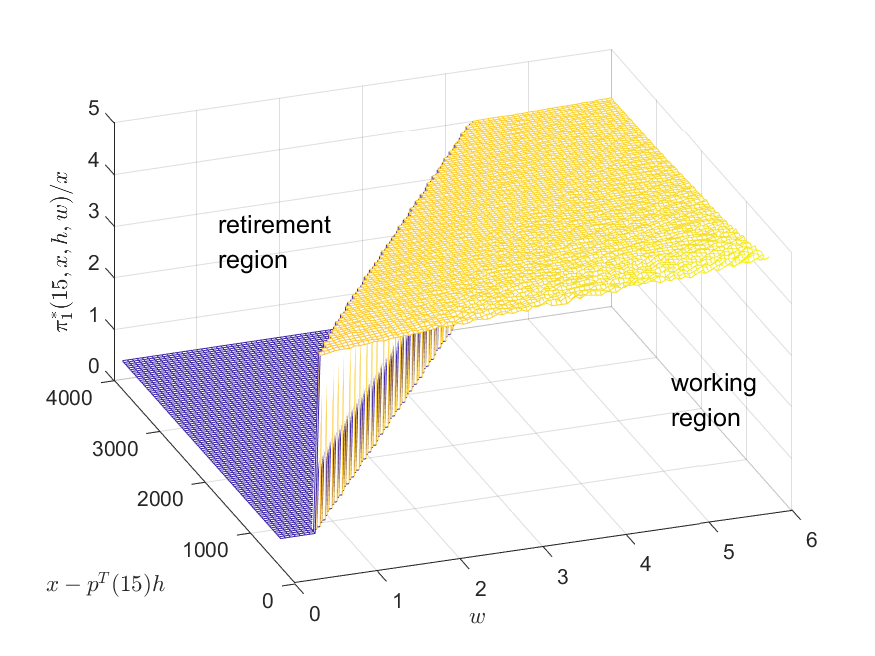}
\par\end{centering}
}
\subfloat[Case II, $\pi_1^*/x$]{
\label{Fig.sub.18.2}\begin{centering}
\includegraphics[width=4.5cm]{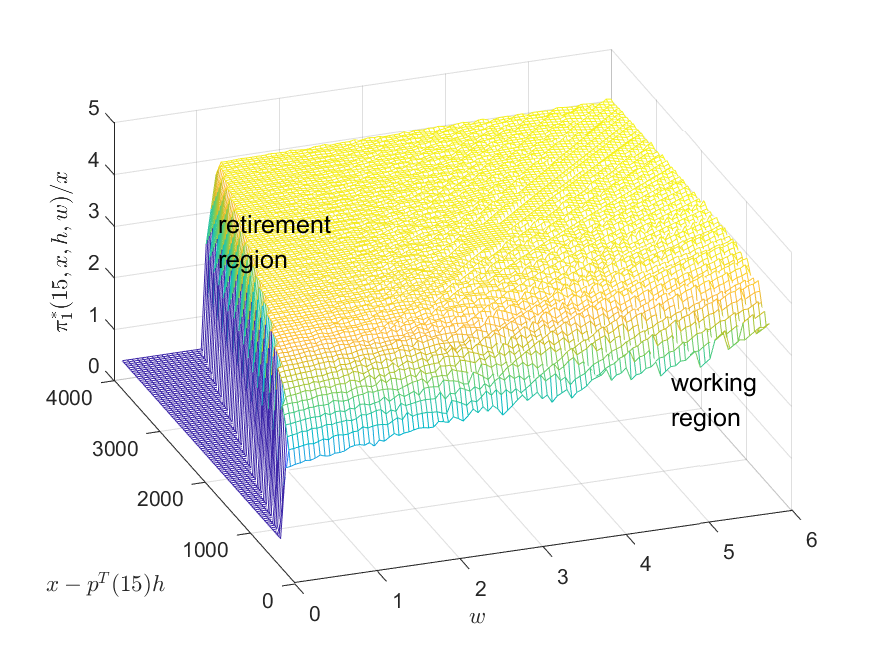}
\par\end{centering}
}

\subfloat[Case I, $\pi_2^*/x$]{
\label{Fig.sub.18.3}\begin{centering}
\includegraphics[width=4.5cm]{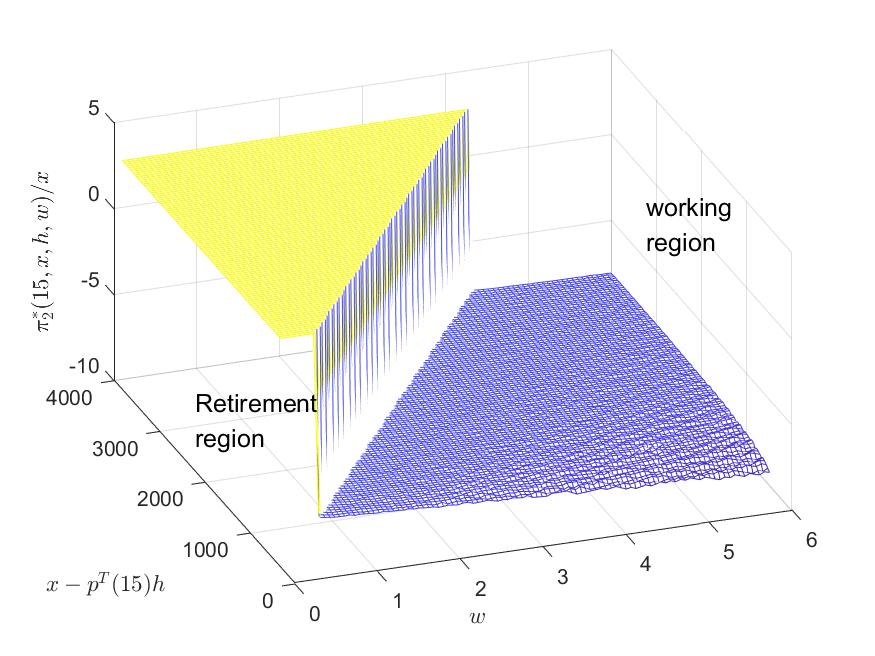}
\par\end{centering}
}
\subfloat[Case II, $\pi_2^*/x$]{
\label{Fig.sub.18.4}\begin{centering}
\includegraphics[width=4.5cm]{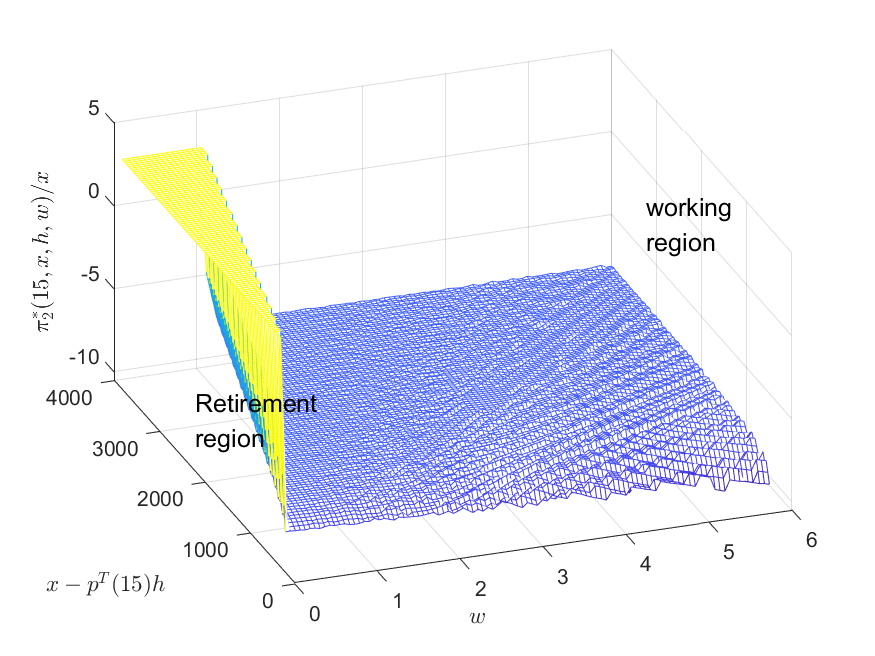}
\par\end{centering}
}
\par\end{centering}
\caption{\label{18} Optimal portfolio $t = 15$,$\gamma = 0.5$.}
\end{figure}
\vskip 5pt

{The optimal portfolios are presented in \Cref{18}. An intuitive conjecture suggests that the agent would use stocks as a hedging tool for wage risk. As the agent's wage increases, the proportion of wealth allocated to stocks for hedging wage risk should also increase. With an increase in wage, the agent should increase the proportion of negatively correlated stocks to hedge against their wage volatility and sell positively correlated stocks. However, \Cref{18} does not support this conjecture. We can observe that although the volatility of the agent's wage is proportional to the second stock, the proportion of the second stock in the investment portfolio increases as the agent's wage increases. From the form of optimal portfolio $\sigma_M^{-1} \big(-\sigma_{wM} q(t) w + D(W_y)(t, y, w))$, we see that the first component of the optimal portfolio is used to hedge against wage fluctuations, which aligns with our intuition. But the second component is brought by the uncertainty in retirement time. In our numerical setup, the second term has a more significant impact than the first term. Thus the proportion of the second stock increases.}

Furthermore, we find that the correlation between wage volatility and stock price processes has a significant impact on the retirement boundary. \Cref{20} illustrates the scenarios in which the correlation coefficient between wage volatility and the volatility of the first stock is 1 and -1, respectively. The parameter settings for these two scenarios are as follows.

(I) The wage risk is positively related to the first stock:
\begin{center}
\begin{tabular}{cccccccc}
   \hline
   $\mu_1$ &$\sigma_{1, 1}$ &$\gamma_{1, 2}$ &$r$     &$\mu_w$ &$\sigma_{w, 1}$ &$\gamma_{w,2}$ &$\gamma$\\
   $0.05$  &$0.2$      &$0.1$      &$0.01$  &$0.02$  &$\frac{0.2}{\sqrt{5}}$      &$\frac{0.1}{\sqrt{5}}$        &$0.5$\\
   \hline
   $\mu_2$ &$\sigma_{2, 1}$ &$\gamma_{2, 2}$ &$\rho$  &$\alpha$&$\beta$    &$T_1$     &$T$\\
   $0.008$  &$-0.06$      &$0.12$      &$0.01$  &$0.2$   &$0.4$      &$30$       &$45$\\
   \hline
\end{tabular}
\end{center}

(III) The wage risk is negatively related to the first stock:
\begin{center}
\begin{tabular}{cccccccc}
   \hline
   $\mu_1$ &$\sigma_{1, 1}$ &$\gamma_{1, 2}$ &$r$     &$\mu_w$ &$\sigma_{w, 1}$ &$\gamma_{w,2}$ &$\gamma$\\
   $0.05$  &$0.2$      &$0.1$      &$0.01$  &$0.02$  &$-\frac{0.2}{\sqrt{5}}$      &$-\frac{0.1}{\sqrt{5}}$        &$0.5$\\
   \hline
   $\mu_2$ &$\sigma_{2, 1}$ &$\gamma_{2, 2}$ &$\rho$  &$\alpha$&$\beta$    &$T_1$     &$T$\\
   $0.008$  &$-0.06$      &$0.12$      &$0.01$  &$0.2$   &$0.4$      &$30$       &$45$\\
   \hline
\end{tabular}
\end{center}

\begin{figure}[ht]
\begin{centering}
\subfloat[Case I, $\pi_1^*/x$]{
\label{Fig.sub.20.1}\begin{centering}
\includegraphics[width=4.5cm]{image/pi1_t=15_gamma=0.5_w_1.eps}
\par\end{centering}
}
\subfloat[Case II, $\pi_1^*/x$]{
\label{Fig.sub.20.2}\begin{centering}
\includegraphics[width=4.5cm]
{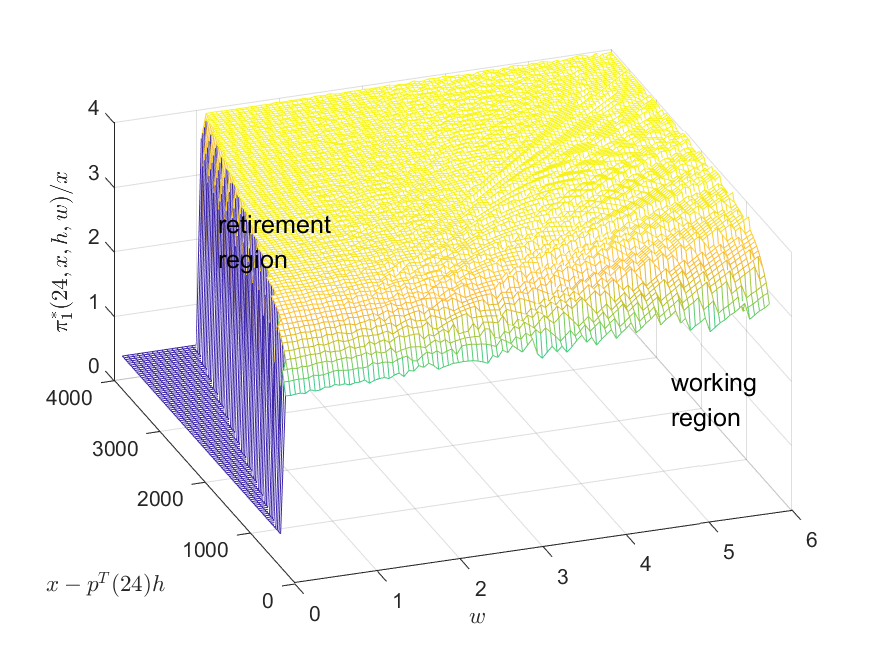}
\par\end{centering}
}

\subfloat[Case I, $\pi_2^*/x$]{
\label{Fig.sub.20.3}\begin{centering}
\includegraphics[width=4.5cm]{image/pi2_t=15_gamma=0.5_w_1.eps}
\par\end{centering}
}
\subfloat[Case II, $\pi_2^*/x$]{
\label{Fig.sub.20.4}\begin{centering}
\includegraphics[width=4.5cm]
{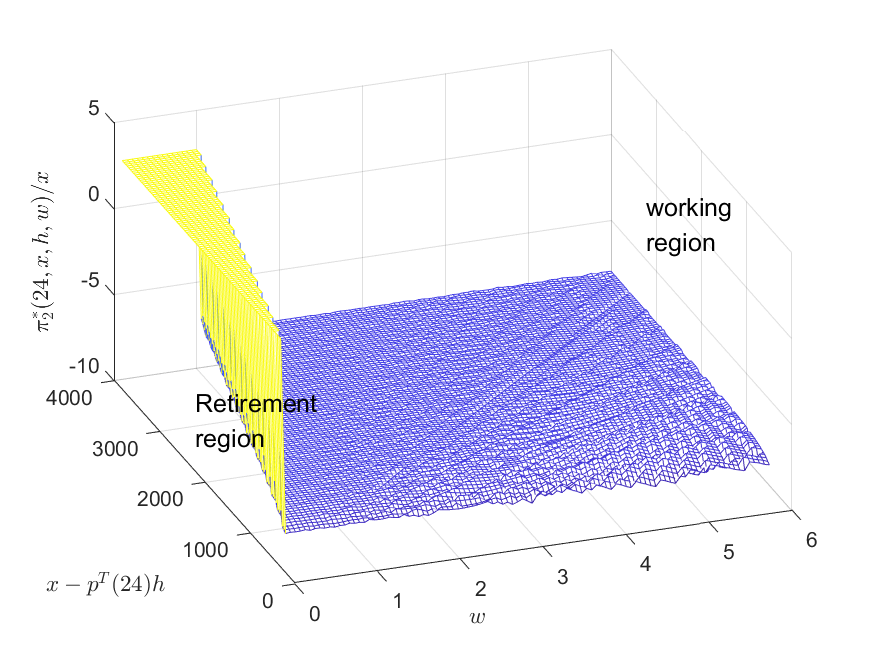}
\par\end{centering}
}
\par\end{centering}
\caption{\label{20} Optimal portfolio $t = 15$,$\gamma = 0.5$.}
\end{figure}
Therefore, we can conclude that the source of wage risk is critical for retirement concerns.

\subsection{Percentage of individual stocks in the optimal portfolio}

Usually, the optimal investment portfolio is determined by the highest Sharpe ratio. However, an intriguing discovery challenges this notion. In our example, we find that even though a portfolio comprising solely of the first stock has the highest Sharpe ratio, it is not the optimal outcome that we compute.

First, the agent needs to utilize the stock market as a means to hedge against wage fluctuations. Additionally, the agent must hedge against variations in retirement timing. This indicates that a diversified portfolio in the market becomes essential as different agents retire at different times, necessitating unique portfolios. Therefore, achieving the optimal portfolio for agents cannot be solely accomplished by hedging wage fluctuations based on the portfolio with the highest Sharpe ratio.

When $d = n$, by \Cref{lm_dual_cal}, we obtain the following equivalent form of the optimal investment strategy before retirement
\begin{align*}
    \Pi^*(t, x, h ,w) =&w  D_y \bigg\{\tilde{\e}\bigg[  e ^{\vartheta (\tauop - t)} q(\tauop) y-q(t)\bigg]\bigg\} \sigma_M^{-1} \sigma_{w, M}+   y W_{yy}(t, y, w) \sigma_M^{-1}\theta_0 .
\end{align*}
We can observe from this form that the optimal investment portfolio comprises two components. The first component is utilized to hedge against the shortfall in wages due to early retirement, while the second component is related to a portfolio associated with the highest Sharpe ratio. 

When $d<n$, we cannot obtain a similar form. However, the optimal portfolio is still not the combination that maximizes the Sharpe ratio.

\section{Conclusions}\label{section_conclude}
In this paper, we present a comprehensive analysis of retirement decision-making and optimal consumption-investment models that incorporate addictive habit formation and jump diffusion. Our model stands out by considering three critical state variables - wealth ($x$), habit ($h$), and wage rate ($w$) - to capture the intricacies of decision-making.

To address the challenges posed by jump diffusion in the optimization problem, we propose a novel dual approach that involves converting it into an optimal stopping problem. Through a measure change approach, we successfully overcome the complexities associated with the three-dimensional optimal stopping problem. This conversion demonstrates the potential of our approach to address a broad range of high-dimensional optimal stopping problems. Furthermore, our analysis has uncovered several intriguing findings, some of which align with empirical studies. These results contribute to a better understanding of individuals' retirement decisions, providing insights into how various factors such as market conditions, wage volatility, and habit formation influence investment strategies and retirement outcomes.

Overall, this research expands on the existing literature on retirement decision models by providing a comprehensive framework for incorporating addictive habit formation and jump diffusion. The insights gained from this study have practical implications for individuals and policymakers who seek to develop effective retirement strategies.



\appendix
\section{Proof of  Theorem \ref{th_dual}}\label{append_pf_th_dual}
To verify the dual relation, we provide some classical arguments from the proof of the main results in Karatzas and Wang (2000) \cite{karatzas2000utility}. Although our financial model with habit formation differs from that of Karatzas and Wang (2000) \cite{karatzas2000utility}, the procedure to prove the theorem is similar but much more complex. Recall that we have already established the inequality (\ref{V_W_ineq}), i.e., we have one side of the inequality in (\ref{dual_v_w})

\begin{equation}
	V(t,x_F, w) - y\big(x_F + q(t)w\big) \leq W(t, y, w).
\end{equation}

Next, we aim to verify the converse inequality of (\ref{dual_v_w}). To do so, it is sufficient to demonstrate that (\ref{dual_v_w}) holds for a particular value of $y$. We present the following lemmas.

\begin{lemma}\label{lm_U}
    We have the following asymptotic results for function $V_{F, p}$
\begin{subequations}
\begin{align}
        &\lim_{y \to 0} \inf_{0 \leq t \leq T_1} I_{V_{F, p}}(t, y) = \infty, \label{eq_in}\\
        &\lim_{y \to \infty} \sup_{0 \leq t \leq T_1} I_{V_{F, p}}(t, y) = 0,  \label{eq_0}\\
        &\sup_{0 \leq t \leq T_1} I_{V_{F, p}} (t, y) \leq C(1 + y ^{- K}), \label{ineq}
\end{align}
\end{subequations}
    where $C$ and $K > 0$ are constants.
\end{lemma}

\begin{proof}
    By the convexity of $V^\mathrm{d}$, we conclude
    \begin{align*}
        I_{V_{F, p}}(t, y) &= -\partial_y V^\mathrm{d}(t, y)\\
        &= \e \bigg[\int_{t}^{T} \xi^t(u) I_{U_{F, 2}}(u, yY^{t, 1}(u)) \D u \Bigg] \\
        &= \e_{\Q} \bigg[\int_{t}^{T} e^{-r u} I_{U_{F, 2}}(u, yY^{t, 1}(u)) \D u \bigg], 
    \end{align*}
     which follows from bounded assumption of $I_{U_{F, 2}}$. \Cref{eq_0} is derived from the asymptotic properties of $I_{U_{F, 2}}$ and the monotone convergence theorem. Furthermore, we can obtain \Cref{eq_in} using similar arguments, as we can estimate it as follows
     \begin{equation}
        \begin{aligned}
          I_{V_{F, p}}(t, y) \geq \e_\Q \left[ \int_{T}^{T_1} e^{-ru}I_{U_{F, 2}}(u, yY ^{t, 1}(u)) \D u\right].
     \end{aligned}
     \end{equation}
\Cref{ineq} is derived from $I_{U_{F, 2}}(t, x) \leq C( 1+x^{-K})$.
\end{proof}        
{ 
\begin{lemma} \label{lm:w_t_y_w_convex}
For fixed $t$, $W(t, y, w)$ is  convex with respect to $(y, w)$.
\end{lemma}
\begin{proof}
Denote 
\begin{equation*}
    \varphi (t, y, w, \tau) = \e^{t, y, w}\bigg[e^{-\rho(\tau - t)} \bigg(V^{\mathrm{d}}\big(\tau, Y(\tau), \big) -  Y(\tau) q(\tau) \mathcal{W}(\tau)\bigg)+ \int_{t}^{\tau}e^{-\rho(\tau - t)} U^{\mathrm{d}}_1\big(u, Y(u)\big) \D u\bigg].
\end{equation*}
It is easy to see that for fixed $(t, \tau)$, $\varphi (t, y, w, \tau)$ is a convex function with respect to $(y, w)$. Therefore, $ \sup_{\tau \in \St}  \varphi (t, y, w, \tau) $ is a convex function with respect to $(y, w)$ for fixed $t$.
\end{proof}}
\begin{lemma}\label{lm_dual_cal}
    For $\tau \in \St $ and fixed $t \in [0, T_1]$, define
    \begin{equation*}
         \widehat{x}_{t, \tau}(y) = \e_\Q \left [ \int_{t}^{\tau} e^{-ru}I_{U_{F, 1}}\big(Y^{t, y}(u)\big) \D u + e^{-r\tau}I_{V_{F, p}}(t, Y^{t, y}(\tau)) \right].
    \end{equation*}
    Then, $\forall x_0 > 0$, there exists $y > 0$ such that 
$ \widehat{x}_{t, \hat{\tau}_y}(y) = x_0.$ Furthermore, $W$ is differentiable in $y\in(0, \infty)$ and
\begin{equation} \label{eq_w_chi}
    W_y(t, y, w) = - \widehat{x}_{t, \hat{\tau}_y}(y) - \e_\Q \left[  e^{-r\hat{\tau}_y}\mathcal{W}^{t, w}(\hat{\tau}_y) q(\hat{\tau}_y)\right].
\end{equation}
\end{lemma}
\begin{proof}
{ 
Notice the asymptotic properties of $I_{U_{F, 1}}$ and $I_{V_{F, p}}$ in \Cref{lm_U},
$I_{U_{F, 1}}(t, x) \leq C(1 + x^{-K})$ and \Cref{ineq}, using the monotone convergence theorem, we obtain
\begin{equation*}
    \lim_{y \to 0}  \widehat{x}_{t, \hat{\tau}_y}(y)  = \infty, \lim_{y \to \infty}  \widehat{x}_{t, \hat{\tau}_y}(y)  = 0.
\end{equation*}
}
We claim that $ \widehat{x}_{t,\hat{\tau}_y} (y)$ 
is continuous in $y$. In fact, the estimation
\begin{equation*}
     \widehat{x}_{t, \hat{\tau}_y}(y) \leq C (2 T_1  + 1) (\sup_{t \leq s \leq T_1} \xi^t(s))\bigg(1 + \delta^{-K} \sup_{t \leq s \leq T_1}\big(Y^{t, 1}(s))^{-K} \bigg)
\end{equation*}
holds uniformly for $y \in [\delta, \infty)$. The right side of the inequality is integrable by Burkholder-Davies-Gundy's inequality. As such, using the dominated convergence theorem, we conclude that $ \widehat{x}_{t,\tauop}
(y)$ is continuous in $y$ on $[\delta, \infty]$. As $\delta > 0$ is arbitrary, $ \widehat{x}_{t,\tauop}
(y)$ is continuous in $(0, \infty)$. Therefore, the mapping $(0, \infty) \to (0, \infty)$ : $y \to  \widehat{x}_{t,\tauop}
(y)$ is surjective. To prove (\ref{eq_w_chi}), we begin with an estimation of the differential of $W$ based on convexity. For $\eta \in \R$ near 0, we obtain the following
{ 
\begin{align*}
    W(t, y + \eta, w) &- W(t, y, w) \\
     \leq &\e  \bigg[\int_{t}^{\hat{\tau}_{y + \eta}} e^{-\rho(u - t)} U^{\mathrm{d}}_1\big(u, (y + \eta) Y^{t, 1}(u)\big) - U^{\mathrm{d}}_1 \big(u, y Y^{t, 1} (u)\big) \D u \bigg] \\
    &+ \e^{-\rho (\hat{\tau}_{y + \eta} - t)} (V^{\mathrm{d}}\big(t, (y + \eta) Y^{t, 1} (\hat{\tau}_{y + \eta})\big)\\
    &- \e  e^{-\rho (\hat{\tau}_{y + \eta} - t)}V^{\mathrm{d}}(t, y Y^{t, 1} (\hat{\tau}_{y + \eta}))\\
    &- \eta w \e e^{-\rho(u - t)}Y^{t, 1}(\hat{\tau}_{y + \eta})\mathcal{W}^{t, 1}(\hat{\tau}_{y + \eta}) \\
    \leq & \eta \e \bigg[ \int_{t}^{\hat{\tau}_{y + \eta}} e^{\rho(u - t)} Y^{t, 1} (u)\partial_yU^{\mathrm{d}}_1(Y^{t, y + \eta}(u)) \D u \bigg] \\
    &+ \eta\e  e^{-\rho (\hat{\tau}_{y + \eta} - t)}Y^{t, 1}(\hat{\tau}_{y + \eta}) \partial_y V^{\mathrm{d}} (t, Y^{t, y + \eta}(\hat{\tau}_{y + \eta}))\\
    &- \eta w \e e^{-\rho (\hat{\tau}_{y + \eta} - t)} Y^{t, 1}(\hat{\tau}_{y + \eta}) \mathcal{W}^{t, 1}(\hat{\tau}_{y + \eta}) q(\hat{\tau}_{y + \eta}) \\
    =& -\eta \e \bigg[\int_{t}^{\hat{\tau}_{y + \eta}} \xi^t(u) U_{F, 1}(Y^{t, y + \eta}(u)) \D u + \xi^t(\hat{\tau}_{y + \eta}) V^{\mathrm{d}}(t, Y^{t, y + \eta}(\hat{\tau}_{y + \eta}))\bigg] \\
    &+ \eta w \e e^{-\rho (\hat{\tau}_{y + \eta} - t)} Y^{t, 1} (\hat{\tau}_{y + \eta} \mathcal{W}^{t, 1} (\hat{\tau}_{y + \eta}))\\
    =& - \eta  \widehat{x}_{t, \hat{\tau}_{y + \eta}}(y + \eta) - \eta  \e_\Q e^{-r\hat{\tau}_{y + \eta}} \mathcal{W}^{t, w}(\hat{\tau}_{y + \eta}) q(\hat{\tau}_{y + \eta}).
\end{align*}}
Dividing by $\eta$ on both sides of the last inequalities and letting $\eta \to 0+$, we have
\begin{equation*}
    W_{y + }(t, y, w) \leq -  \widehat{x}_{t,\tauop}
(y) - \e_\Q e^{-r\hat{\tau}_{y + \eta}}
\mathcal{W}^{t, 1} (\hat{\tau}_y) q(\hat{\tau}_y).
\end{equation*}
Conversely, letting $\eta \to 0^-$, we get
\begin{equation*}
    W_{y - }(t, y, w) \geq - \widehat{x}_{t,\tauop}
(y)- \e_\Q e^{-r\hat{\tau}_{y + \eta}}
\mathcal{W}^{t, 1} (\hat{\tau}_y) q(\hat{\tau}_y).
\end{equation*}
Combining the last two inequalities and \Cref{lm:w_t_y_w_convex} implies that (\ref{eq_w_chi}) holds.
\end{proof}

\begin{proof}[Proof of (\ref{dual_v_w})]
    First, $\e [b(\hat{\tau}_y) \xi^t(\hat{\tau}_y)]$ is continuous in $y$, which can be proved similarly as the continuity of $ \widehat{x}_{t,\tauop}
(y)$. 

As $ \widehat{x}_{t,\tauop}
(y)$ maps $(0, \infty)$ onto $(0, \infty)$,  {and $\lim\limits_{y \to \infty}\e [b(\hat{\tau}_{y}) \xi (\hat{\tau}_{y})] = 0$, then  $ \widehat{x}_{t,\tauop}
(y) + \e b(\tauop) \xi^t(\tauop)$ maps
$(0, \infty)$ onto $(0, \infty)$.}

We choose $y^* > 0$ such that
\begin{equation*}
     \widehat{x}_{t, \tauy} (y^*) + \e b(\tauy) \xi^t(\tauy) = x_F + q(t) w,
\end{equation*}
which is equal to
\begin{equation*}
    \e_\Q \left[ \int_{t}^{\tauy} e^{-ru}c_F^*(u) \D u + e^{-r \tauy}(D^*+ b(\tauy) ) \right] = x_F + b(t),
\end{equation*}
with $c_F^*(u) = I_{U_{F, 1}} (u, Y^{t, y^*}(u))$ and $D^* = I_{V_{F, p}}(\tauy)$. This implies that $c_F^*$ and $\hat{\tau}_{y^*}$ satisfy the equation in \Cref{lm_rep}.  Besides, all inequalities in (\ref{dual_inq}) become equalities. Thus, the proof is completed.
{ 
\begin{remark}
    Here, $y^*$ may not be unique in the proof of this theorem, but the proof of the theorem only requires the existence of $y^*$. We only need to prove that the inequality holds with equality. If $W(t, y, w)$ is strictly convex, $y^*$ is unique.
\end{remark}}
\end{proof}
\section{Proof of Theorem \ref{th:m_conti}} \label{AB}
\begin{lemma}\label{lem_m1}
   Define
    \begin{equation*}
        g_{L, t, m}(s) \define \bar{F}\big(t - s, m(1 - Ls)\big),
    \end{equation*}
    where $0 \leq s \leq \frac{1}{L}$. Then there exists an $L$ large enough such that, for any $(t, m) \in \mathcal{C}$, we have $g_{L, t, m}(s) > g_{L, t, m}(0)$.
\end{lemma}
\begin{proof}
    Define $\psi(t, m, \tau) = \tilde{\e}^{t, m} \bigg[\int_{t}^{\tau} e^{\vartheta (u - t)}\bigg(\Delta (u)  M(u) + 1\bigg) \D u\bigg]$ and we have $\bar{F}(t, m) = \sup\limits_{\tau} \psi(t, m, \tau)$.

  There exists a sufficiently large $L$ such that for any $(t, m) \in \mathcal{C}$, we can ensure $\Delta (u - s) M(u - sL) > \Delta (u) M(u)$. Consequently, we obtain $\psi(t - s, m(1 - Ls), \tau) > \psi(t, m, \tau)$, leading to the desired conclusion.
\end{proof}
Using \Cref{lem_m1}, we can get some properties of the curve $m^*(t)$.

\begin{lemma}\label{lem_m2}
    For $L$ in \Cref{lem_m1} and any $m^*(t) < \infty, s > 0$, we have $m^*(t - s) > m^*(t) (1 - 2Ls)$.
\end{lemma}
\begin{proof}
    According to \Cref{th_mon}, we have
    \begin{equation*}
        \bar{F}\big(t, m^*(t)(1 - Ls)\big) > \bar{F}\big(t, m^*(t)\big).
    \end{equation*}    
    From \Cref{lem_m1}, we get
        \begin{equation*}
            g_{L, t, m^*(t)(1 - Ls)}(s) \geq g_{L, t, m^*(t)(1 - Ls)}(0),
        \end{equation*}
    which means
    \begin{equation*}
        \bar{F}\big(t - s, m^*(t)(1 - 2Ls)\big) \geq \bar{F}\big(t, m^*(t)(1 - Ls)\big) .
    \end{equation*}
    Further, we have the following inequality from the definition of $m^*(t)$
    \begin{equation*}
        m^*(t - s) > m^*(t)(1 - 2Ls).
    \end{equation*}
\end{proof}

\begin{lemma}\label{lm_m}
    $m^*(t)$ is bounded on $[0, T_1)$.
\end{lemma}
\begin{proof}
    Suppose that there exists  a $t_0$ such that $m^*(t_0) < \infty$. According to \Cref{lem_m2}, we see $\forall ~ 0 < s \leq \frac{1}{4L},$
    \begin{equation*}
        m^*(t + s) < \frac{m^*(t)}{1 - 2Ls}.
    \end{equation*}
    By iteration, we obtain $\forall t \geq t_0$,
    \begin{equation*}
        m^*(t) < \infty .
    \end{equation*}
    Consider the set $\{t: m^*(t) = \infty\}$. If the set is empty, this theorem has been proved. Otherwise, let 
    \begin{equation*}
        t_1 \define \sup\big\{t: m^*(t) = \infty \big\},
    \end{equation*}
    and using \Cref{lem_m2}, we have $\forall ~ t < t_1$,
    \begin{equation*}
        m^*(t) = \infty . 
    \end{equation*}
    By the dynamic programming principle and the definition of $m^*(t)$, we have
    \begin{align*}
         \lim\limits_{m \to \infty}\bar{F}(t, m) &=  
         \lim\limits_{m \to \infty}\tilde{\e}^{t, m}\bigg[\int_{t}^{t_1} e^{\vartheta (u - t)}\big(\Delta (u)  M(u) + 1\big) \D u + \bar{F}\big(t_1, M(t_1)\big)
		\bigg] \\
		&\leq \lim\limits_{m \to \infty}\tilde{\e}^{t, 1}\bigg[\int_{t}^{t_1} e^{\vartheta (u - t)}\big(\Delta (u)  mM(u) + 1\big) \D u\bigg] + \bar{F}(t_1, 0)
		= -\infty,
    \end{align*}
which contradicts $\bar{F} \geq 0$.
Thus the set $\{t: m^*(t) = \infty\}$ is empty, and $m^*(t) < \infty$ holds for all $t \in [0, T_1)$. By \Cref{lem_m2}, $m^*(t)$ is bounded.
\end{proof}
{ \begin{lemma}
$m^*(t)$ is right continuous.
\end{lemma}
}
{ 
\begin{proof}
From \cref{lem_m2}, we see that $m^*(t)$ is right-upper semi-continuous and $m^*(t_0) \geq \liminf\limits_{t \to t_0^+} m^*(t)$.  For fixed $t_0$, there exist a sequence of $t_n > t_0$ and $\liminf\limits_{t \to t_0^+} m^*(t) = \lim\limits_{n \to \infty} m^*(t_n)$. Thus, 
\begin{equation*}
    \bar{F}(t_0, \liminf\limits_{t \to t_0^+} m^*(t)) = \bar{F}(t_0, \lim\limits_{n \to \infty} m^*(t_n)) = \lim\limits_{n \to \infty} \bar{F} (t_0, m^*(t_n)) = 0.
\end{equation*}
By the definition of $m^*(t)$, $m^*(t_0) \leq \liminf\limits_{t \to t_0^+} m^*(t).$ Combining $m^*(t_0) \geq \liminf\limits_{t \to t_0^+} m^*(t)$, we see that $m^*(t)$ is right continuous.
\end{proof}

Before proving the left continuity of $m^*(t)$, we need the following continuity of the value function.
\begin{lemma}\label{lm:f_bar_lip}
For $t_1 < t_2$, we have 
\begin{equation*}
    \bar{F}(t_1, m) - \bar{F}(t_2, m) \leq t_2 - t_1.
\end{equation*}
\end{lemma}
\begin{proof}
As
\begin{equation*}
    \bar{F}(t_1, m) = \tilde{\e}^{t_1, m} \bigg[\int_{t_1}^{\tau^*_{t_1, m}} e^{\vartheta (u - t)}\big(\Delta (u)  M(u) + 1\big) \D u\bigg],
\end{equation*}
$\big(\tau^*_{t_1, m} + t_2 - t_1\big)\wedge T_1$ is an admissible stopping time, and
\begin{align*}
     \bar{F}(t_2, m) &\geq \tilde{\e}^{t_2, m} \bigg[\int_{t_2}^{(\tau^*_{t_1, m} + t_2 - t_1)\wedge T_1} e^{\vartheta (u - t)}\big(\Delta (u)  M(u) + 1\big) \D u\bigg]\\
     &= \tilde{\e}^{t_1, m} \bigg[\int_{t_2}^{\tau^*_{t_1, m} \wedge (T_1 - t_2 + t_1)} e^{\vartheta (u - t)}\big(\Delta (u + t_2 - t_1)  M(u) + 1\big) \D u\bigg]\\
     &\geq \tilde{\e}^{t_1, m} \bigg[\int_{t_2}^{\tau^*_{t_1, m} \wedge (T_1 - t_2 + t_1)} e^{\vartheta (u - t)}\big(\Delta (u)  M(u) + 1\big) \D u\bigg],
\end{align*}
where the last inequality is derived from the fact that $\Delta(u)$ is negative and increasing,  
we have
\begin{align*}
     \bar{F}(t_1, m) - \bar{F}(t_2, m) \leq &\tilde{\e}^{t_1, m} \bigg[\int_{\tau^*_{t_1, m}\wedge(T_1 - t_2 + t_1)}^{\tau^*_{t_1, m}} e^{\vartheta (u - t)}\big(\Delta (u)  M(u) + 1\big) \D u\bigg]\\
     \leq & \tilde{\e}^{t_1, m} \bigg[\int_{\tau^*_{t_1, m}\wedge(T_1 - t_2 + t_1)}^{\tau^*_{t_1, m}} 1 \D u\bigg]\\
     \leq &t_2 - t_1.
\end{align*}
Thus, the proof follows.
\end{proof}
\begin{lemma}
$m^*(t)$ is left continuous.
\end{lemma}
\begin{proof}
From \cref{lem_m2}, we see that $m^*(t)$ is left-lower semi-continuous. 

For fixed $t_0$, assume $\limsup\limits_{t \to t_0^-} m^*(t) > m^*(t_0)$. There exists a $\delta > 0$ such that for any $t \in (t_0 - \delta, t_0)$, we have $m^*(t) > c_0 = \frac{1}{2}\big[\big(\limsup\limits_{t \to t_0^+} m^*(t)\big) + m^*(t_0)\big]$. We see $\bar{F}(t_0, m) = 0$ for $m \in (m^*(t_0), c_0)$.

Denote $m' = \sqrt{m^*(t_0)c_0}, \phi(t) = m' + \sqrt{(t_0 - t)}$, and consider $\bar{F}(t_0, m') - \bar{F}(t, \phi(t))$ . 
\begin{align*}
     \bar{F}(t, \phi(t)) =&\tilde{\e}^{t, \phi(t)}\bigg[\int_{t}^{\tau^*_{t, \phi(t)}\wedge t_0} e^{\vartheta (u - t)}\big(\Delta (u)  M(u) + 1\big) \D u + \chi_{\{\tau^*_{t, \phi(t)} > t_0\}}\bar{F}\big(t_0, M(t_0)\big)\bigg]. 
\end{align*}

As $(\tau^*_{t, m'} + t_0 - t)\wedge (2t_0 - t)$ is a stopping time for problem $\bar{F}(t_0, m)$, we have 
\begin{align*}
    \bar{F}(t_0,m') \geq& \tilde{\e}^{t_0, m'}\bigg[\int_{t_0}^{(\tau^*_{t, m'} + t_0 - t)\wedge (2t_0 - t)} e^{\vartheta (u - t_0)}\big(\Delta (u)  M(u) + 1\big) \D u  \bigg] \\
    &+ \tilde{\e}^{t_0, m'} \bigg[ \chi_{\{\tau^*_{t, \phi(t)} > t_0\}}\bar{F}\big(2t_0 - t, M(2t_0 - t)\big)\bigg]\\
    = & \tilde{\e}^{t, m'}\bigg[\int_{t}^{\tau^*_{t, m'}\wedge t_0} e^{\vartheta (u - t)}\big(\Delta (u + t_0 - t)  M(u) + 1\big) \D u \bigg] \\
    &+ \tilde{\e}^{t, m'}\bigg[\chi_{\{\tau^*_{t, \phi(t)} > t_0\}}\bar{F}\big(2t_0 - t, M(t_0)\big)\bigg].
\end{align*}
Combining the last two inequalities above yields 
\begin{align*}
    \bar{F}(t_0, m') - \bar{F}(t, \phi(t)) \geq & \tilde{\e}\bigg[\int_{t}^{\tau^*_{t, m'}\wedge t_0} e^{\vartheta (u - t)}\big(\Delta (u + t_0 - t)M^{t, m'}(u)- \Delta(u)M^{t, \phi(t)}(u)\big) \D u \bigg] \\
    & - \tilde{\e} \bigg[ \chi_{\{\tau^*_{t, \phi(t)} > t_0\}}\bigg(\bar{F}\big(t_0, M^{t, \phi(t)}(t_0)\big) - \bar{F}\big(2 t_0 - t, M^{t, m'}(t_0)\big)\bigg)\bigg]\\
    =& \tilde{\e}\bigg[\int_{t}^{\tau^*_{t, m'}\wedge t_0} \left(O(t_0 - t)M^{t, m'}(u) + O(\phi(t) - m')\right) \D u \bigg] \\
    &- \tilde{\e} \bigg[ \chi_{\{\tau^*_{t, \phi(t)} > t_0\}}\bigg(\bar{F}\big(t_0, M^{t, \phi(t)}(t_0)\big) - \bar{F}\big(2 t_0 - t, M^{t, m'}(t_0)\big)\bigg)\bigg]\\
    \geq& \tilde{\e}\bigg[\int_{t}^{t_0} \chi_{\{\tau^*_{t, m'} = t_0\}}\left(O(t_0 - t)M^{t, m'}(u) + O(\phi(t) - m')\right) \D u \bigg] \\
    &- \tilde{\e} \bigg[ \chi_{\{\tau^*_{t, \phi(t)} > t_0\}}\bigg(\bar{F}\big(t_0, M^{t, m'}(t_0)\big) - \bar{F}\big(2 t_0 - t, M^{t, m'}(t_0)\big)\bigg)\bigg]\\
    \geq & \tilde{\mathbb{P}} \big(\tau^*_{t, m'} = t_0\big) O\big((t_0 - t)^2 + (\phi(t) - m')(t_0 - t)\big)\\
    &- \tilde{\mathbb{P}}\big(M^{t, m'}(t_0) < m^*(t_0)\big) O(t_0 - t).
\end{align*}
Here, $f(t) = O(g(t))$ means $f(t) = \alpha(t) g(t)$, where $\alpha(t) > 0$ and is bounded. The equality in the last formula is derived from the Lipschitz property. The second inequality in the last formula is derived from $\bar{F}\big(t_0, M^{t, \phi(t)}(t_0)\big) \geq \bar{F}\big(t_0, M^{t, m'}(t_0)\big)$. The final inequality relies on \Cref{lm:f_bar_lip}. 

Consider the composition  formula  $M(u) = M_C(u) + M_D(u)$, where $M_C(\cdot)$ is a continuous process and $M_D(u)$ is the sum of all jumps before time $u$.  It is easy to see that
 \begin{equation*}
\tilde{\mathbb{P}}\bigg[\bigcup\limits_{t \leq u \leq t_0} (M^{t, m'}_D(u) \neq  0)\bigg] = O(t_0 - t).
 \end{equation*}
Then, we have 
\begin{align*}
    \tilde{\mathbb{P}}\big(M^{t, m'}(t_0) < m^*(t_0)\big) &\leq \tilde{\mathbb{P}}\bigg[\bigcup\limits_{t \leq u \leq t_0} (M^{t, m'}_D(u) \neq  0)\bigg] + \tilde{\mathbb{P}} \big(M_C^{t, m'}(t_0) < m^*(t_0)\big) \\
    &\leq O(t_0 - t) + \tilde{\mathbb{P}} \big(M_C^{t, m'}(t_0) < m^*(t_0)\big).
\end{align*}
 
As $\log M_C^{t, m'}(t_0)$ is a normally distributed random variable with variance $O(t_0-t)$, using the result of the tail bound of normal distribution, we get 
\begin{equation*}
    \tilde{\mathbb{P}}\big(\log M_C^{t, m'}(t_0) \notin (m^*(t_0), c_0)\big) = O(\frac{1}{\sqrt{t_0 - t}} e^{-\frac{1}{t_0 - t}}).
\end{equation*}

Combining the two inequalities above, we get 
\begin{align*}
     \tilde{\mathbb{P}}\big(M^{t, m'}(t_0) < m^*(t_0)\big) &\leq O(t_0 - t) + \tilde{\mathbb{P}} \big(M_C^{t, m'}(t_0)
     < m^*(t_0)\big) \\
     &\leq O(t_0 - t) + O(\frac{1}{\sqrt{t_0 - t}} e^{-\frac{1}{t_0 - t}}) \\
     &\leq O(t_0 - t).
\end{align*}
Additionally, using the continuity of $m^*(t)$, we can easily get $\tilde{\mathbb{P}} \big(\tau^*_{t, m'} = t_0\big) = O(1)$.
Thus, 
\begin{align*}
    \bar{F}(t_0, m') - \bar{F}(t, m') \geq & \tilde{\mathbb{P}} \big(\tau^*_{t, m'} = t_0\big) O\big((t_0 - t)^2 + (\phi(t) - m')(t_0 - t)\big) \\
    &- \tilde{\mathbb{P}}\big(M^{t, m'}(t_0) < m^*(t_0)\big) O(t_0 - t) \\
    \geq & O(1)O\big((t_0 - t)^2 + (\phi(t) - m')(t_0 - t)\big) -  O((t_0 - t)^2)\\
    = & O\big((t_0 - t)^2 + (t_0 - t)^{3/2}\big) -  O((t_0 - t)^2).
\end{align*}

Therefore, there exists some time $t$ such that $\bar{F}(t_0, m') - \bar{F}(t, m') > 0$ or $\bar{F}(t_0, m') > 0$. This contradicts to $m' \geq m^*(t_0)$ and the definition of $m^*(\cdot)$, which indicates that the  assumption   $\limsup\limits_{t \to t_0^-} m^*(t) > m^*(t_0)$ is incorrect. For any $t_0$, we have $\limsup\limits_{t \to t_0^-} m^*(t) = m^*(t_0)$, and the left-lower semi-continuous property results in $\liminf\limits_{t \to t_0^-} m^*(t) \geq m^*(t_0)$. Combining these two limits, we see $\lim\limits_{t \to t_0^-} m^*(t) = m^*(t_0)$ ,
 thus we complete the proof.
\end{proof}

\begin{lemma}
    $\lim_{t \to T_1} m^*(t)= -(\Delta (T_1))^{-1}$.
\end{lemma}
\begin{proof}
    We prove the lemma in the following two steps.

\underline{\textbf{Step 1}:} Show $m^*(t) \geq -(\Delta (t))^{-1}$.

\underline{\textbf{Step 2}:} Show $\lim_{t \to T_1} m^*(t) \leq -(\Delta (T_1))^{-1}$.

\begin{proof}[\textbf{Proof of \underline{Step 1}}]

It is sufficient to show $\bar{F}(t, m) > 0$ for $t \in [0, T_1), m \in (0, -(\Delta (t))^{-1})$.

For fixed $(t, m)$, let $\tilde{\tau}_{t, m} = \text{inf}\{t: M^{s, m}(t) \geq -(\Delta (t))^{-1}\}$. We have \begin{equation*}
    \bar{F}(t, m) \geq \tilde{\e}^{t, m} \bigg[\int_{t}^{\tilde{\tau}_{t, m}} e^{\vartheta (u - t)}\big(\Delta (u)  M(u) + 1\big) \D u\bigg] \geq 0.
\end{equation*}
The last inequality holds because $(\Delta (u)  M(u) + 1)\chi_{\{u \leq \tilde{\tau}_{t, m}\}} \geq 0$ (note $\Delta (u)  < 0$).
\end{proof}

\begin{proof}[\textbf{Proof of \underline{Step 2}}]
Suppose  $\lim_{t \to T_1} m^*(t) > -(\Delta (T_1))^{-1}$. 

There exist $\epsilon > 0$ and $\delta > 0$ such that 
\begin{align*}
    m^*(t) &> -(\Delta (T_1))^{-1} + 2\epsilon,\\
     -(\Delta (t))^{-1} &< -(\Delta (T_1))^{-1} + \epsilon,
\end{align*} for $t \in (T_1 - \delta, T_1)$ because of the continuity of $m^*(t)$ and $\Delta (T_1)$.

For $t \in [T_1 - \delta, T_1)$ and $m \in (-(\Delta (T_1))^{-1} + \epsilon, -(\Delta (T_1))^{-1} + 2\epsilon)$, let 
\begin{align*}
    \tilde{\tau}_{t, m} &= \text{inf}\big\{u: M^{t, m}(u) \notin  (-(\Delta (T_1))^{-1} + \epsilon, -(\Delta (T_1))^{-1} + 2\epsilon)\big\},\\
    \bar{\tau}_{t, m} &= \text{inf}\big\{u:M^{t, m}_D(u) \neq 0\big\}.
\end{align*}
We have $\tilde{\tau}_{t, m} \leq \tau^*_{t, m}$ because of $-(\Delta (T_1))^{-1} + 2\epsilon) \leq m^*(t)$. Here $\tau^*_{t, m}$ is the optimal stopping time for $(t, m)$; see \Cref{df:op_stop_time}. Using dynamic programming principle, we have
\begin{equation}\label{ineq:f_bar_t_1_bound_pre}
    \begin{aligned}
    \bar{F}(t, m) &= \tilde{\e}^{t, m} \bigg[\int_{t}^{\tilde{\tau}_{t, m} \wedge \bar{\tau}_{t, m}} e^{\vartheta (u - t)}\big(\Delta (u)  M(u) + 1\big) \D u + \bar{F}(\tilde{\tau}_{t, m} \wedge \bar{\tau}_{t, m}, M(\tilde{\tau}_{t, m} \wedge \bar{\tau}_{t, m}))\bigg] \\
    &\leq \tilde{\e}^{t, m} \bigg[\int_{t}^{\tilde{\tau}_{t, m} \wedge \bar{\tau}_{t, m}}e^{\vartheta (u - t)}\left(-\frac{-(\Delta (T_1))^{-1} + 2\epsilon}{-(\Delta (T_1))^{-1} + \epsilon} + 1\right) \D u + T_1 - \tilde{\tau}_{t, m} \wedge \bar{\tau}_{t, m}\bigg]\\
     &\leq \tilde{\e} \bigg[-e^{\vartheta(T_1 - t)}\frac{\epsilon}{-(\Delta (T_1))^{-1} + \epsilon} \big(\tilde{\tau}_{t, m} \wedge \bar{\tau}_{t, m} - t\big) +  T_1 - \tilde{\tau}_{t, m} \wedge \bar{\tau}_{t, m}\bigg].
\end{aligned}
\end{equation}
The second line is derived by \Cref{lm:f_bar_lip}. Using the fact that $M^{t, m}_{D}$ is a sum of weighted Poisson jump, we have 
\begin{equation}\label{eq:tau_tilde_ineq}
\tilde{\mathbb{P}}\bigg[\bar{\tau}_{t, m} \leq t + u\bigg] \leq \sum_{j = d + 1}^n \lambda_j u.
\end{equation} 
Using the property of Brownian motion, we have
\begin{equation}\label{eq:tau_bar_ineq}
    \tilde{\mathbb{P}}\bigg[\tilde{\tau}_{t, m} = T_1\bigg] \geq 1 - o(T_1 - t).
\end{equation}
Combining \Cref{eq:tau_tilde_ineq,eq:tau_bar_ineq}, we have 
\begin{equation*}
    \tilde{\mathbb{P}}\bigg[ \bar{\tau}_{t, m} \wedge \tilde{\tau}_{t, m} = T_1 \bigg] \geq 1 - o(T_1 - t) - \sum_{j = d + 1}^n \lambda_j (T_1 - t) = 1 - o(T_1 - t) - O(T_1 - t).
\end{equation*}
Using $\tilde{\tau}_{t, m} \wedge \bar{\tau}_{t, m} - t \geq (T_1 - t) \chi_{\{\bar{\tau}_{t, m} \wedge \tilde{\tau}_{t, m} = T_1 \}}$, we have
\begin{equation}\label{ineq:f_bar_t_1_bound_1}
    \tilde{\e} \bigg[-e^{\vartheta(T_1 - t)}\frac{\epsilon}{-(\Delta (T_1))^{-1} + \epsilon} \big(\tilde{\tau}_{t, m} \wedge \bar{\tau}_{t, m} - t\big) \bigg] \leq  -\epsilon (T - t)\big(1 - o(T_1 - t) - O(T_1 - t)\big).
\end{equation}
Using \Cref{eq:tau_bar_ineq},  we have $\tilde{\e} \big[T_1 - \bar{\tau}_{t, m}\big] \leq \frac{1}{2}\sum_{j = d + 1}^n \lambda_j (T_1 - t)^2 = O((T_1 - t)^2)$, which results in
\begin{equation}\label{ineq:f_bar_t_1_bound_2}
    \begin{aligned}
        \tilde{\e} \bigg[T_1 - \tilde{\tau}_{t, m} \wedge \bar{\tau}_{t, m}\bigg] &\leq \tilde{\e} \bigg[T_1 - \tilde{\tau}_{t, m} + T_1 -\bar{\tau}_{t, m}\bigg] \\
        &\leq  \tilde{\e} \bigg[(T_1 - t) \chi_{\{\tilde{\tau}_{t, m} < T_1\}}  \bigg] + O((T_1 - t)^2)\\
        &= O((T_1 - t)^2).
\end{aligned}
\end{equation}
Combining \Cref{ineq:f_bar_t_1_bound_pre,ineq:f_bar_t_1_bound_1,ineq:f_bar_t_1_bound_2}, we have 
\begin{align*}
   \bar{F}(t, m) &=  \tilde{\e} \bigg[-e^{\vartheta(T_1 - t)}\frac{\epsilon}{-(\Delta (T_1))^{-1} + \epsilon} \big(\tilde{\tau}_{t, m} \wedge \bar{\tau}_{t, m} - t\big) +  T_1 - \tilde{\tau}_{t, m} \wedge \bar{\tau}_{t, m}\bigg] \\
   &\leq -\epsilon (T - t)\big(1 - o(T_1 - t) - O(T_1 - t)\big) + O((T_1 - t)^2).
\end{align*}
It is easy to see that there exists a $t$ such that 
\begin{equation*}
    \bar{F}(t, m)\leq -\epsilon (T - t)\big(1 - o(T_1 - t) - O(T_1 - t)\big) + O\big((T_1 - t)^2\big) \leq 0.
\end{equation*}
which contradicts to $m^*(t) > -(\Delta (T_1))^{-1} + 2\epsilon > m$ or $\bar{F}(t, m) > \bar{F}(t, -(\Delta (T_1))^{-1} + 2\epsilon) > \bar{F}(t, m^*(t)) = 0$.
\end{proof}
\end{proof}

\section{Proof of Theorem \ref{taucontinuous}}\label{append_pf_taucontinuous}
Denote $\tau_n=\tau^*_{t_n, x_n}$ and $M_n(u) = M^{t_n, m_n}(u)$.
We prove the theorem in the following two steps.

\underline{\textbf{Step 1}:} Show $\liminf\limits_{n \to \infty}\tau_n \geq \tau_0$.

\underline{\textbf{Step 2}:} Show $\limsup\limits_{n \to \infty}\tau_n \leq \tau_0$.

\begin{proof}[\textbf{Proof of \underline{Step 1}}]
Based on  the definition of optimal stopping time \cref{df:op_stop_time}, we have
\begin{equation*}
    \tau_{t, m}^*  = \inf\big\{s: M^{t, m}(s) \geq m^*(s)\big \}.
\end{equation*}

  Denote $\tau' = \liminf\limits_{n \to \infty}\tau_n$. It is sufficient to prove $\tilde{\Pmeasure}(\{\tau' < \tau_0\}) = 0$.
For any $\omega \in \Omega$, if $\tau'(\omega) = s_0 < \tau_0(\omega)$, then we can infer $M_0(s_0^-)(\omega) = m^*(s_0)$ and $M_0(s)(\omega) < m^*(s)$ for $t \leq s \leq s_0$. Moreover, $M_0$ has a jump at $(s_0, \omega)$ when $s_0 < \tau_0$. Denote  $s_0$ by $t(\omega)$. For other $\omega$, define $t(\omega) = \infty$. Then we obtain the probability measure $\nu$ of $t(\omega)$ and
\begin{align*}
    \tilde{\Pmeasure}(\{\tau' < \tau_0\}) &\leq \int_{0}^{T_1}\tilde{\Pmeasure}(\{ \tau'(\omega) < \tau_0(\omega) | t(\omega) = u\}) \nu(\D u)\\
    &= \int_{0}^{T_1} \tilde{\Pmeasure}(\{ M_0(u)(\omega) < m^*(u) | M_0(u^-)(\omega) = m^*(u)\}) \nu(\D u)\\
    &= 0.
\end{align*}
$M_0(u)(\omega) < M_0(u^-)(\omega)$ means that $M_0(u)(\omega)$ jumps at $(u, \omega)$, and the conditional probability of it is 0. Thus the last equation holds.
\end{proof}
\begin{proof}[\textbf{Proof of \underline{Step 2}}]

We illustrate this step by considering two cases:
    The property is equal to
    \begin{equation*}
        \limsup_{n \to \infty}\tau^*_{\tau_0, M_n(\tau_0)} \leq \tau_0, \text{ a.s.},
    \end{equation*}
    or 
    \begin{equation*}
        \limsup_{n \to \infty}\tau^*_{\tau_0,  m_nM_0(\tau_0)/m_0 } \leq \tau_0, \text{ a.s.}.
    \end{equation*}
    As $M_0(\tau_0)/m_0 = m^*(\tau_0)$, it is sufficient to show 
    \begin{equation*}
        \limsup\limits_{n \to \infty} \tau^*_{t, m_n m^*(t)/ m_0 } \leq t. \text{ a.s.},
    \end{equation*}
    or
        \begin{equation*}
           \limsup_{\epsilon \to 0} \tau^*_{t, m^*(t) - \epsilon } = t. \text{ a.s.}. 
        \end{equation*}
    As $\lim\limits_{\epsilon \to 0^+}\tau^*_{t, m^*(t) - \epsilon} = \tau'$ and $B_k = \{\omega : \tau' > \frac{1}{k} + t \}$,  we only need to prove  $\tilde{\mathbb{P}}[B_k] = 0$. Consider $B_{k, \epsilon} = \{\omega : \tau^*_{t, m^*(t) - \epsilon} > \frac{1}{k} + t \}$. For $\omega \in B_{k ,\epsilon}$, we have
    \begin{equation}\label{m_bd}
    (1 - \frac{\epsilon}{m^*(t)}) M^{t, m^*(t)}(t + s)(\omega) < m^*(t + s)(\omega), \quad \forall s \in (0, \frac{1}{k}).
\end{equation}
For $k$ large enough, using \Cref{lem_m2}, we can infer $m^*(t + s) < \frac{m^*(t)}{1 - 2 L s},\quad \forall s\in (0, \frac{1}{k})$.

Consider the composition $M = M_C + M_D$, where $M_C$ is a continuous process and $M_D$ is the sum of all jumps before $t$.  It is easy to see that
 \begin{equation*}
\tilde{\mathbb{P}}\bigg[\bigcup\limits_{0 \leq s \leq \frac{1}{k}} (M_D(s) \neq  0)\bigg] = O(\frac{1}{k}).
 \end{equation*}

Define $C_{k, \epsilon}$ as the set of $\omega$ satisfying the following constraint
\begin{equation}
    (1 - \frac{\epsilon}{m^*(t)}) M_{C}^{t, m^*(t)}(t + s)(\omega) < m^*(t + s)(\omega), \quad \forall s \in (0, \frac{1}{k}).
\end{equation}
Then 
\begin{equation*}
    \tilde{\mathbb{P}}(B_{k, \epsilon}) \leq \tilde{\mathbb{P}}[C_{k, \epsilon}] + \tilde{\mathbb{P}}\bigg[\bigcup\limits_{0 \leq s\leq\frac{1}{k}} (M_D(s) \neq  0)\bigg] \leq \tilde{\mathbb{P}}[C_{k, \epsilon}] + O(\frac{1}{k}).
\end{equation*}

Define $D_{k, \epsilon}$ as the set of $\omega$ satisfying the following constraint \begin{equation}\label{m_bd3}
    (1 - \frac{\epsilon}{m^*(t)}) M_{C}^{t, m^*(t)}(t + s)(\omega) < \frac{m^*(t)}{1 - 2 L s} (\omega), \quad \forall s \in (0, \frac{1}{k}).
\end{equation}
Then $\tilde{\mathbb{P}}[C_{k, \epsilon}] \leq \tilde{\mathbb{P}}[D_{k, \epsilon}]$. As $\log M_{C}^{t, m^*(t)}(t + s)$ is a Brownian motion,  we can easily prove $\tilde{\mathbb{P}}[D_{k, \epsilon}] = O(\epsilon)$  using Girsanov's theorem. Thus $\tilde{\mathbb{P}}(B_{k, \epsilon}) \leq O(\epsilon) + O(\frac{1}{k})$. Letting $\epsilon \to 0$, we have $\tilde{\mathbb{P}}(B_k) \leq O(\frac{1}{k})$. Besides, $\tilde{\mathbb{P}}(B_{k}) < \tilde{\mathbb{P}}(B_j) \quad \forall j > k$, and $\tilde{\mathbb{P}}(B_k) < O(\frac{1}{j})$. Letting $j \to \infty$, we have $\tilde{\mathbb{P}}(B_k) = 0$. 
\end{proof}

\section{Proof of Theorem \ref{th:F_c_1}}\label{append_pf_th_F_c_1}
 It suffices to show that the value function has continuous partial derivatives across the stopping boundary, that is,
\begin{align}
& \lim _{n \rightarrow \infty} \bar{F}_x\left(t_n, x_n\right)=0, \label{eq:f_x_conver}\\
& \lim _{n \rightarrow \infty} \bar{F}_t\left(t_n, x_n\right)=0,\label{eq:f_t_conver}
\end{align}
for any sequence $\left(t_n,  x_n\right)$ in $\mathcal{C}$ converging to $\left(t_0, x_0\right) \in \partial \mathcal{C}$ as $n \rightarrow \infty$. Fix this sequence and denote $\tau_n=\tau^*_{t_n, x_n}$.

\begin{enumerate}
    \item Convergence of $\bar{F}_x$.
    
    Similar to the derivation of $W_y$ in \Cref{lm_dual_cal}, we have 
\begin{align*}
    \bar{F}_m(t_n, m_n) &= \tilde{\e}^{t_n, 1}\bigg[\int_{t_n}^{\tau_n} e^{\vartheta (u - t_n)}\Delta (u)  M(u) \D u\bigg]\\
    &= \tilde{\e}\bigg[\int_{0}^{T} e^{\vartheta u}\Delta (u + t_n)  M^{0, 1}(u) \chi_{\{u \leq \tau_n - t_n\}} \D u\bigg].
\end{align*}
Using \Cref{taucontinuous}, we have $\lim\limits_{n \to \infty} \tau_n = \tau_0 = 0$. Therefore $\lim\limits_{n \to \infty}\chi_{\{u \leq \tau_n - t_n\}} = 0$.
By the dominated convergence theorem and the boundedness of $\Delta(\cdot)$ as well as \Cref{taucontinuous}, we complete the proof of  \Cref{eq:f_x_conver}.
\item Convergence of $\bar{F}_t$.

 Denote $\tau_{n,\epsilon} = \tau^*_{t_n + \epsilon, x_n}$. 
 Similar to the derivation of $W_y$ in \Cref{lm_dual_cal}, we have
 \begin{align*}
     \bar{F}(t_n - \epsilon, m_n) &\leq \tilde{\e}^{t_n- \epsilon, m_n }\bigg[\int_{t_n - \epsilon}^{\tau_{n} - \epsilon} e^{\vartheta (u - t_n + \epsilon)}\big(\Delta (u)  M(u) + 1\big) \D u\bigg] \\
     &=\tilde{\e}\bigg[\int_{0}^{\tau_{n} -t_n} e^{\vartheta u}\big(\Delta (u + t_n - \epsilon)  M^{0, m_n}(u) + 1\big) \D u\bigg], \\
     \bar{F}(t_n , m_n) - \bar{F}(t_n - \epsilon, m_n)
     &\geq \tilde{\e}\bigg[\int_{0}^{\tau_{n} -t_n} e^{\vartheta u}\big(\Delta (u + t_n)\! - \! \Delta (u + t_n - \epsilon)\big)  M^{0, m_n}(u)\D u\bigg] ,\\
      \bar{F}_t(t_n , m_n) &\geq \tilde{\e}\bigg[\int_{0}^{\tau_{n} -t_n} e^{\vartheta u}\Delta' (u + t_n)   M^{0, m_n}(u)\D u\bigg],
 \end{align*}
 where the last inequality is derived from the dominated convergence theorem. Using the dominated convergence theorem, we have 
 \begin{equation}\label{eq:f_t_geq}
      \lim_{n \to \infty }\bar{F}_t(t_n , m_n) \geq 0.
 \end{equation}
 Using \Cref{lem_m1}, we have 
 \begin{align*}
     \bar{F}_t(t_n, m_n) + L m_n \bar{F}_m(t_n, m_n) \leq 0.
 \end{align*}
As
 \begin{equation*}
     \lim_{n \to \infty }\bar{F}_m(t_n , m_n) = 0,
 \end{equation*}
we have 
  \begin{equation}\label{eq:f_t_leq}
     \lim_{n \to \infty }\bar{F}_t(t_n , m_n) \leq 0.
 \end{equation}
 Combining \Cref{eq:f_t_geq,eq:f_t_leq}, we complete the proof of \Cref{eq:f_t_conver}.
\end{enumerate}
\section{Proof of Theorem \ref{th:integral_eq}}\label{append_pf_th_integral_eq} Letting $\mathcal{K}_n$ be an increasing sequence of compact subsets of $\mathcal{C} \cup \mathcal{D}$ such that $\cup_{n \in \mathbb{N}} \mathcal{K}_n=\mathcal{C} \cup \mathcal{D}$.  For $n$ large enough such that $\frac{1}{n} \leq T_1 - t$, 
define 
\begin{equation*}
    \tau_n=\inf \left\{s \in[0, T_1-t]:\left(t+s,  M^{t, m}(t + s)\right) \notin \mathcal{K}_n\right\} \wedge\big(T-t-\frac{1}{n}\big).
\end{equation*}
We apply a version of It\^o formula in Theorem 2.1 of Cai and Angelis (2023) \cite{cai2023change}. We delay the
verification of the assumptions required until the end of the proof. Define an adjusted process $\widehat{M}(u) \define e^{-2Lu}M(u)$. Here the $L$ is the constant defined in \Cref{lem_m1} and $e^{-2Lu} m^*(u)$ is decreasing with respect to $u$. Denote $\widehat{\mu}_m = -2L + \mu$ to simplify our equation and the corresponding operator $\widehat{A}$ and $\widehat{B}$.

For this adjusted process, the following equations still hold,
\begin{equation}\label{stop7}
	\begin{cases}
	&\widehat{F}(t, \widehat{m}) \define \supp{\tau \in \St} \tilde{\e}^{t,  \widehat{m}}\bigg[\int_{t}^{\tau} e^{\vartheta (u - t)}(\Delta (u)  e^{2Lu} \widehat{M}(u) + 1) \D u\bigg],\\
	&\widehat{F}(t,  \widehat{m}) = \bar{F}(t, e^{2Lt}  \widehat{m}).\\
	\end{cases}\end{equation}
\begin{equation}\label{pde:ad}
		\begin{aligned}
		\text{min}& \big\{\!\!-\!\!\vartheta \widehat{F}(t,  \widehat{m}) \!\!-\!\! D_t \widehat{F}(t,  \widehat{m}) \!\!-\!\! (\Delta (t)e^{2Lu}  \widehat{m} + 1)
		\!\!-\!\!\big[A'(t,  \widehat{m}, D_{m}\widehat{F}, D_{mm}\widehat{F})\!\! +\!\! B'(t,  \widehat{m}, \! \widehat{F})\big],\widehat{F}\big\} = 0.
		\end{aligned}
\end{equation}
Then,
\begin{align*}
    &e^{\vartheta (s  \wedge \tau_n)}\widehat{F}\big(t + s \wedge \tau_n, \widehat{M}^{t,  \widehat{m}}(t + s \wedge \tau_n)\big) \\
    &= \widehat{F}(t ,  \widehat{m}) - \int_{0}^{s \wedge \tau_n} e^{\vartheta u}\big(\Delta(u) e^{2Lu} \widehat{M}^{t,  \widehat{m}}(u) + 1\big) \chi_{\{ \widehat{M}^{t,  \widehat{m}}(u) < e^{-2Lu}m^*(u)\}} \D u \\
    &~+ \int_{0}^{s \wedge \tau_n} \sum_{j = 1}^{d}e^{\vartheta u} \sigma_{m, j} \widehat{M}^{t,  \widehat{m}}(u) \widehat{F}_m (t + u, \widehat{M}^{t,  \widehat{m}}(u)) \D B_j(u) \\
     &~+ \int_{0}^{s \wedge \tau_n}  \sum_{j = d+1}^{n}e^{\vartheta u} \gamma_{m, j}  \widehat{M}^{t,  \widehat{m}}(u) \widehat{F}_m(t + u, \widehat{M}^{t,  \widehat{m}}(u)) \D \Tilde{N}_j(u).
\end{align*}
Taking expectations and applying the optional sampling theorem yield
\begin{align*}
        \widehat{F}(t,  \widehat{m})=&\mathrm{E}^{t,  \widehat{m}}\bigg[\int_0^{\tau \wedge \tau_n} e^{\vartheta u}\big(\Delta(u) e^{2Lu}\widehat{M}(u) + 1\big) \chi_{\{ \widehat{M}(u) < e^{-2Lu}m^*(u)\}} \D u \bigg]\\
    &+\mathrm{E}^{t,  \widehat{m}}\bigg[e^{\vartheta u} \widehat{F}\left(t+\left(\tau \wedge \tau_n\right), \widehat{M}({\tau \wedge \tau_n})\right)\bigg].
\end{align*}
Using the adjusted process and corresponding notes, we have
\begin{equation*}\hspace{-3mm}\bar{F}(t, m)=\mathrm{E}^{t, m}\bigg[\int_0^{\tau \wedge \tau_n} e^{\vartheta u}\big(\Delta(u) M(u) + 1\big) \chi_{\{ M(u) < m^*(u)\}} \D u +e^{\vartheta u} \bar{F}\left(t+\left(\tau \wedge \tau_n\right), M({\tau \wedge \tau_n})\right)\bigg].
\end{equation*}
 $\Delta(u) M(u) + 1$ is bounded when $M(u) < m^*(u)$. Besides, $\bar{F}(t + s, \cdot) \leq T - (t + s) \leq T - t$ is also bounded. Letting $\tau_n \to T - t$ and using  the dominated convergence theorem derive
\begin{equation*}
    \bar{F}(t, m)=\mathrm{E}^{t, m}\bigg[\int_0^{\tau} e^{\vartheta u}\big(\Delta(u) M(u) + 1\big) \chi_{\{ M(u) < m^*(u)\}} \D u +e^{\vartheta u} \bar{F}\left(t+\tau, M_{\tau }\right)\bigg],
\end{equation*}
for any stopping time $\tau \in [0, T_1 - t]$. Since $\tau = T_1-t$ and $m=m^*(t, m)$, this theorem is proved.

It remains to verify the assumption in Theorem 2.1 of Cai and Angelis (2023) 
 \cite{cai2023change}. 

 Identifying $X_t = \widehat{M}(t), b(t) = e^{-2Lu}m^*(t)$, $U(t,  \widehat{m}) = \widehat{F}(t,  \widehat{m})$.

 Assumption A.1 : $\beta^{11}(t,  \widehat{m}) = \frac{1}{2}\big(\sum_{j = 1}^{d}\sigma_{m, j}^2\big)  \widehat{m}^2$ is locally Lipschitz. 
 
 $\mathrm{P}\left(\left(t, \boldsymbol{X}_{t-}\right) \in \partial \mathcal{C}\right)=0$ can be derived from the continuity of $m^*(t)$ proved by \Cref{th:m_conti}. 

 Assumption A.2 : The smoothness of $\widehat{F}(t,  \widehat{m})$ ($\bar{F}(t, m)$) is proved by \Cref{th:F_c_1}. Using \Cref{pde:ad}, we have 
 \begin{align*}
     \bigg(\frac{1}{2} \sum_{j=1}^d \sigma_{m, j}^2\bigg)  \widehat{m}^2\widehat{F}_{mm}(t,  \widehat{m}) = &-\vartheta \widehat{F}(t,  \widehat{m}) - D_t \widehat{F}(t,  \widehat{m}) - (\Delta (t)e^{2Lu}  \widehat{m} + 1)\\
		&-\big[\mu'_m  \widehat{m} D_m \bar{F}(t, m) + B'(t,  \widehat{m},  \widehat{F})\big].
 \end{align*}
Each component on the right side is continuous. Thus, the right side is bounded in the compact set $\mathcal{K}_n$.

 Assumption A.3 : From the decreasing property of boundary $e^{-2Lu}m^*(u)$ with respect to $u$, we see that $b(t, \cdot)$ and $b(\cdot,  \widehat{m})$ are  monotonic. 
 } 
\section{Proof of results in Section \ref{section_primal}}\label{append_c}
\begin{proof}[\textbf{Proof of \Cref{lm_bd_pr}}]
The optimality of $\tilde{\tau}_y$ for Problem (\ref{stop6}) has been established. Additionally, based on the findings in \Cref{lm_mc}, it is evident that $\hat{\tau}_y$ is also optimal for Problem (\ref{stop1}). Subsequently, it is essential to demonstrate the continuity, as highlighted in \Cref{th_dual}, which has been proven in \Cref{taucontinuous}.

{  Notice that
\begin{align*}
    \e [b(\hat{\tau}_{y}) \xi (\hat{\tau}_{y})] &= \e_{\mathbb{Q}}  [e^{-r\hat{\tau}_{y}}b(\hat{\tau}_{y})],\\
    \lim_{y \to \infty} \hat{\tau}_{y}& = T_1,\\
    b(T_1) &= 0.
\end{align*}
 Using the dominated convergence theorem, we have 
\begin{equation*}
    \lim_{y \to \infty}  \e [b(\hat{\tau}_{y}) \xi (\hat{\tau}_{y})] =  \lim_{y \to \infty} \e_{\mathbb{Q}}  [b(\hat{\tau}_{y})] = 0 .
\end{equation*}
Thus, \Cref{eq:e_b_limit} holds.}
\end{proof}
\begin{proof}[\textbf{Proof of \Cref{lm_primal_map}}]
 Based on \Cref{th_dual} and the convexity of $W$, we have
 \begin{equation*}
     V(t, x, h, w) = (x - p^T(t) h + q(t) w) y + W(t, y, w), y = \mathcal{P}^{-1}_{t, w}(x - p^T(t) h + q(t) w).
 \end{equation*}
 If $y \in \mathcal{C}^d_{t,w}$, then
 \begin{align*}
 (x - p^T(t)h+ q(t) w) y^* + W(t, y^*, w) &>  (x + q(t) w) y^* + V^\mathrm{d} (t,y^*) - q(t) y^*w\\
&= y ^ *(x - p^T(t)h) + V^\mathrm{d}(t, y^*)\\
&= V_{F, p}(t, x),
 \end{align*}
which shows  $(x, h) \in \mathcal{C}^p_{t, w}.$ As such, if $(x, h) \in \hat{\mathcal{C}^p_{t, w}}$, it is easy to see that $\mathcal{P}^{-1}_{t, w}(x - p^T(t) h + q(t) w) \in \mathcal{C}^p_{t, w}$. Thus, $(x, h) \in \mathcal{C}^p_{t,w}$ and $\mathcal{C}^p_{t,w} \subset \mathcal{C}^p_{t,w}$.

On the other hand, if $(x, h) \in \mathcal{C}^p_{t,w}$, then for $y^* = \mathcal{P}^{-1}_{t,w}(x - p^T(t)h + q(t)w)$, based on the last part of proof in \Cref{lm_dual_cal}, we know that the optimal stopping time $\hat{\tau}_{y^*}$ for $W$ is also optimal for $V$. As such, $\hat{\tau}_{y^*} > t$, which implies $y^* \in \mathcal{C}^d_{t,w}$. Therefore $x - p^T(t)h + q(t)w = \mathcal{P}_{t,w} y^* \in \mathcal{P}_{t,w} \mathcal{C}^{d}_{t, w}$ and the proof is completed. 
\end{proof}
\begin{proof}[\textbf{Proof of \Cref{th_bd_pr}}]
    Based on the proof of \Cref{lm_dual_cal}, we know $\mathcal{P}_{t,w}\R_+ = (0, \infty)$. As such, if $(x, h, w) \in \G_t$, then $x - p^T(t)h + q(t)w \notin \mathcal{P}_{t,w}\mathcal{C}^d_{t,w}$ if and only if $x - p^T(t)h +q(t)w \in \mathcal{P}_{t,w}(\R_+ \backslash \mathcal{C}^d_{
t,w}).$ 
\end{proof}
\begin{proof}[\textbf{Proof of \Cref{th_op_st}}]
\quad
\vspace{6pt}
 Let $\hat{\tau}_{t, y, w}$ be as in \Cref{th_bd_pr}. Based on the proof of \Cref{th_dual}, defining $c^*(u) = I_{U_1}(u, Y^{t, y}(u))$ and $B^* = I_{V_{F, p}}(\hat{\tau}_{t, y, w}, Y^{t, y}(\hat{\tau}_{t, y, w})) + b(\hat{\tau}_{t, y, w})$, (\ref{op_c}) is directly verified.

Next, we prove (\ref{op_pi}). Considering $(s, y', w') \in \mathcal{C}_p$, given $\big\{s < \hat{\tau}_{t, y, w}, Y^{t, y}(s) = y', \mathcal{W}^{t, w}(s)\\ = w'\big\}$, $\hat{\tau}_{s, y', w'}$ and $\hat{\tau}_{t, y, w}$ have the same distribution. Furthermore, utilizing the Markovian property of $Y$, we have
{\small
\begin{align*}
     \e&\bigg[e^{-\rho(\hat{\tau}_{t, y, w} - t)}Y^{t, y}(\hat{\tau}_{t, y, w})I_{V_{F, p}}(\hat{\tau}_{t, y, w}, Y^{t, y}(\hat{\tau}_{t, y, w})) \\
    &+ \e^{-\rho(\hat{\tau}_{t, y, w} - t)}Y^{t, y}(\hat{\tau}_{t, y, w})q(\hat{\tau}_{t, y, w})\mathcal{W}(\hat{\tau}_{t, y, w}) \\
     &+ \int_{s}^{\hat{\tau}_{t, y, w}}e^{-\rho(\hat{\tau}_{t, y, w} - t)}Y^{t, y}(u)I_{U_1}(u, Y^{t, y}(u)) \D u \bigg| Y^{t, y}(s) = y', \mathcal{W}^{t, w}(s) = w'\bigg]\\
    =&\e \bigg[e^{-\rho(\hat{\tau}_{s, y', w'} - t)}Y^{s, y}(\hat{\tau}_{s, y', w'})I_{V_{F, p}}(\hat{\tau}_{s, y', w'}, Y^{s, y'}(\hat{\tau}_{s, y', w'})) \\
    &+ e^{-\rho (\hat{\tau}_{s, y', w'} - t)}Y^{s, y'}(\hat{\tau}_{s, y', w'})q(\hat{\tau}_{s, y', w'})\mathcal{W}^{s, w'}(\hat{\tau}_{s, y', w'})\\
    &+ \int_{s}^{\hat{\tau}_{s, y', w'}} e^{-\rho (\hat{\tau}_{s, y', w'} - t)}Y^{s, y'}(u)I_{U_1}(u, Y^{s, y'}(u)) \D u\bigg]\\
    =& -W_{y}(s, y', w')y'.
\end{align*}
}
Integrating with respect to the distribution of $Y^{t, y}(s)$ and $\mathcal{W}^{t, w}(s)$, for $s \in [t, \hat{\tau}_{s, y, w})$, we have
\begin{align*}
    \e_s[&e^{-\rho(\hat{\tau}_{s, y, w} - t)}Y^{t,y }(\hat{\tau}_{s, y, w})I_{V_{F, p}}(\hat{\tau}_{s, y, w}, Y^{t, y}(\hat{\tau}_{s, y, w})) \\
    &+ e^{-\rho(\hat{\tau}_{s, y, w} - t)}Y^{t, y}(\hat{\tau}_{s, y, w})q(\hat{\tau}_{s, y, w})\mathcal{W}^{t, w}(\hat{\tau}_{s, y, w})\\
    &+ \int_{s}^{\hat{\tau}_{s, y, w}} e^{-\rho (\hat{\tau}_{s, y, w} - t)} Y^{t, y}(u) I_{U_1}(u, Y^{t, y}(u)) \D u]\\
    =& -W_y(s, Y^{t, y}(s), \mathcal{W}^{t, w}(s))Y^{t, y}(s).
\end{align*}
Combining the last equation with \Cref{lm_comp}, we have 
\begin{equation*}
    \xi^t(s)(X_F(s) + q(s)\mathcal{W}^{t, w}(s)) = - W_y(s, Y^{t, y}(s), \mathcal{W}^{t, w}(s)) \xi^t(s).
\end{equation*}
Dividing both sides of the last equation by $\xi^t$, taking the differential, and applying It\^o's formula, we can derive the stochastic differential equation of $\xi^t(s)(X_F(s)+q(s) \mathcal{W}_{t,w})$. Comparing the coefficients of diffusion terms in this SDE with those obtained from (\ref{eq_x}), we obtain the following
\begin{align*}
    \pi^* (s) = \sigma_M^{-1}(-\sigma_{w,M} \mathcal{W}^{t, w}(s) - D(W_y)(t, Y^{t, y}(s), \mathcal{W}^{t, w}(s))), 
\end{align*}
which shows that the feed-back form of $\pi^*$ in $(y, w)$ is
\begin{equation*}
    \Pi^*(t, y, w) = \sigma_M^{-1}(-\sigma_{w,M} w - D(W_y)(t, y, w)).
\end{equation*}

\end{proof}

\bibliographystyle{siamplain}
\bibliography{references}

\end{document}